# Thermoreflectance-Based Techniques for Micro- and Nanoscale Thermophysical Property Measurements: Principles, Methods, and Recent Advances

Puqing Jiang[1*], Shanshan Chen[2], Xiaobo Li[1*], Ronggui Yang[1,3*]

1. School of Energy and Power Engineering, Huazhong University of Science and Technology, Wuhan 430074, China;
2. School of Physics, Renmin University of China, Beijing 100872, China;
3. Department of Energy and Resources Engineering, College of Engineering, Peking University, Beijing 100871, China

* E-mail: jpq2021@hust.edu.cn (P.J); xbli35@hust.edu.cn (X.L); ronggui@pku.edu.cn (R.Y)

**Abstract** As the characteristic dimensions of semiconductor devices continue to approach the nanoscale, heat transport and thermal energy storage behavior in materials and interfaces exhibit increasingly pronounced dependencies on length scale, structure, and interfacial properties. Accurate characterization of key thermophysical properties, including thermal conductivity, interfacial thermal conductance, and volumetric heat capacity, is therefore essential for the rational design of functional materials and the thermal management of advanced electronic devices. Thermoreflectance techniques based on photothermal excitation combine noncontact operation, high spatial and temporal resolution, and a broad measurement range, and have emerged as powerful tools for micro- and nanoscale thermophysical characterization. This review systematically surveys the physical principles and technical characteristics of transient thermoreflectance (TTR), time-domain thermoreflectance (TDTR), frequency-domain thermoreflectance (FDTR), steady-state thermoreflectance (SSTR), spatial-domain thermoreflectance (SDTR), and the square-pulsed source (SPS) method. Within a unified heat-diffusion framework, these techniques are compared with respect to parameter sensitivity, measurement uncertainty, and applicable measurement regimes. Representative applications are further used to delineate the conditions under which thermoreflectance techniques are suitable for characterizing low-thermal-conductivity materials, anisotropic films, crystalline materials, and multilayer heterostructures, as well as to elucidate the complementary roles of different methods in quantifying thin-film thermal conductivity, interfacial thermal conductance, and in-plane/cross-plane heat transport. Finally, emerging directions are discussed, including ultrahigh spatiotemporal resolution, multiphysics coupling, in-line industrial inspection, intelligent data processing, and natural-language-driven analysis, highlighting opportunities to advance both fundamental studies of micro- and nanoscale heat transport and thermal management technologies for advanced devices.

---

With the rapid development of functional materials and high-performance electronic devices, accurate characterization of micro- and nanoscale heat transport has become essential for elucidating fundamental heat-transport mechanisms, optimizing device thermal management, and assessing operational reliability [1, 2]. At the material level, heat transport in low-dimensional materials, anisotropic crystals, polymer films, multilayer heterostructures, and thermally functional materials commonly exhibits pronounced dependence on characteristic length scales, interfaces, and microstructure. Transport parameters such as thermal conductivity ($k$) and interfacial thermal conductance ($G$) are strongly influenced by boundary and interface scattering, crystallographic orientation, and processing conditions; volumetric heat capacity ($C$) may also vary with composition, phase state, and crystallinity. Consequently, conventional bulk thermophysical properties often do not adequately represent the actual thermal response of micro- and nanostructures [3-5]. At the device level, advances in high-power and high-frequency electronics, including 5G communications and gallium nitride (GaN) high-electron-mobility transistors (HEMTs), have driven local power densities and heat fluxes to ever higher levels. Channel hot spots, thin-film thermal resistance, and thermal resistance at heterogeneous interfaces have consequently become important constraints on device performance and lifetime [6]. These effects are particularly pronounced when the characteristic dimensions of a material or device become comparable to phonon mean free paths, such that boundary scattering, interfacial thermal resistance, and anisotropic heat diffusion become increasingly significant [7, 8]. Under such conditions, conventional macroscopic measurement techniques and homogeneous-medium assumptions become inadequate, imposing increasingly stringent requirements on the accuracy and spatial resolution of micro- and nanoscale thermophysical-property measurements [1, 9].

Compared with macroscopic measurements, experimental determination of micro- and nanoscale thermophysical properties is not simply a matter of reducing the measurement dimensions; rather, it introduces simultaneous constraints associated with spatial resolution, temporal resolution, and coupled multiparameter inversion. First, temperature and heat-flux fields in micro- and nanoscale materials and devices are often highly localized. Hot spots may be concentrated within thin films, heterogeneous interfaces, transistor channels, or microscopic defects, requiring micrometer or even submicrometer spatial resolution to resolve the local thermal response of the region of interest. Second, processes such as phonon relaxation, interfacial energy transfer, and transient heat diffusion in thin films span a broad range of timescales, from picoseconds to milliseconds, requiring sufficient temporal resolution and dynamic range. More importantly, the experimental signal depends not only on unknown thermal conductivity, volumetric heat capacity, and interfacial thermal conductance, but also on structural and experimental parameters such as film thickness and laser-spot size. Strongly correlated parameter sensitivities can render the inverse problem ill-posed, amplifying the effects of experimental noise, model simplifications, and errors in independently determined input parameters, thereby degrading measurement accuracy and reliability [10, 11]. There is therefore a strong need for micro- and nanoscale thermophysical measurement techniques that combine high spatiotemporal resolution, noncontact detection, and robust parameter-estimation frameworks to support fundamental investigations of heat transport in complex materials and the thermal design of advanced electronic devices.

Given these measurement challenges, existing thermophysical characterization methods remain subject to substantial limitations at the micro- and nanoscale [12]. Among steady-state methods, the guarded-hot-plate approach is well established and physically intuitive, but its requirements on sample size and geometry make it unsuitable for local thermophysical measurements. Microbridge methods can characterize heat transport in micro- and nanostructures, but are susceptible to contact thermal resistance, edge heat losses, and parasitic heat flow through sensor leads [13]. The transient plane source (TPS) method is highly reliable for bulk materials, yet its thermal penetration depth is coupled to sensor dimensions, and its spatial resolution is typically limited to the millimeter scale, making it difficult to characterize micrometer- or submicrometer-scale films, interfaces, and local hot spots [14]. Moreover, for low-dimensional materials, anisotropic films, and multilayer heterostructures, contact-based methods and techniques requiring integrated sensors may struggle to distinguish in-plane and cross-plane heat-transport contributions or to resolve interlayer thermal boundary conductance without perturbing the system under study. Precise micro- and nanoscale thermophysical measurements therefore require characterization techniques capable of combining localized excitation, noncontact detection, and highly sensitive temperature readout.

Against this background, thermoreflectance (TR), a photothermal technique based on optical excitation and detection, has emerged as a powerful tool for micro- and nanoscale thermophysical characterization [15]. A pump beam produces a localized temperature rise at the sample surface or within a metallic transducer layer, while a probe beam detects the small reflectivity change induced by the resulting temperature variation. When interpreted using an appropriate heat-diffusion model, the measured thermoreflectance response provides access to heat-transport dynamics and thermophysical properties. By adjusting experimental parameters such as laser-spot size, modulation frequency, or pump-probe delay time, different thermoreflectance techniques can achieve micrometer-scale or submicrometer spatial resolution and collectively span a broad temporal range from picoseconds to seconds, making them well suited for characterizing heat transport in thin films, heterogeneous interfaces, multilayer structures, and anisotropic materials. After decades of development and methodological refinement [16-18], thermoreflectance has diversified into several distinct approaches (Fig. 1), including transient thermoreflectance (TTR), time-domain thermoreflectance (TDTR), frequency-domain thermoreflectance (FDTR), steady-state thermoreflectance (SSTR), spatial-domain thermoreflectance (SDTR), and the square-pulsed source (SPS) method. These approaches differ in their excitation schemes, signal-acquisition domains, and parameter-inversion strategies and thus offer complementary capabilities across different material systems, spatiotemporal regimes, and thermophysical quantities of interest.

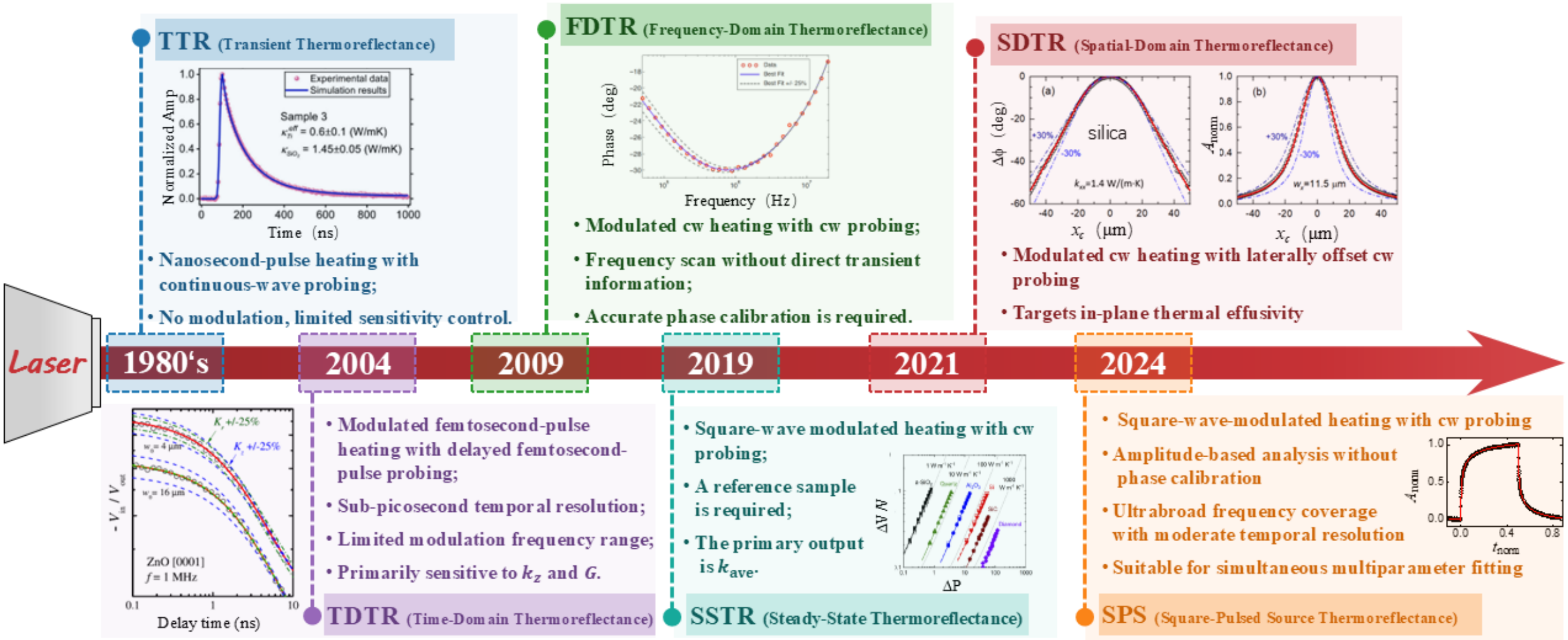

**Figure 1 Development timeline and mainstream branches of thermoreflectance measurement**

Modern pump-probe thermoreflectance can be traced back to the 1980s. In 1986, Paddock and Eesley used a picosecond pulsed-laser pump-probe system with a mechanical delay stage to quantitatively measure the thermal diffusivity of metal films using transient thermoreflectance, establishing the experimental basis for TTR-based thin-film thermophysical characterization [19]. Subsequently, the Maris group conducted a series of studies on picosecond ultrasonics, coherent phonons, and ultrafast pump-probe processes, thereby extending the technique to the characterization of thin-film acoustics and heat transport phenomena [20-26]. Kading et al. further extended transient photothermal measurements to the characterization of heat transport in dielectric films and multilayer structures [27]. Building on these developments, Cahill and co-workers systematically established an axisymmetric heat-diffusion model for multilayer structures under Gaussian-beam heating and integrated this framework with femtosecond pump-probe optics, pump modulation, and lock-in detection. Together, these advances established the experimental and analytical framework for TDTR [28] and enabled increasingly mature quantitative measurements of thin-film thermal conductivity and interfacial thermal conductance [29-37].

As research targets expanded from relatively simple single-layer films to low-thermal-conductivity materials, anisotropic crystals, and complex multilayer heterostructures, the limitations of relying on a single experimental variable for achieving sufficient parameter sensitivity, extending the measurable range, and decoupling multiple thermophysical parameters became increasingly apparent. These challenges motivated the development of frequency-domain, steady-state, and spatial-domain thermoreflectance implementations, as well as strategies that combine information from multiple measurement domains. FDTR varies the pump modulation frequency to tune the thermal penetration depth and redistribute sensitivity among different thermophysical parameters, thereby providing frequency-domain information for parameter identification in multilayer structures [10]. SSTR measures the steady-state temperature rise under continuous or low-frequency heating and enables thermophysical characterization over a broad range of thermal conductivities [38]. Spatially scanned thermoreflectance and beam-offset methods introduce the relative lateral position of the pump and probe beams as an additional measurement variable, thereby enhancing sensitivity to in-plane heat diffusion and thermal-conductivity anisotropy [39]. More recently, SPS has employed square-wave modulation and full-period temperature-waveform analysis, utilizing the thermal responses during both the heating and cooling portions of each cycle to introduce an additional measurement dimension for the simultaneous estimation of thermal conductivity, volumetric heat capacity, and related interfacial parameters. Its applications have subsequently been extended to low-thermal-conductivity materials and multilayer heterostructures [40-43].

Although thermoreflectance has been widely used for the quantitative characterization of micro- and nanoscale thermophysical properties, most existing reviews and studies focus on a single technique or a specific class of materials [44-46]. A comprehensive comparison of the underlying physical connections, performance limits, and regimes of applicability among the different thermoreflectance methods remains limited. Because the various thermoreflectance approaches differ markedly in excitation mode, experimental configuration, and signal-acquisition format, they are often treated as independent techniques. At the fundamental level, however, they are governed by the same sequence of physical processes: photothermal excitation generates a temperature field in the sample; the temperature-induced reflectivity change provides the thermal signal; and a heat-diffusion model is then used to infer the relevant thermophysical properties. Their principal differences lie in the choice of experimental observable or control variable—such as pump–probe delay time, modulation frequency, steady-state temperature rise, or spatial position—as well as in the corresponding signal-processing and parameter-estimation strategies. Without a unified heat-diffusion framework, researchers working with complex systems such as multilayer films, heterogeneous interfaces, low-thermal-conductivity materials, and anisotropic crystals may find it difficult to identify the most appropriate measurement approach and experimental conditions, or to quantitatively assess parameter sensitivity, inversion stability, and measurement uncertainty.

Motivated by these considerations, this review begins with the thermoreflectance effect and the governing heat-diffusion equations and provides a systematic examination of the physical principles, experimental configurations, and parameter-sensitivity characteristics of representative thermoreflectance techniques, including TTR, TDTR, FDTR, SSTR, SDTR, and SPS. Their common physical basis and distinguishing characteristics are then established within a unified heat-diffusion framework. The methods are systematically compared from four complementary perspectives: characteristic heat-diffusion scales, observables and control variables, capabilities for resolving in-plane and cross-plane heat transport, and the stability of joint multiparameter inversion. Representative examples involving low-thermal-conductivity polymers, anisotropic films, crystalline materials, and GaN/Si heterostructures are used to illustrate the capabilities and limitations of different thermoreflectance approaches for characterizing complex materials and devices. Finally, emerging opportunities and challenges are considered, including higher spatiotemporal resolution, in situ multiphysics characterization, rapid parameter estimation, industrial in-line nondestructive inspection, natural-language-driven intelligent data analysis, and adaptive experimentation, with the broader goal of advancing both fundamental micro- and nanoscale heat-transport studies and thermal-management strategies for next-generation microelectronic devices.

# 1 Physical Foundations and Theoretical Framework

## 1.1 Thermoreflectance Effect and Optical Thermometry

The fundamental physical basis of thermoreflectance measurement is the temperature dependence of the optical reflectivity of a material surface. From an optical perspective, a temperature change modifies the complex refractive index and, consequently, the optical response of the material, thereby changing its surface reflectivity. By expanding the reflectivity ($R(T)$) in a Taylor series about a reference temperature ($T_0$) and neglecting second- and higher-order terms, the relative reflectivity change can be written as [47]:

$$\frac{\Delta R}{R} = C_{\mathrm{TR}}\,\Delta T \tag{1}$$

where $R$ is the material reflectivity, $\Delta R$ is the temperature-induced change in reflectivity, $\Delta T$ is the temperature change in the probed region, and $C_{\mathrm{TR}}$ is the thermoreflectance coefficient evaluated at the reference temperature $T_0$, with units of $\mathrm{K}^{-1}$. Both the magnitude and sign of $C_{\mathrm{TR}}$ depend not only on the material itself but also on the optical and structural conditions of the measurement, including probe wavelength, incidence angle, polarization, film thickness, and microstructure. For commonly used metallic transducers such as Al and Au, $|C_{\mathrm{TR}}|$ is typically on the order of $10^{-5}$ to $10^{-4}$ $\mathrm{K}^{-1}$, meaning that a 1 K temperature change produces only a relative reflectivity change of approximately $10^{-5}$ to $10^{-4}$ [48-50]. Because thermoreflectance signals are inherently small, lock-in amplification, balanced detection, and periodic ensemble averaging are commonly used to improve the signal-to-noise ratio. Under optimized detection bandwidth and integration conditions, equivalent temperature resolutions at the millikelvin level or better can be achieved [51].

In practice, a metallic film with a thickness ranging from several tens of nanometers to approximately 100 nm is commonly deposited on the sample surface to serve as an optical transducer. This metal layer performs two principal functions: photothermal excitation and optical thermometry. It absorbs the pump beam and thereby generates a controlled, localized heat source, while its relatively strong thermoreflectance response converts the resulting temperature variation into a measurable optical signal. A suitably selected transducer can also reduce complications arising from the intrinsic optical properties of the underlying sample, thereby simplifying interpretation of the measured signal within a multilayer heat-diffusion framework.

The transducer material, thickness, and probe wavelength should therefore be selected in a coordinated manner according to the optical and thermal requirements of the experiment. The probe wavelength is generally chosen to provide a relatively large thermoreflectance coefficient while maintaining sufficient optical reflectivity and detection sensitivity. Common configurations include a 532 nm probe with an Au transducer and a 785 nm probe with an Al transducer, although the optimal combination depends on the specific material system and experimental configuration. The transducer should be sufficiently thick to be effectively optically opaque at the pump and probe wavelengths, thereby suppressing parasitic reflections from underlying layers, but sufficiently thin to minimize perturbations arising from its own heat capacity and lateral heat spreading. Consequently, metallic transducers with thicknesses of approximately 50–100 nm are widely used, although the optimal thickness should be determined by considering the optical absorption depth, transducer/sample interfacial properties, sensitivity to the target thermophysical parameters, and experimental signal-to-noise ratio.

It is important to note that thermoreflectance measurements employing a metallic transducer do not directly measure the surface temperature of the uncoated material. Rather, they probe the temperature response of the transducer, which is thermally coupled to the underlying structure. The measured signal therefore represents the combined thermal response of the multilayer system comprising the transducer, the target film, intervening interfaces, and the substrate. Although the transducer enhances measurement sensitivity, reproducibility, and applicability across diverse material systems, it also introduces additional thermal mass, lateral heat conduction, and an additional thermal interface. Accordingly, quantitative parameter estimation must account for the transducer thickness and thermophysical properties, together with the thermal conductance of the transducer/sample interface. Uncertainties in these quantities should also be incorporated into sensitivity and uncertainty analyses because they can propagate directly into the extracted thermophysical properties.

### 1.2 Governing Heat-Diffusion Equation

Thermoreflectance measurements commonly use modulated or pulsed laser irradiation to generate localized photothermal heating. For samples coated with a metallic transducer, the pump energy is absorbed primarily by the transducer and converted into heat, giving rise to a spatially and temporally varying temperature field near the sample surface. In most micro- and nanoscale thermoreflectance experiments, provided that the relevant thermal length scales remain sufficiently larger than the characteristic nonequilibrium transport lengths of the dominant heat-carrying phonons, heat conduction within the sample can be described using the classical Fourier heat-diffusion model [28]. For an anisotropic layered system heated by an axisymmetric pump spot, the governing equation for each layer can be written in cylindrical coordinates as:

$$C\frac{\partial T}{\partial t}=\frac{k_r}{r}\frac{\partial}{\partial r}\left(r\frac{\partial T}{\partial r}\right)+k_z\frac{\partial^2 T}{\partial z^2} \tag{2}$$

where $T$ is the temperature rise relative to the initial temperature, $C$ is the volumetric heat capacity, and $k_r$ and $k_z$ are the in-plane (radial) and cross-plane (axial) thermal conductivities, respectively. For an isotropic material, $k_r = k_z = k$.

In most thermoreflectance models, optical absorption in the metallic transducer can be represented either by a surface heat-flux boundary condition or by a volumetric heat-generation term distributed over a finite optical absorption depth, depending on the transducer thickness and optical absorption depth. For a sample whose dimensions are much larger than the thermally perturbed region, the substrate is commonly treated as semi-infinite, with far-field boundary conditions requiring the temperature rise to vanish sufficiently far from the heated region: $T(r, z \to \infty, t) \to 0,\ T(r \to \infty, z, t) \to 0$.

For an ideally bonded interface, both temperature and normal heat flux are continuous across the interface. When a finite interfacial thermal resistance is present, the heat flux remains continuous, but a temperature discontinuity is allowed; the relation can be written as:

$$q = G(T_1 - T_2) \tag{3}$$

where $G$ is the interfacial thermal conductance, and $T_1$ and $T_2$ are the temperatures on the two sides of the interface. Equivalently, the temperature discontinuity across the interface is proportional to the transmitted heat flux through the inverse conductance $(1/G)$. For a multilayer structure consisting of a metallic transducer, the film under test, and a substrate, the layer-specific heat-diffusion equations, together with the associated interfacial and boundary conditions, define the forward thermal model from which thermoreflectance signals are calculated.

The governing equations above also make clear that the measured thermoreflectance response generally depends on multiple coupled quantities rather than on a single thermophysical parameter. In addition to thermal conductivity, volumetric heat capacity, and interfacial thermal conductance, the response is influenced by structural and experimental parameters such as film thickness and laser-spot size. The degree to which each parameter affects the measured signal depends on the characteristic time, frequency, and spatial scales probed by the experiment. All thermoreflectance techniques excite and observe this controlled heat-diffusion process; they differ primarily in the temporal or spatial form of the heat input, the measured observable, and the domain in which the signal is acquired and analyzed. Thus, the various techniques can be understood within a unified heat-diffusion framework as different experimental projections of the same underlying thermal transport problem in the time, frequency, and spatial domains, or in combinations of these domains.

### 1.3 Thermal Penetration Depth and Characteristic Length Scales

To quantify the effective heat-diffusion region in different thermoreflectance measurements and to identify the spatial regions over which the measured signal is sensitive to specific thermophysical properties, the thermal penetration depth is introduced as a central characteristic length scale for describing the spatial decay of a periodically driven thermal disturbance [52]. For a periodically modulated heat source under the Fourier-diffusion approximation, the characteristic penetration depth of the thermal disturbance can be expressed as:

$$d_p = \sqrt{\frac{k}{\pi f C}} \tag{4}$$

where $d_p$ is the thermal penetration depth, $k$ is the material thermal conductivity, $C$ is the volumetric heat capacity, and $f$ is the pump modulation frequency. For a layered anisotropic system, the appropriate directional thermal conductivity should be used; for example, $k_z$ governs the characteristic penetration depth associated with cross-plane diffusion. Equation (4) shows that the thermal penetration depth depends on both intrinsic thermophysical properties and the experimental modulation frequency. The quantities $k$ and $C$ are material properties, whereas f is an adjustable experimental variable, and $d_p$ scales as $f^{-1/2}$. Increasing the modulation frequency therefore decreases the thermal penetration depth and localizes the periodically varying thermal field more strongly near the heated surface.

It should be emphasized that $d_p$ in Eq. (4) represents a characteristic decay length of the thermal field rather than a sharply defined boundary of the heated region. The material volume and heat-flow pathways contributing to the actual thermoreflectance signal are therefore not determined solely by the absolute magnitude of $d_p$, but also by its relation to the sample thickness and laser-spot size. Although film thickness and spot size do not appear explicitly in Eq. (4), they provide geometric length scales that constrain cross-plane and lateral heat diffusion, respectively. In thermoreflectance experiments, it is therefore more meaningful to compare $d_p$ with the film thickness h and with the pump-spot diameter $2r_0$. The former indicates the extent to which the thermal disturbance samples successive layers and interfaces in the cross-plane direction, whereas the latter governs the relative importance of cross-plane and radial heat spreading.

First, for a multilayer film stack, comparison of the thermal penetration depth $d_p$ with the target film thickness $h$ provides a useful estimate of the depth over which the periodically driven temperature field samples the structure. When $d_p \ll h$, the disturbance is largely confined to the near-surface region of the target layer and has limited interaction with the underlying interface, so the measured signal mainly reflects the thermal response of the top layer. When $d_p \sim h$, the disturbance reaches the film/substrate interface, and the interfacial thermal conductance and thermal-transport properties of lower layers begin to make appreciable contributions. When $d_p \gg h$, the disturbance penetrates farther into the substrate, and the measured signal increasingly reflects substrate transport and the combined thermal response of the multilayer stack. The ratio $d_p/h$ therefore provides a useful scale parameter for assessing which portions of the multilayer stack are thermally sampled and how sensitivity is distributed among film, interface, and substrate properties.

Second, comparing the thermal penetration depth $d_p$ with the lateral pump-spot scale $2r_0$ indicates the relative importance of cross-plane and radial heat transport. As shown in Fig. 2, when $d_p \ll 2r_0$, the characteristic cross-plane diffusion length is much smaller than the lateral dimension of the heated region, so heat flow beneath the center of the pump spot is dominated by transport through the sample thickness and approaches a quasi-one-dimensional configuration; the signal is then only weakly sensitive to in-plane thermal conductivity. As $d_p$ approaches and becomes comparable to or larger than $2r_0$, radial diffusion contributes an increasingly large fraction of the heat flow, the temperature field exhibits more pronounced three-dimensional spreading, and sensitivity to in-plane thermal conductivity increases [53]. The relative magnitude of $d_p/(2r_0)$ can therefore serve as a useful dimensionless indicator of the importance of lateral heat diffusion, although the precise transition between cross-plane- and radial-diffusion-dominated regimes also depends on the multilayer geometry and thermophysical-property contrasts.

The thermal penetration depth thus provides a direct connection among intrinsic thermophysical properties, experimentally adjustable variables, and sample geometry. Changing the modulation frequency $f$ directly tunes $d_p$ and therefore the effective propagation range of the thermal disturbance in the thickness direction. Changing the spot size $r_0$ does not alter $d_p$ as defined in Eq. (4), but it changes the ratio $d_p/(2r_0)$ and thereby modifies the relative contributions of cross-plane and radial heat diffusion. Experimental design should therefore consider scale ratios such as $d_p/h$ and $d_p/(2r_0)$ in combination rather than independently, so that the dominant region contributing to the measured signal can be controlled and sensitivity to the thermophysical property or material layer of interest can be optimized.

It should also be noted that $d_p$ is not an independently controllable quantity determined solely by the experimental conditions. As Eq. (4) shows, it depends on both the adjustable variable $f$ and the thermophysical properties $k$ and $C$, which may themselves be the quantities to be determined. In practical measurements of an unknown material, the thermal penetration depth can therefore only be estimated initially from literature values, nominal properties, or other prior information, and this estimate can be used to select an appropriate modulation frequency and spot size. The resulting experimental design should subsequently be evaluated and, when necessary, refined using parameter-sensitivity analysis, uncertainty propagation, and iterative fitting. In this sense, thermal penetration depth is most appropriately regarded as a design and interpretation metric rather than an independently prescribed experimental quantity.

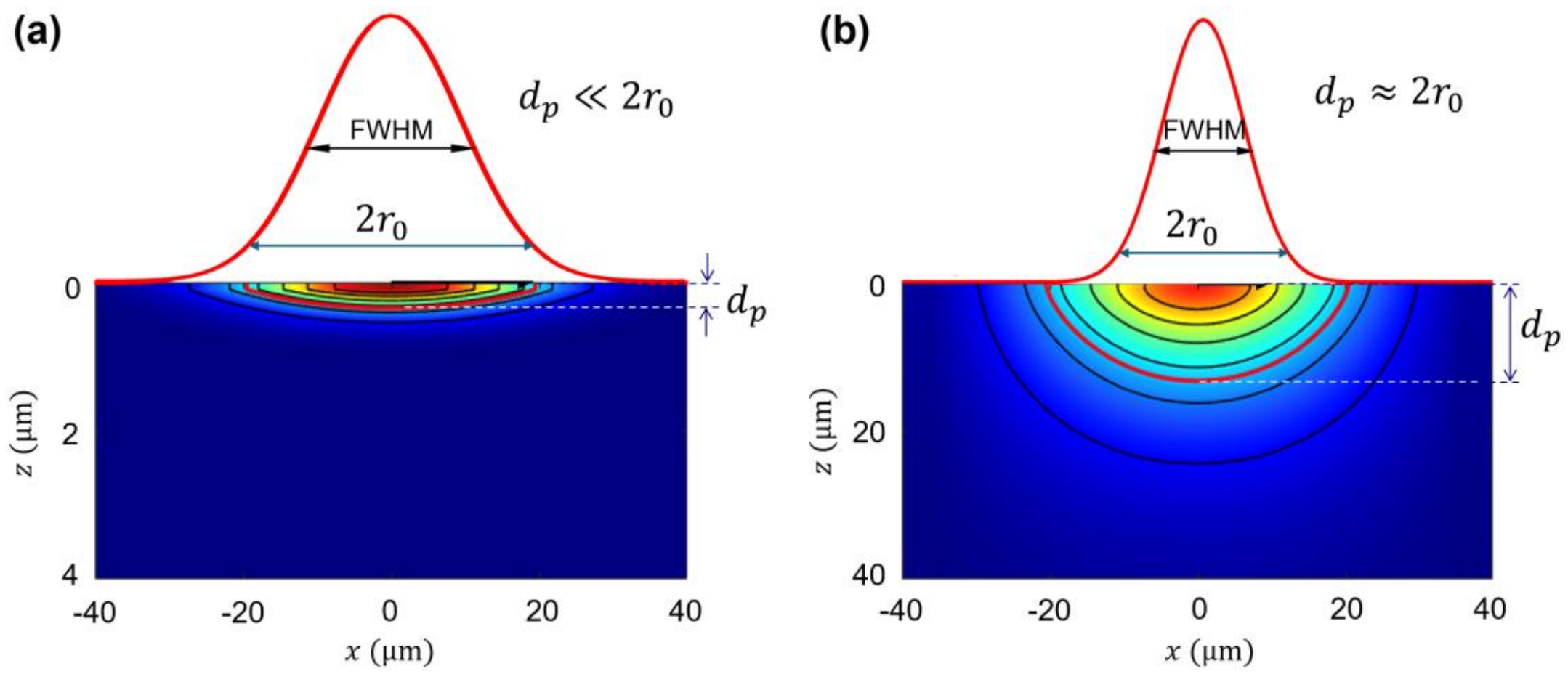

**Figure 2  Effects of thermal penetration depth and laser spot size on heat-diffusion pathways**

## 1.4 Parameter Sensitivity and Inversion Strategy

Thermoreflectance measurement is fundamentally an inverse heat-conduction problem: the thermal response at the sample surface is measured and interpreted through a heat-diffusion model to infer unknown thermophysical properties. To quantify the response of the measured signal to a parameter $x$, a dimensionless sensitivity coefficient is introduced [54]:

$$S_x = \frac{\partial \ln A}{\partial \ln x} \tag{5}$$

Here, $A$ denotes the experimental observable used by a particular measurement technique, such as amplitude, phase, a lock-in signal ratio, or normalized temperature rise. $S_x$ quantifies the relative change in the measured signal produced by a small relative change in parameter $x$. In general, a larger magnitude $|S_x|$ indicates a stronger response of the signal to that parameter and therefore greater potential for parameter identification. If $|S_x|$ is close to zero, the signal is only weakly affected by changes in that parameter, and a reliable estimate is unlikely to be obtained under that experimental condition alone.

A large single-parameter sensitivity, however, does not necessarily mean that the parameter can be independently estimated. In multiparameter problems, the more important issue is whether the signal responses associated with different parameters are sufficiently distinct. When the sensitivity curves of two or more parameters exhibit highly similar shapes or dependencies in the time, frequency, or spatial domain, their effects on the signal become strongly correlated, leading to parameter coupling and an ill-conditioned inverse problem. Under these circumstances, even when the raw experimental signal has a high signal-to-noise ratio, small measurement errors, model discrepancies, or uncertainties in prior parameters can be greatly amplified during inversion, increasing the uncertainty in the extracted thermophysical properties. Sensitivity analysis must therefore address two complementary aspects: first, whether the target parameter produces a sufficiently large response in the signal—that is, whether it is detectable under the chosen measurement conditions; and second, whether its response can be distinguished from those of other unknown parameters—that is, whether it is independently identifiable.

This parameter coupling is closely related to the fact that thermoreflectance signals often depend on combinations of thermophysical properties rather than on individual parameters independently. Under many representative experimental conditions, the measured signal does not respond independently to $k_r$, $k_z$, and $C$, but is instead governed mainly by certain parameter combinations, such as the cross-plane thermal effusivity $\sqrt{k_z C}$, the in-plane thermal diffusivity $k_r/C$, and the areal heat capacity of a film $hC$ [55], where $h$ is the film thickness. Consequently, a single time window, modulation frequency, or spot size generally does not provide sufficient independent information to resolve multiple strongly correlated parameters simultaneously. The key to thermoreflectance parameter estimation is therefore not merely the selection of a numerical fitting algorithm, but also the deliberate design of experimental variables so that different parameters exhibit distinct sensitivity signatures across multiple measurement conditions, thereby improving multiparameter identifiability.

Sensitivity analysis is therefore a central component of both experimental design and parameter decoupling. By calculating $S_x$ as a function of experimental variables such as delay time, modulation frequency, spot size, or spatial offset, one can identify regions of high sensitivity to the target parameter and assess the degree of coupling among different parameters. Specifically, sensitivity analysis can be used to: (1) select appropriate time windows, frequency ranges, or spatial measurement ranges to enhance sensitivity to the parameter of interest; (2) compare the magnitudes and shapes of sensitivity curves to assess multiparameter separability; (3) combine experimental noise and prior-parameter uncertainty to estimate the expected uncertainty in the extracted properties; and (4) guide the design of experimental conditions and sample structures, for example by optimizing transducer thickness, spot size, modulation frequency, and substrate selection, thereby improving inversion stability. In this sense, control of the thermal penetration depth and geometric length scales can be directly mapped onto control of parameter sensitivity: changing modulation frequency and spot size modifies the temperature field and heat-flow pathways and thereby redistributes sensitivity among the relevant thermophysical parameters.

After the experimental configuration has been selected and the measurement window optimized for sensitivity and parameter separability, the unknown thermophysical properties are generally obtained by fitting the experimental signal to a heat-diffusion model. Nonlinear least squares is widely used in thermoreflectance data analysis; algorithms such as Levenberg–Marquardt iteratively update the unknown parameters to minimize the residual between the measured and modeled signals [56]. For problems involving only a few parameters, reasonable initial estimates, and weak parameter correlations, such local optimization methods are computationally efficient and generally robust. When parameter coupling is strong, however, or when the objective function contains multiple local extrema or the admissible parameter range is broad, the inferred solution can become sensitive to the initial guess, experimental noise, and errors in prior parameters. In practice, numerical optimization should therefore be combined with physical constraints, sensitivity analysis, and uncertainty quantification to improve both the stability of the inversion and the credibility of the extracted parameters.

For problems involving a broad parameter space or strong sensitivity of local optimization to the initial guess, global search methods such as particle swarm optimization and genetic algorithms, as well as hybrid approaches that combine global and local optimization, have increasingly been applied to thermoreflectance inversion [57]. These methods can explore a wider region of parameter space and reduce the risk of convergence to suboptimal local solutions, but generally require substantially more model evaluations, creating a trade-off between computational efficiency and optimization robustness. In recent years, machine-learning methods have also been introduced for rapid thermophysical-property estimation and high-throughput thermoreflectance data processing [58-60]. When the training data adequately cover the relevant experimental and thermophysical parameter space, these methods can substantially reduce the computational cost of repeated inversion. Their predictive accuracy and transferability, however, remain dependent on the quality and coverage of the training data, the generalization capability of the model, and the consistency between the training framework and the underlying heat-transport physics.

# 2 Classification and Evolution of Thermoreflectance Measurement Techniques

Building on the thermoreflectance effect, heat-diffusion equation, and parameter-sensitivity analysis introduced above, this section discusses the classification and technical evolution of representative thermoreflectance methods, including TTR, TDTR, FDTR, SSTR, SDTR, and SPS. Although these approaches differ in optical layout, photothermal excitation, and signal acquisition, their underlying physical process is the same: laser-induced photothermal excitation generates a temperature response in the sample; the temperature-induced reflectivity change is measured as a thermal signal; and a heat-diffusion model is used to estimate the relevant thermophysical properties.

The principal distinctions among the methods therefore do not lie in the governing heat-conduction equation itself, but in the excitation scheme, observable, and signal-acquisition dimension used to interrogate the same heat-diffusion process.

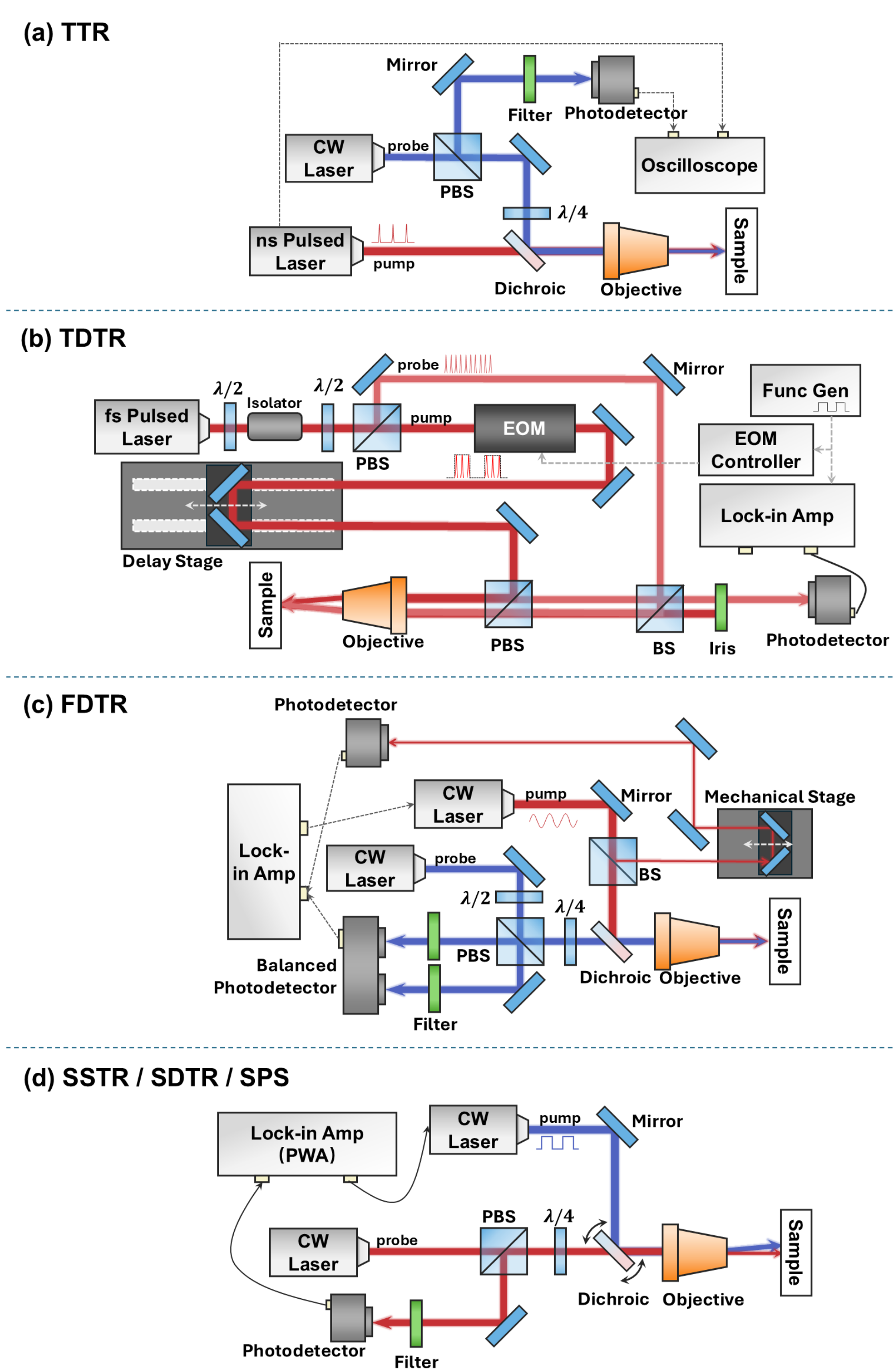

**Figure 3 Schematic diagrams of experimental systems for different thermoreflectance techniques**

Figure 3 summarizes the experimental configurations of several mainstream thermoreflectance methods in approximately chronological order of their development. Early TTR extracted heat-diffusion information by recording the transient temperature decay following pulsed-laser heating. TDTR subsequently incorporated pump modulation, lock-in detection, and pump–probe delay scanning within an ultrafast optical architecture, substantially improving the quantitative characterization of heat transport in thin films and interfaces. FDTR, by contrast, uses modulation-frequency scanning as the primary measurement variable, thereby introducing an independent control dimension through which the thermal penetration depth and associated parameter sensitivities can be tuned for the identification of thermophysical properties in multilayer structures. Later approaches, including SSTR, SDTR, and SPS, further expanded the accessible information by measuring quasi-steady temperature rise, spatially resolved thermal fields, and full-cycle thermal waveforms, respectively. Taken together, these developments have extended thermoreflectance from predominantly single-domain measurements toward increasingly multidimensional and complementary characterization strategies.

The historical sequence shown in Fig. 3 is not adopted directly in the discussion that follows, because a purely chronological presentation would obscure the more fundamental distinctions among the methods in their characteristic heat-diffusion scales, observables, control variables, and parameter sensitivities. We therefore organize the following discussion according to the sequence "quasi-steady response → single-pulse transient response → frequency-domain response → ultrafast pump–probe time-domain response → full-period waveform response → spatial-domain response", corresponding to SSTR, TTR, FDTR, TDTR, SPS, and SDTR, respectively. SSTR is considered first not because it was developed earliest, but because the steady-state response represents the long-time asymptotic limit of diffusive heat transport and thus provides a useful physical baseline for interpreting the non-steady techniques. Building on this framework, we then compare the methods in terms of characteristic heat-diffusion length scales, parameter sensitivity and identifiability, multiparameter estimation capability, and applicable measurement regimes. Figure 4 provides a complementary overview of their approximate measurement ranges and representative application spaces.

### 2.1 Steady-State Limit: SSTR

Steady-state thermoreflectance (SSTR) [38] generally uses a low-frequency-modulated continuous-wave laser as the pump heat source. When the period of the heating modulation is sufficiently long compared with the characteristic thermal relaxation time of the system, the temperature field approaches a quasi-steady state during each heating interval; SSTR can therefore be viewed as the long-time limiting response of the heat-diffusion process. Rather than resolving the transient temperature decay following pulsed excitation, SSTR characterizes the quasi-steady temperature rise generated under a prescribed absorbed power and laser-spot size. The underlying physical picture is straightforward: under otherwise identical heating conditions, a material with higher thermal conductivity transports heat away from the heated region more efficiently and therefore exhibits a smaller surface temperature rise, whereas a material with lower thermal conductivity produces stronger local heat accumulation and, consequently, a larger quasi-steady temperature rise.

As illustrated in Fig. 3d, a typical SSTR system consists of a low-frequency-modulated continuous-wave pump beam, a continuous-wave probe beam, and a reflected-signal acquisition module [38]. The pump generates localized heating at the sample surface or within a metallic transducer, while the probe detects the resulting temperature-induced change in reflectivity. During data analysis, the effective thermal conductivity of the material or the effective thermal resistance of the measured structure can be inferred either by comparing the temperature rise of the sample with that of a reference sample under equivalent heating conditions or by interpreting the measured temperature rise using the absorbed pump power, laser-spot size, and an appropriate steady-state heat-diffusion model.

SSTR offers an intuitive physical interpretation and a relatively simple experimental configuration. Because the steady-state temperature field is governed by the balance between heat input and heat conduction and therefore does not explicitly depend on volumetric heat capacity, the method is well suited to characterizing the effective thermal conductivity of bulk or relatively thick materials, as well as the effective thermal resistance of layered structures. For multilayer films, buried interfaces, or heterostructures with complex heat-flow pathways, however, the quasi-steady temperature rise generally contains coupled contributions from the thermal resistances of individual layers and interfaces together with substrate heat spreading. A single steady-state temperature-rise amplitude is therefore often insufficient to independently resolve the thermophysical properties of multiple constituent layers and interfaces. This limitation reflects the reduced dimensionality of the steady-state observable rather than insufficient signal sensitivity alone. When layer- or interface-specific characterization, multiparameter decoupling, or time-dependent heat-transport information is required, additional observables in the time, frequency, or spatial domains are needed. A natural departure from this long-time limit is to resolve the temperature evolution following a discrete heating event, which leads to the transient thermoreflectance (TTR) approach discussed next.

### 2.2 Single-Pulse Transient Response: TTR

Transient thermoreflectance (TTR) [61] uses a pulsed laser to generate an impulsive or short-duration thermal excitation at the sample surface and records the subsequent temporal decay of the surface temperature, making it one of the most direct time-domain thermoreflectance methods. Whereas SSTR primarily uses the quasi-steady temperature-rise amplitude, the thermophysical information in TTR is encoded mainly in the characteristic timescale and temporal shape of the transient temperature response: faster heat diffusion produces a more rapid decay of the surface temperature, whereas slower heat diffusion leads to a more persistent thermal response. The transient decay curve therefore provides time-resolved information on thermal diffusion and associated thermophysical properties.

As shown in Fig. 3a, TTR commonly uses a nanosecond or picosecond pulsed laser as the pump [27, 62] to induce a rapid temperature rise at the sample surface or within a metal transducer, while a continuous-wave probe monitors the heated region. A high-speed photodetector records the temperature-induced change in probe reflectivity, yielding the temporal evolution of the surface thermal response. During data analysis, the transient temperature-rise signal is often normalized to its peak value to reduce sensitivity to multiplicative amplitude factors, such as absorbed pump energy, the thermoreflectance coefficient, and system gain. Parameter extraction can then rely more strongly on the characteristic decay timescale and waveform shape rather than on the absolute signal amplitude.

TTR provides a clear physical picture and directly resolves the temporal relaxation of a thermal disturbance following pulsed excitation. It is particularly suitable for relatively thick films, bulk materials, and samples whose characteristic thermal relaxation times lie within the accessible temporal bandwidth of the measurement system. In multilayer heterostructures, low-thermal-conductivity materials, or systems with appreciable interfacial thermal resistance, however, the transient decay curve is generally influenced simultaneously by film thermal conductivity, volumetric heat capacity, interfacial thermal conductance, and substrate properties. The corresponding sensitivity signatures may overlap substantially within a single temporal window, leading to strong parameter correlations and limiting independent multiparameter estimation from a single decay trace. In addition, detector bandwidth and signal-to-noise ratio constrain the experimentally accessible time window, while the progressive decrease in temperature-rise amplitude at long delay times makes the signal increasingly susceptible to experimental noise. These factors can restrict the characterization of slow thermal relaxation and long-range heat diffusion.

Overall, TTR establishes a fundamental time-domain measurement framework based on the relaxation of an impulsively generated thermal disturbance, but a single transient response provides limited active control over the characteristic heat-diffusion length and the distribution of parameter sensitivities. Improved parameter identifiability therefore requires additional experimental degrees of freedom that systematically alter the sampled heat-diffusion regime. One such variable is modulation frequency, which continuously tunes the thermal penetration depth and redistributes sensitivity among film, interface, and substrate properties. This concept forms the basis of the frequency-domain thermoreflectance (FDTR) method discussed next.

### 2.3 Frequency-Domain Response: FDTR

Frequency-domain thermoreflectance (FDTR) periodically modulates the pump intensity to generate a periodic thermal response in the sample and measures the amplitude and phase of the thermoreflectance signal relative to the modulated heat source over a range of modulation frequencies. Unlike TTR, which directly records transient temperature decay after a pulse, FDTR uses modulation frequency as the primary measurement variable and probes the complex thermal response in the frequency domain. A central feature of FDTR is the ability to tune the characteristic thermal penetration depth by varying the modulation frequency: at higher frequencies, the thermal disturbance is more strongly confined to the near-surface region, whereas at lower frequencies, the penetration depth increases, allowing deeper portions of the multilayer structure, including buried interfaces and the substrate, to contribute more strongly to the measured response.

As shown in Fig. 3c, a typical FDTR system uses continuous-wave pump and probe beams. The pump is periodically modulated using an electro-optic modulator, an acousto-optic modulator, or direct intensity modulation, while the probe is focused on the heated region to detect the temperature-induced oscillation in reflectivity. A lock-in amplifier extracts the amplitude and phase of the thermoreflectance response at the pump modulation frequency [10, 63]. Scanning the modulation frequency over a selected range yields a complex frequency-dependent thermal response, which can be interpreted using a multilayer heat-diffusion model to estimate parameters such as film thermal conductivity, interfacial thermal conductance, substrate thermophysical properties, and the areal heat capacity of the transducer.

Physically, different modulation frequencies correspond to different characteristic thermal penetration depths and, consequently, different distributions of parameter sensitivity within the multilayer structure. Frequency scanning therefore allows FDTR to vary the relative contributions of different layers and interfaces, providing a degree of thermal depth discrimination. At relatively high modulation frequencies, the thermal disturbance is concentrated near the surface, and the signal tends to exhibit greater sensitivity to the transducer and near-surface thermal properties. As the modulation frequency decreases and the thermal penetration depth increases, buried layers, interfaces, and the substrate increasingly influence the measured response. This behavior should not, however, be interpreted as a sharply defined depth resolution or strict spatial isolation of individual layers; rather, frequency variation continuously redistributes the sensitivity weights associated with different thermophysical parameters and regions of the multilayer stack. Accordingly, there is no universal boundary separating high-, intermediate-, and low-frequency sensitivity regimes, because the relevant frequency range depends jointly on thermal diffusivity, layer thicknesses, laser-spot size, interfacial thermal conductance, and the overall multilayer architecture. FDTR experiments should therefore employ sensitivity analysis to identify a modulation-frequency range that provides sufficient sensitivity to the target parameters while improving their separability from other unknown quantities.

FDTR offers a relatively simple optical configuration and a readily adjustable frequency-domain control variable, providing multifrequency thermal-response information without requiring a mechanical delay line. It is therefore well suited to characterizing heat transport in films, interfaces, and multilayer structures. By varying modulation frequency, the characteristic heat-diffusion scale can be tuned without altering the sample geometry, providing substantial flexibility for experimental design and redistribution of parameter sensitivity. Nevertheless, FDTR remains subject to the fundamental challenge of parameter correlation: if multiple thermophysical properties exhibit similar sensitivity signatures over the selected frequency range, their joint estimation in a complex multilayer system can remain ill-conditioned. Low-frequency measurements can also be more susceptible to environmental drift and low-frequency noise, whereas high-frequency measurements impose increasingly stringent requirements on modulation and detection bandwidth, phase calibration, and signal-to-noise ratio.

Overall, FDTR extends thermoreflectance characterization into the frequency domain, with modulation frequency serving as the primary scan variable for tuning the characteristic heat-diffusion scale and redistributing parameter sensitivity. Frequency scanning alone, however, provides only one experimental dimension for interrogating the thermal response. Introducing pump–probe delay time provides an additional temporal coordinate through which different stages of heat diffusion can be resolved. The simultaneous use of modulation frequency and delay time is a defining feature of time-domain thermoreflectance (TDTR), discussed next.

## 2.4 Ultrafast Pump-Probe Time-Domain Response: TDTR

Time-domain thermoreflectance (TDTR) is a representative ultrafast thermoreflectance technique based on a pump–probe architecture. Unlike TTR, which directly records the continuous temperature decay following a single heating pulse, TDTR typically uses a high-repetition-rate femtosecond laser to generate pump and probe pulse trains, periodically modulates the pump intensity, and scans the relative arrival time of the pump and probe pulses using a mechanical delay line to measure the thermoreflectance response as a function of pump–probe delay. Periodic pump modulation enables phase-sensitive lock-in detection of the weak thermoreflectance signal while also influencing the periodic thermal response and characteristic heat-diffusion scale. Thus, modulation frequency provides an important experimental control parameter, whereas pump–probe delay time serves as the primary measurement coordinate for resolving and fitting the transient thermal response.

As shown in Fig. 3b, a typical TDTR system consists of a femtosecond pulsed laser, a pump-modulation module, a mechanical delay stage, a pump-probe optical path, and a lock-in detection module [64]. The high-repetition-rate laser pulses are split into pump and probe beams. The pump is intensity-modulated, typically in the MHz range using an electro-optic modulator or similar device, and focused onto the sample surface to produce a train of heating pulses with a periodically modulated envelope in the metal transducer. The probe passes through the mechanical delay line, which varies its optical path length and hence its arrival time relative to the pump, and is focused onto the pump-heated region to sample the temperature-dependent reflectivity at different delay times. The lock-in amplifier extracts the in-phase ($V_{\mathrm{in}}$) and out-of-phase ($V_{\mathrm{out}}$) components at the modulation frequency, and their ratio $-V_{\mathrm{in}}/V_{\mathrm{out}}$ is commonly used as the fitting signal [28]. This ratio reduces sensitivity to multiplicative amplitude factors, including the thermoreflectance coefficient, optical gain, and laser power, thereby improving the robustness of quantitative parameter extraction.

From a physical perspective, the TDTR signal represents the periodically modulated thermal response generated by a high-repetition-rate sequence of ultrafast heating pulses. Scanning the pump–probe delay samples different stages of the transient thermal evolution on picosecond-to-nanosecond timescales, such that different delay-time ranges can exhibit distinct sensitivities to transducer properties, film thermal transport, interfacial thermal conductance, and substrate properties. At the same time, the pump modulation frequency sets an additional periodic timescale and modifies the characteristic thermal penetration depth and heat accumulation, thereby redistributing the relative contributions of the film, interfaces, and substrate to the measured response. Joint selection of the delay-time window and modulation frequency can therefore reshape the sensitivity functions of different parameters, improve their separability, and enhance inversion stability. TDTR thus combines delay-resolved transient information with frequency-dependent control of the periodically driven thermal field, providing complementary dimensions for experimental design and sensitivity optimization.

With ultrafast pump–probe timing, well-established multilayer heat-diffusion models, and sensitive lock-in detection, TDTR has become a widely used method for measuring thin-film thermal conductivity and interfacial thermal conductance, particularly in multilayer structures consisting of a metallic transducer, functional film, and substrate. For nanometer- to micrometer-scale films, the delay-dependent response can be interpreted using a multilayer heat-diffusion model to extract quantities such as cross-plane thermal conductivity and interfacial thermal conductance [65]. However, TDTR signals may still respond simultaneously to multiple thermophysical and structural parameters, and several unknowns generally cannot be determined independently from a single experimental condition when their sensitivities are highly correlated. In practice, some quantities - such as volumetric heat capacity, film thickness, transducer properties, or selected interface parameters - are therefore commonly prescribed or measured independently to reduce parameter correlations and improve inversion stability [46]. The usable ranges of modulation frequency and delay time are also constrained by laser repetition rate, pulse heat accumulation, modulator bandwidth, and system signal-to-noise ratio.

Overall, TDTR uses pump–probe delay time as its principal measurement coordinate while exploiting periodic modulation for phase-sensitive detection and additional control of the thermal response. It has become a mature tool for characterizing film thermal conductivity, interfacial thermal conductance, and heat transport in multilayer structures. Its parameter sensitivity can be further tailored through modulation frequency, laser-spot size, and the selected delay-time fitting window, providing considerable flexibility for experimental optimization. Nevertheless, for low-thermal-conductivity materials, deeply buried interfaces, or systems involving strongly correlated unknown parameters, reliable multiparameter inversion may still require independent prior information, multiple measurement conditions, or additional observables.

### 2.5 Full-Period Waveform Response: SPS

The square-pulsed source (SPS) method [40] is a thermoreflectance technique based on periodic square-wave photothermal excitation and measurement of the complete temperature response over each modulation period. Unlike TDTR, which primarily resolves ultrafast thermal dynamics through pump–probe delay scanning, or FDTR, which measures the amplitude and phase of the periodic response as a function of modulation frequency, SPS uses a square-wave-modulated continuous-wave pump to periodically heat the sample while recording the time-resolved temperature response throughout each heating–cooling cycle. SPS therefore not only tunes the characteristic heat-diffusion scale through modulation frequency but also retains the intra-period temporal evolution associated with heating and cooling. The resulting data provide thermophysical constraints through both frequency-dependent changes in the thermal field and the detailed shape of the full-period temperature waveform.

As illustrated in Fig. 3d, the SPS optical configuration resembles that of SSTR in that both can employ continuous-wave pump–probe optics, but the two methods differ substantially in excitation protocol, signal-acquisition strategy, and information content. SSTR primarily uses the quasi-steady temperature rise obtained under low-frequency heating to infer effective thermal conductivity or effective thermal resistance. SPS instead applies periodic square-wave excitation and records the time-dependent temperature response throughout each modulation period, typically over multiple modulation frequencies. At each frequency, SPS therefore yields not merely a single temperature-rise amplitude or a pair of amplitude and phase values, but a resolved heating–cooling waveform that contains information on the evolution of the thermal field within the modulation cycle. Reported SPS systems span modulation frequencies of approximately 1 Hz to 20 MHz [40], although the experimentally accessible range is ultimately constrained by the modulation, detection, and data-acquisition bandwidths, as well as by the required signal-to-noise ratio.

Variation of modulation frequency provides SPS with control over the characteristic heat-diffusion length scale. According to the thermal-penetration-depth relation discussed above, higher modulation frequencies correspond to shorter characteristic diffusion lengths and weight the thermal response more strongly toward the transducer and near-surface region. As the modulation frequency decreases, the thermal penetration depth increases, and buried films, interfaces, and the substrate can contribute more substantially to the measured response. At sufficiently low frequencies, the response approaches the quasi-steady limit and increasingly reflects the combined thermal resistance and heat-spreading characteristics of the multilayer system. Thus, varying modulation frequency continuously redistributes the relative contributions of different regions and heat-transport processes within the same experimental configuration. Importantly, however, individual frequencies do not map uniquely onto specific material layers or thermophysical properties. As in FDTR, frequency-dependent depth discrimination arises from a continuous redistribution of parameter sensitivities rather than from sharply defined spatial selectivity. The relevant sensitivity ranges remain jointly determined by thermal diffusivity, layer thicknesses, laser-spot size, interfacial thermal conductance, and multilayer geometry, and appropriate frequency ranges should therefore be selected using sensitivity analysis.

In addition to frequency-dependent control of the heat-diffusion scale, SPS retains the intra-period evolution of the temperature response during the heating and cooling portions of each cycle. Different portions of this waveform can exhibit distinct sensitivity signatures for the transducer, constituent layers, interfaces, and substrate, thereby providing additional information for parameter estimation. Compared with reducing the response at each frequency to a single amplitude or phase quantity, analysis of the full-period waveform can, under suitable experimental conditions, increase the number and diversity of useful sensitivity signatures and thereby improve the identifiability of selected thermophysical parameters. This capability can be particularly valuable for low-thermal-conductivity materials and multilayer systems in which several thermophysical parameters contribute simultaneously to the measured response.

Although SPS expands the accessible frequency-dependent heat-diffusion scales while retaining intra-period temporal information, its measurement capability remains constrained by experimental bandwidth, signal quality, and parameter coupling. At low modulation frequencies, measurements can become increasingly susceptible to ambient-temperature drift, long-term laser-power fluctuations, and low-frequency noise. At high frequencies, the finite rise and fall times of the modulator and the bandwidths of the detector and data-acquisition system can distort both the applied heat-source waveform and the measured thermal response, causing the experimental excitation to deviate from an ideal square wave. Such instrumental transfer characteristics should therefore be accounted for when they become comparable to the relevant modulation timescales. Moreover, multifrequency full-period measurements provide additional inversion information but do not by themselves guarantee unique multiparameter estimation. If several unknown parameters retain similar sensitivity signatures across the selected frequency and intra-period time ranges, reliable extraction still requires appropriate prior constraints, sensitivity-based experimental design, and rigorous uncertainty analysis.

Overall, SPS combines broadband square-wave modulation with full-period temperature-waveform acquisition, integrating frequency-dependent control of the characteristic heat-diffusion scale with time-resolved information within each modulation cycle. This combination can provide richer sensitivity information for multiparameter estimation than measurements based solely on a single response metric at each frequency. Compared with TDTR, SPS emphasizes periodic thermal evolution over timescales set by the square-wave modulation and detection bandwidth, whereas TDTR provides ultrafast pump–probe access to picosecond-to-nanosecond thermal dynamics; the two methods therefore probe complementary temporal regimes and offer different routes for tailoring parameter sensitivity. SSTR, TTR, FDTR, TDTR, and SPS primarily interrogate the thermal response through steady-state, temporal, or frequency-related observables and control variables. An additional and fundamentally different source of information is obtained by resolving the spatial distribution of the temperature field, particularly when in-plane transport or thermal-conductivity anisotropy is of interest. This motivates the spatial-domain thermoreflectance approaches discussed next.

### 2.6 Spatial-Domain Response: SDTR and Beam-Offset Methods

Spatial-domain thermoreflectance (SDTR) [39] systematically varies the lateral position of the probe relative to the pump and measures the thermoreflectance response as a function of spatial offset, thereby resolving the lateral spreading of the temperature field. Unlike the methods discussed above, which primarily interrogate the thermal response through steady-state, temporal, frequency-dependent, or full-period waveform information, SDTR introduces lateral position as an independent measurement coordinate, making the spatial evolution of in-plane heat diffusion directly accessible to parameter estimation. Because the lateral temperature field is particularly sensitive to in-plane heat transport, this approach is well suited to characterizing anisotropic crystals, oriented films, and other material systems exhibiting pronounced directional heat transport.

In a typical SDTR experiment, a modulated pump generates a localized periodic heat source at the sample surface, while the probe is scanned laterally relative to the pump center and the thermoreflectance response is recorded at each offset. As the probe moves away from the heated region, the amplitude of the temperature oscillation generally decreases, while its phase lag relative to the periodic heat source evolves as heat diffuses laterally through the sample. Analysis of the spatial dependence of amplitude or phase, together with an appropriate heat-diffusion model, enables estimation of in-plane thermal conductivity or diffusivity. To enhance sensitivity to lateral heat diffusion, the modulation frequency, laser-spot size, and spatial scan range should be jointly selected so that the characteristic lateral diffusion length is comparable to the relevant optical and geometric length scales. If the thermal-diffusion length is much smaller than the spot size, the measured spatial profile is dominated more strongly by the optical heating and probing profiles, reducing sensitivity to intrinsic in-plane transport. Reducing the modulation frequency can increase the lateral diffusion length and broaden the temperature field, thereby enhancing sensitivity to in-plane thermal properties, although this change simultaneously modifies cross-plane heat penetration.

Beam-offset TDTR (BO-TDTR) [66, 67] provides a representative example of incorporating spatial scanning into an ultrafast TDTR platform. A pump–probe delay time with high sensitivity to in-plane heat transport is typically selected, after which the lateral offset between the pump and probe is systematically varied to map the spatial dependence of the lock-in response. The in-plane thermal conductivity or diffusivity can be inferred from the width and shape of the resulting spatial profile, for example from the full width at half maximum of the out-of-phase component ($V_{\mathrm{out}}$), in conjunction with an anisotropic heat-diffusion model. Whereas SDTR commonly analyzes the spatial decay and phase evolution of a periodically driven thermal field, BO-TDTR combines pump–probe timing with lateral scanning and often extracts in-plane transport information from the spatial broadening of a selected lock-in signal. Despite differences in excitation, optical configuration, and signal processing, both approaches treat lateral pump–probe displacement as a measurement coordinate and exploit the resulting spatial dependence to enhance sensitivity to in-plane heat transport.

Introducing a spatial measurement dimension is particularly valuable for characterizing anisotropic heat transport. For crystals or oriented films with strongly direction-dependent thermal conductivity, spatial scans can be performed along different in-plane directions to resolve directional variations in thermal spreading. Combined with an anisotropic heat-diffusion model, these measurements can provide direction-dependent in-plane thermal conductivities and, when sufficient angular information and appropriate model constraints are available, enable reconstruction of the principal directions and components of the in-plane thermal-conductivity tensor. Relative to measurements relying primarily on temporal or frequency variation, spatial scanning provides an observable directly associated with lateral redistribution of the temperature field and can substantially alter the relative sensitivities to in-plane conductivity, cross-plane conductivity, interfacial thermal conductance, and other parameters. Spatial information can therefore improve parameter identifiability when its sensitivity signature is sufficiently distinct from those provided by other measurement dimensions.

The performance of spatial-domain methods is nevertheless governed by the interplay among thermal-diffusion length scales, optical-spot dimensions, and multilayer geometry. Increasing lateral thermal spreading generally requires reducing the modulation frequency, but this simultaneously increases the cross-plane thermal penetration depth, enhancing sensitivity to the substrate, buried interfaces, and deeper layers and potentially complicating isolation of in-plane transport within the target film. Low-frequency measurements are also more susceptible to environmental-temperature drift, laser-power fluctuations, and low-frequency noise. Conversely, if the modulation frequency is too high and the lateral diffusion length becomes small relative to the spot size, the measured spatial profile is increasingly governed by the convolution of the pump and probe intensity distributions and localized cross-plane heat flow, thereby reducing sensitivity to intrinsic in-plane thermophysical properties [39]. Spatial-domain experiments therefore require coordinated selection of modulation frequency, pump and probe spot sizes, and scan range, guided by characteristic-length-scale considerations and quantitative sensitivity analysis.

It is also important to distinguish the spatial-domain methods defined here from other thermoreflectance approaches that incorporate a nonzero pump–probe offset without using spatial position as the primary measurement coordinate. To avoid ambiguity in classifying beam-offset techniques, we adopt the criterion of whether lateral offset is systematically varied and explicitly enters the parameter inversion as an independent measurement variable. When pump–probe offset is scanned and in-plane heat-transport properties are inferred from the resulting spatial variation in amplitude, phase, or another thermal observable, the method is classified here as spatial-domain thermoreflectance. In contrast, when a lateral offset is fixed and used primarily to modify the heat-flow geometry or redistribute parameter sensitivity, while parameter estimation continues to rely principally on frequency scanning or full-period waveform analysis—as in BO-FDTR [68, 69] and BO-SPS [43] —the approach is more appropriately regarded as a frequency-domain or full-period waveform method incorporating a beam-offset geometry. The defining criterion is therefore not the presence of a pump–probe offset itself, but whether lateral position is systematically scanned and constitutes an explicit measurement and inversion dimension.

Overall, SDTR and related spatial-scanning methods extend thermoreflectance characterization by resolving the lateral structure of the temperature field and introducing spatial position as an additional measurement coordinate. This capability provides a particularly direct route to characterizing in-plane thermal conductivity and direction-dependent heat transport. Nevertheless, the measured spatial response remains jointly influenced by cross-plane heat diffusion, interfacial thermal conductance, substrate properties, and optical-beam geometry; introducing a spatial observable therefore redistributes and potentially improves parameter sensitivity but does not inherently eliminate parameter correlations. For complex multilayer structures and multiparameter inverse problems, spatial-domain measurements should therefore be designed using sensitivity and uncertainty analyses and, where appropriate, combined with independent prior information or complementary temporal, frequency-domain, and full-period measurements such as TDTR, FDTR, and SPS. Such combinations provide a pathway toward multidimensional thermoreflectance characterization in which complementary observables are used to improve parameter identifiability across increasingly complex material systems.

### 2.7 Section Summary

In summary, all thermoreflectance methods described above are built on the same thermoreflectance transduction mechanism and underlying heat-diffusion framework, while differing primarily in heat-source excitation, measurement coordinate, thermal observable, and signal-acquisition strategy. SSTR characterizes the quasi-steady temperature rise; TTR records the transient thermal relaxation following pulsed excitation; FDTR varies modulation frequency and measures the corresponding amplitude and phase response; TDTR uses pump–probe delay time as its principal measurement coordinate while employing periodic modulation for lock-in detection and additional sensitivity control; SPS records the full-period temperature waveform under periodic square-wave excitation, with modulation frequency providing an additional means of tuning the heat-diffusion scale; and SDTR and related spatial-scanning methods resolve the thermal response as a function of lateral position to probe in-plane heat diffusion.

These techniques should therefore not be viewed as a simple chronological progression in which one method supersedes another, but rather as complementary measurement paradigms that interrogate the same heat-diffusion problem through quasi-steady, transient, frequency-dependent, ultrafast delay-resolved, full-period waveform, and spatial responses. This framework classifies each technique according to its primary measurement coordinate and observable while recognizing that experimental control variables—including modulation frequency, pump–probe delay, laser-spot size, and beam offset—can be combined or varied across different implementations to redistribute parameter sensitivity. The following section therefore compares these methods in terms of characteristic heat-diffusion scales, parameter sensitivity and identifiability, capabilities for resolving in-plane and cross-plane transport, practical experimental constraints, and representative application regimes.

## 3 Measurement Ranges and Applicable Conditions of Different Thermoreflectance Methods

The preceding section introduced representative thermoreflectance methods—SSTR, TTR, FDTR, TDTR, SPS, and SDTR—from the perspective of their primary measurement coordinates and observables. Although these techniques share the same thermoreflectance transduction mechanism and underlying heat-diffusion physics, differences in excitation, observable, and experimental conditions lead to distinct characteristic heat-diffusion scales, parameter-sensitivity distributions, capabilities for resolving in-plane and cross-plane transport, and suitable application regimes. Rather than discussing each technique individually again, this section provides a comparative assessment based on several considerations central to experimental design: the effective heat-diffusion scale and its experimental control, sensitivity and identifiability of thermophysical parameters, characterization of anisotropic heat transport, and method selection for different material systems. These comparisons establish a practical framework for matching measurement conditions and observables to the characteristic dimensions and target properties of a given sample and provide a basis for interpreting the representative applications discussed later.

### 3.1 Effective Heat-Diffusion Scale and Experimental Control

The effective probing range of a thermoreflectance measurement is fundamentally constrained by the characteristic length scales over which heat diffuses during the relevant experimental timescale. For a periodically modulated heat source, the thermal penetration depth $d_p = \sqrt{\alpha/\pi f}$ introduced above can be used to characterize the propagation length of the periodic thermal disturbance, where $\alpha$ is the material thermal diffusivity and $f$ is the pump modulation frequency. For a single-pulse transient process, $d_p = \sqrt{\alpha\tau}$ can be used to estimate the characteristic diffusion length over an observation time $\tau$. These characteristic scales should not be interpreted as sharp measurement boundaries or uniquely defined probing depths. Their practical significance lies in comparison with geometric scales such as film thickness, interface depth, and spot size. The ratios among these thermal and geometric scales determine the dominant heat-flow regime and strongly influence the sensitivity of the measured signal to different regions and thermophysical parameters. Modulation frequency, delay time, pulse-repetition period, spot size, and signal-acquisition window can therefore alter the effective region sampled by the thermal field and redistribute the relative sensitivities to surface layers, buried interfaces, and the substrate.

For quasi-steady or long-timescale measurements, heat can spread over comparatively large distances before the measured response is established. SSTR uses quasi-steady temperature-rise information and is therefore governed by the combined thermal resistance and heat-spreading behavior of the sample–substrate system. It is particularly useful when the desired quantity is an effective thermal conductivity or equivalent thermal resistance and independent layer-by-layer resolution is not required. When films, interfaces, and the substrate all contribute substantially to the measured temperature rise, however, a single quasi-steady observable generally cannot distinguish their individual contributions. TTR instead uses the transient temperature decay following pulsed excitation, with the effective diffusion scale determined by the thermal diffusivity and the portion of the transient captured within the accessible observation window. It is therefore suitable for relatively thick films, bulk materials, and systems whose characteristic thermal relaxation times fall within the effective temporal range of the instrument. For ultrathin films or interface-dominated structures, very rapid thermal relaxation can approach the temporal-resolution or detection-bandwidth limits, whereas slowly decaying long-time signals can become increasingly noise-limited. In either regime, separation of different thermophysical contributions becomes more difficult, and inversion uncertainty can increase [62, 70].

FDTR provides direct experimental control over the characteristic periodic heat-diffusion length by varying the modulation frequency. Increasing the modulation frequency reduces the thermal penetration depth and weights the response more strongly toward the near-surface region, whereas decreasing the frequency increases the penetration depth and enhances contributions from buried layers, interfaces, and the substrate. Frequency scanning therefore changes both the effective heat-diffusion scale and the distribution of parameter sensitivities throughout the multilayer structure. This tunability does not, however, imply that individual frequencies correspond uniquely to particular depths or layers. In multilayer systems, the sensitivity ranges associated with different layers and interfaces can overlap substantially, such that simply extending the frequency range does not necessarily enable independent multiparameter estimation. The useful frequency window must instead be determined by considering the sample geometry together with quantitative sensitivity and parameter-correlation analyses [68].

TDTR resolves the thermal response primarily through pump–probe delay scanning, thereby sampling different stages of heat diffusion on picosecond-to-nanosecond timescales. The delay-time window determines which portions of the transient response contribute to fitting, while pump modulation frequency modifies the periodic thermal penetration scale and interpulse heat accumulation and can therefore be used to redistribute parameter sensitivity. Measurements performed at multiple modulation frequencies, spot sizes, or delay-time windows can provide complementary sensitivity signatures and improve parameter separability relative to a single experimental condition. The accessible range of thermal scales nevertheless remains constrained by the laser repetition rate, mechanical-delay range, heat accumulation, modulation and detection bandwidths, and signal-to-noise ratio. TDTR should therefore be viewed as combining ultrafast delay-resolved sampling with additional experimental variables that tune the spatial and periodic characteristics of the thermal field, rather than as possessing a single fixed probing depth.

SPS combines frequency-dependent control of the thermal-diffusion scale with intra-period temporal information. Through square-wave modulation and acquisition of the full-period temperature waveform, SPS not only changes the characteristic heat-diffusion length with modulation frequency but also retains the thermal evolution during different portions of the heating and cooling cycle. Across an appropriate frequency range, the relative contributions of the near-surface region, buried structure, substrate, and overall thermal-resistance network can therefore vary substantially, while different portions of the waveform can provide additional and potentially distinct sensitivity signatures. This combination allows multifrequency, full-period measurements to place multiple constraints on parameter estimation within a common experimental configuration. As with FDTR, however, neither a particular modulation frequency nor a particular portion of the waveform corresponds uniquely to an individual layer or thermophysical property. The effective sensitivity range must therefore be established from the multilayer geometry, characteristic diffusion scales, and sensitivity analysis.

Unlike the preceding approaches, which primarily manipulate or sample heat diffusion through temporal and frequency-related variables, SDTR and related spatial-scanning methods introduce lateral position as an explicit measurement coordinate. Coordinated selection of modulation frequency, laser-spot size, and spatial scan range can match the characteristic lateral diffusion length to the optical geometry and thereby enhance sensitivity to in-plane heat transport. When the lateral diffusion length is much smaller than the spot size, the measured spatial profile is influenced more strongly by the local optical intensity distribution and localized cross-plane heat flow, reducing sensitivity to intrinsic in-plane transport. When the thermal and optical length scales become comparable, lateral spreading produces a more pronounced modification of the measured spatial profile, increasing sensitivity to in-plane thermal properties. The distinguishing feature of spatial-domain methods is therefore not simply access to a larger diffusion length, but direct measurement of the lateral evolution of the thermal field, extending characteristic-scale analysis from cross-plane penetration to in-plane thermal spreading.

In summary, the representative thermoreflectance techniques provide different experimental routes for selecting or interrogating characteristic heat-diffusion scales. SSTR probes the long-time, quasi-steady thermal field; TTR samples transient diffusion through the observation-time window; FDTR tunes the periodic thermal penetration depth through modulation frequency; TDTR resolves ultrafast thermal evolution through pump–probe delay while using modulation frequency, spot size, and fitting window to redistribute sensitivity; SPS combines frequency-dependent diffusion scales with full-period temporal information; and SDTR directly resolves lateral thermal spreading through spatial scanning while using modulation frequency and spot size to control the relevant in-plane diffusion scale. Accordingly, no thermoreflectance technique possesses a universal measurement depth or fixed applicable length scale. Experimental conditions should instead be selected by matching the characteristic thermal-diffusion lengths to the sample dimensions and by verifying, through sensitivity analysis, that the resulting observable contains sufficient and distinguishable information about the target thermophysical properties.

### 3.2 Parameter Sensitivity and Identifiability

The definitions of parameter sensitivity and correlation, together with the distinction between a parameter being detectable in the measured response and independently identifiable from other unknowns, were introduced in Section 1.4. Rather than repeating these concepts, this section compares how the different measurement dimensions available to thermoreflectance techniques generate distinct sensitivity signatures and thereby influence parameter identifiability.

From a comparative perspective, SSTR and TTR rely primarily on a quasi-steady response or a single transient decay, respectively, and are therefore most effective for systems with relatively few unknown parameters or with sufficient prior information to constrain the inverse problem. FDTR acquires frequency-dependent thermal responses through modulation-frequency scanning and can exploit the resulting redistribution of parameter sensitivities across frequency to constrain film, interface, and substrate properties. Its ability to separate multiple unknowns depends not only on the magnitude of sensitivity to each parameter, but more importantly on whether their frequency-dependent sensitivity signatures are sufficiently distinct over the experimentally accessible range.

TDTR uses pump–probe delay time as its primary measurement coordinate and can generate complementary sensitivity conditions by varying modulation frequency, laser-spot size, and the selected fitting window. SPS retains the complete intra-period temperature waveform at different modulation frequencies, thereby providing both frequency-dependent and intra-period temporal sensitivity information. These approaches can improve parameter identifiability when the additional experimental conditions produce sufficiently different sensitivity signatures for the target unknowns. An increased number of measured data points alone, however, does not increase the effective dimensionality of the inverse problem if those data carry largely redundant parameter information. Strongly correlated parameters can therefore remain difficult to estimate independently even from large datasets.

SDTR and related spatial-scanning or beam-offset methods introduce lateral spatial information, which is particularly valuable for distinguishing sensitivity to in-plane heat transport from that associated with cross-plane diffusion and interfaces. In anisotropic or multilayer systems, spatial measurements can be combined with frequency- or time-domain information to generate complementary sensitivity signatures and modify the correlation structure among in-plane thermal conductivity, cross-plane thermal conductivity, interfacial thermal conductance, and other unknown parameters. The benefit of introducing a spatial dimension therefore depends on whether the spatial response provides information that is sufficiently distinct from that already contained in the temporal or frequency-domain observables.

The central issue in a multiparameter inverse problem is therefore not to identify a single universally superior thermoreflectance technique, but to determine whether the available observations provide sufficiently distinct information to resolve the target parameter set with acceptable uncertainty. This condition can be evaluated using a sensitivity matrix and related tools such as the Fisher information matrix, singular-value decomposition, parameter-correlation metrics, or posterior-correlation analysis. Poorly conditioned sensitivity matrices or nearly collinear sensitivity vectors indicate that different parameter combinations can produce similar changes in the measured response, even when the individual parameter sensitivities are large. Additional modulation frequencies, delay-time windows, spot sizes, spatial offsets, or independently constrained parameters can then be introduced to improve the conditioning of the inverse problem rather than merely increase the number of observations.

For complex systems, diversity of parameter sensitivity across complementary measurement dimensions is generally more valuable than a larger number of highly redundant data points acquired under nearly identical conditions. Experimental design should therefore begin with the target parameter set and sample geometry and then select measurement coordinates and control variables that maximize useful sensitivity while minimizing parameter correlation within realistic experimental constraints. Available prior information and expected measurement uncertainty should also be incorporated into this assessment. Identifiability should ultimately be evaluated both before measurement, to guide experimental design, and after inversion, to quantify parameter covariance and determine whether the extracted properties are uniquely constrained by the available data.

### 3.3 Capability for Distinguishing In-Plane and Cross-Plane Heat Transport

For thin films, layered structures, and anisotropic materials, in-plane and cross-plane thermal conductivities can differ substantially, making directional sensitivity and identifiability important considerations in method selection. From the perspective of characteristic length scales, when the lateral thermal-diffusion length is small relative to the laser-spot size, heat flow beneath the heated region approaches a predominantly cross-plane regime, and the measured response is generally more sensitive to cross-plane thermal conductivity, interfacial thermal conductance, and the associated cross-plane thermal resistances. As lateral thermal spreading becomes appreciable relative to the optical-spot size or pump–probe offset, radial or directional in-plane heat flow contributes more strongly to the measured response, thereby increasing sensitivity to in-plane thermal conductivity.

In conventional coaxial pump–probe configurations, TTR, FDTR, and TDTR often operate under conditions designed to emphasize cross-plane heat transport. FDTR uses modulation frequency to tune the characteristic thermal penetration depth, while TDTR additionally exploits pump–probe delay time and modulation frequency to redistribute sensitivity among quantities such as cross-plane thermal conductivity and interfacial thermal conductance. Decreasing the laser-spot size or modulation frequency can increase the relative contribution of radial heat spreading and thereby introduce sensitivity to in-plane thermal conductivity. However, an axisymmetric coaxial geometry primarily probes an azimuthally averaged lateral response and therefore provides limited information for distinguishing thermal conductivities along different in-plane directions, particularly in strongly anisotropic materials.

When the target parameters include in-plane thermal conductivity or in-plane anisotropy, introducing a lateral pump–probe displacement or spatial-scanning coordinate can substantially redistribute parameter sensitivity toward lateral heat transport. SDTR and BO-TDTR explicitly use spatially resolved measurements, while fixed-offset implementations such as BO-FDTR and BO-SPS modify the heat-flow geometry to enhance sensitivity to in-plane transport [66, 68, 69]. For anisotropic materials, spatial measurements performed along different in-plane directions and interpreted using an anisotropic heat-diffusion model can provide direction-dependent thermal-conductivity information and, with sufficient directional measurements, constrain the in-plane conductivity tensor. Introducing a spatial dimension or beam offset does not, however, isolate in-plane transport from cross-plane conductivity, interfacial thermal conductance, or substrate properties. The modulation frequency, pump and probe spot sizes, offset or scan range, and sample geometry must therefore be jointly optimized so that the resulting sensitivity signatures provide adequate directional discrimination while limiting parameter correlation.

Overall, coaxial TDTR, FDTR, and SPS can be configured to provide strong sensitivity to cross-plane film thermal conductivity, interfacial thermal conductance, and multilayer thermal resistance, whereas spatial-scanning and beam-offset configurations provide additional leverage for characterizing in-plane transport and anisotropy. The distinction is not absolute: coaxial measurements can retain appreciable sensitivity to in-plane transport when lateral heat spreading is significant, while spatially resolved measurements remain coupled to cross-plane transport and interfaces. For complex anisotropic systems in which multiple quantities such as $k_r$, $k_z$, and $G$ must be determined simultaneously, a single measurement condition may not provide sufficiently distinct sensitivity signatures. Measurements employing multiple spot sizes, modulation frequencies, pump–probe offsets, or spatial directions can therefore be combined to exploit complementary sensitivities and improve directional parameter identifiability.

### 3.4 Material Systems and Application Scenarios

In practical experiments, selection of a thermoreflectance method and measurement configuration is governed jointly by sample architecture, target thermophysical properties, characteristic heat-diffusion scales, available prior information, and acceptable experimental complexity. The different thermoreflectance techniques therefore do not form an absolute hierarchy of performance; rather, their applicability depends on whether the accessible observables provide sufficient sensitivity and identifiability for the properties of interest under the relevant sample dimensions and heat-flow geometry. Table 1 summarizes the representative thermoreflectance techniques in terms of primary measurement dimension, characteristic heat-diffusion scale and control variables, principal thermophysical sensitivities, and representative application scenarios.

For bulk materials or relatively thick and structurally simple samples, SSTR provides a conceptually straightforward approach when the target quantity is an effective thermal conductivity or overall thermal resistance. In such systems, the quasi-steady temperature rise reflects the combined thermal resistance and heat-spreading behavior within the sampled region, and the inverse problem can remain relatively well constrained when contributions from multiple internal interfaces or thin layers are absent or independently known. SSTR is therefore particularly useful when effective rather than layer-resolved thermal properties are desired and the long-time thermal response can be described reliably by the corresponding heat-diffusion model. For more complex multilayer structures, however, the quasi-steady response contains coupled contributions from individual layers, interfaces, and the substrate, limiting the number of properties that can be independently extracted from a single measurement condition.

For nanometer- to micrometer-scale films and interfaces, TDTR and FDTR are widely used for quantitative characterization of thin-film thermal transport and interfacial thermal conductance. TDTR resolves rapid thermal evolution through pump–probe delay scanning combined with modulated lock-in detection, whereas FDTR varies modulation frequency to redistribute the characteristic thermal penetration depth and parameter sensitivities across the multilayer structure. The resulting frequency dependence represents a change in sensitivity weighting rather than strict depth profiling or layer-selective measurement. These characteristics make TDTR and FDTR particularly useful for metal/film/substrate structures and other multilayer systems in which cross-plane film transport and interface effects are important. When several film, interface, and substrate properties are simultaneously unknown, however, both methods generally require careful experimental design, sensitivity and correlation analyses, and independently constrained parameters to achieve stable multiparameter inversion.

**Table 1 Primary measurement information, representative applications, and limitations of different thermoreflectance techniques**

| Method | Primary observable | Control and range of heat-diffusion scale | Representative applications | Limitations |
|---|---|---|---|---|
| **SSTR** | Quasi-steady temperature-rise amplitude | Quasi-steady | Effective thermal conductivity and overall thermal resistance of bulk/thicker samples | Thermal-resistance contributions can be strongly coupled in multilayer systems; sensitive to absorbed power and steady-state boundary conditions |
| **TTR** | Temperature-decay curve | Set by pulse spacing/observation window | Relatively thick films (>1 μm) and bulk materials | Limited by detection bandwidth, long-time signal-to-noise ratio, and parameter correlations |
| **FDTR** | Response versus modulation frequency | Frequency tunable | Frequency-dependent thermal response and parameter estimation in films, interfaces, and multilayer structures | Sensitivity bands of different parameters may overlap; changing frequency does not provide strict depth profiling |
| **TDTR** | Delay-time response and lock-in signal ratio | Moderately tunable (typically 1-10 μm) | Cross-plane thermal conductivity of films, interfacial thermal conductance, and ultrafast thermal processes | Often requires known film thickness, heat capacity, or some interfacial parameters; usable frequency range is limited |
| **SDTR** | Spatially scanned response | Dominated by lateral diffusion | In-plane anisotropy and crystallographic orientation | Low sensitivity to cross-plane thermal conductivity and interfacial thermal conductance |
| **SPS** | Full-period waveform at different modulation frequencies | Broad (reported systems: 1 Hz-20 MHz) | Low-thermal-conductivity materials, multilayer heterostructures, and strongly coupled parameter systems | Low-frequency drift and high-frequency waveform fidelity limit the effective range; not intended for picosecond-scale ultrafast processes |

Low-thermal-conductivity materials, porous media, and multilayer heterostructures often exhibit relatively long thermal relaxation times or strongly coupled responses from multiple layers and interfaces, requiring measurement conditions that access longer heat-diffusion timescales or provide complementary parameter sensitivities. Depending on the sample dimensions, target parameters, and available instrument bandwidth, suitable strategies can include low-frequency FDTR, transient measurements with extended observation windows, TDTR measurements employing multiple modulation frequencies or spot sizes, SPS, or joint analysis of complementary measurement configurations or techniques. SPS provides one such route by combining frequency-dependent control of the heat-diffusion scale with full-period temporal information. Its practical benefit in a given system, however, depends on whether the additional waveform and frequency information produces sufficiently distinct parameter sensitivities, as well as on low-frequency stability, waveform fidelity, available prior information, and experimental signal-to-noise ratio.

For two-dimensional materials, oriented films, and anisotropic crystals, the measurement configuration generally needs to enhance or explicitly resolve lateral heat diffusion when in-plane thermal properties are the quantities of interest. SDTR and BO-TDTR introduce spatially resolved information through lateral scanning, whereas fixed-offset implementations such as BO-FDTR and BO-SPS modify the heat-flow geometry to increase sensitivity to in-plane transport. Measurements performed along different in-plane directions can further provide direction-dependent sensitivity to thermal-conductivity anisotropy. Nevertheless, cross-plane thermal conductivity, interfacial thermal conductance, substrate properties, and optical-beam geometry can remain coupled to the measured response. Reliable multidirectional or three-dimensional thermal characterization may therefore require measurements at multiple spatial directions, frequencies, spot sizes, or offsets, together with appropriate prior constraints and anisotropic heat-diffusion modeling.

The purpose of Table 1 is therefore not to establish a performance hierarchy among thermoreflectance techniques, but to provide qualitative guidance for matching experimental methods and conditions to specific thermophysical characterization problems. Method selection should account for film thickness, thermal-conductivity range, interface depth, thermal anisotropy, relevant thermal-diffusion length scales, instrument bandwidth, and the availability and uncertainty of independently known parameters. Sensitivity and uncertainty analyses should then be used to determine whether the target properties can be resolved with sufficient identifiability and acceptable uncertainty under the selected conditions. Accordingly, the measurement scales and application categories summarized in Table 1 should be interpreted as representative operating regimes rather than fixed applicability boundaries; substantial overlap among techniques is expected, and the appropriate method ultimately depends on the combined sample geometry, target parameter set, and experimental configuration.

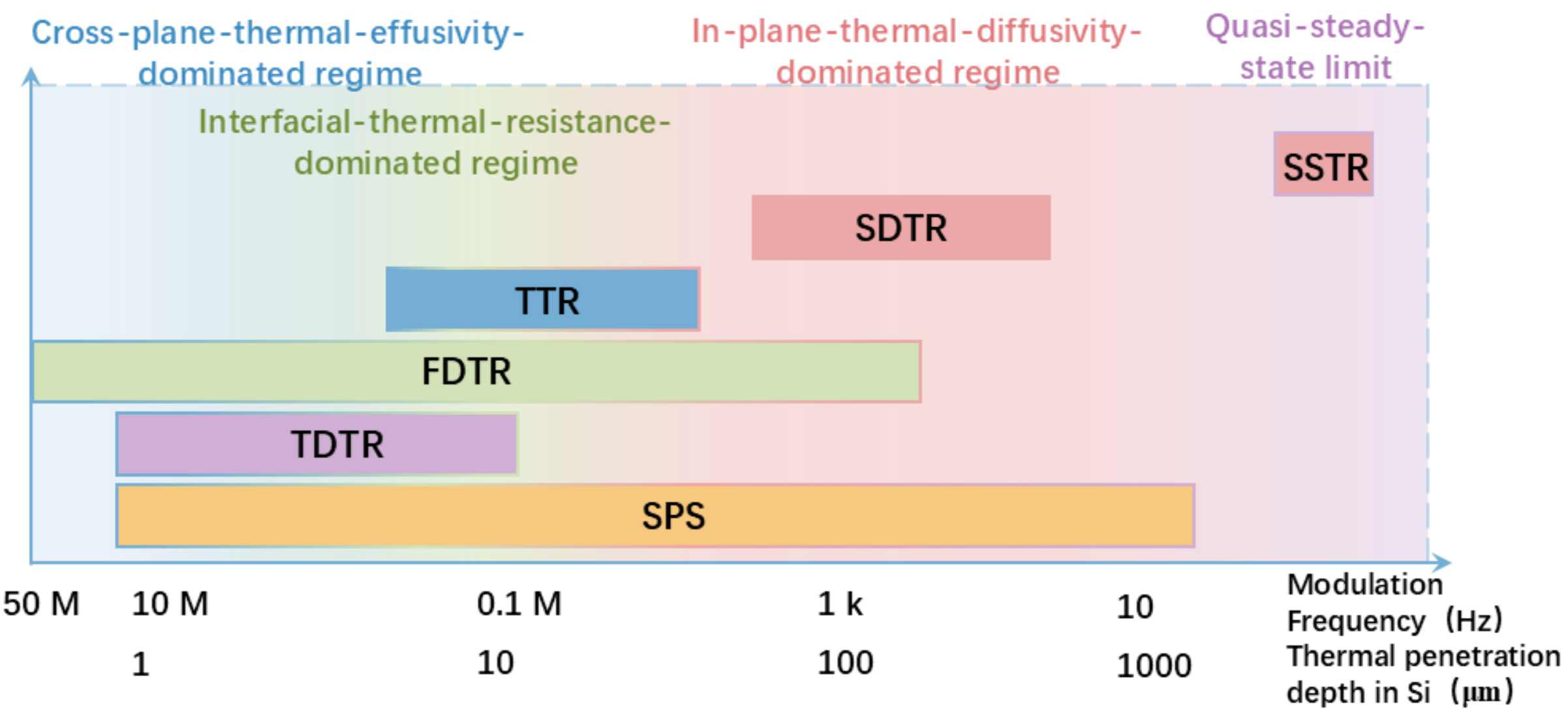

**Figure 4 Accessible effective thermal-diffusion-length ranges of different thermoreflectance methods in single-crystal silicon**

Figure 4 uses single-crystal silicon as a representative reference material to illustrate the approximate characteristic thermal-diffusion-length ranges sampled by different thermoreflectance techniques under representative experimental conditions. The indicated ranges should not be interpreted as rigid thresholds or fixed probing depths; rather, they are reference scales estimated from representative modulation frequencies, time-domain observation windows, and optical-spot dimensions associated with the respective methods. The actual thermal-diffusion length depends on the intrinsic thermal diffusivity of the material, while the effective region contributing to the measured response is further governed by the multilayer architecture, optical geometry, and experimental operating conditions.

Within the submicrometer-to-tens-of-micrometers characteristic length-scale regime, high-frequency TDTR, FDTR, and SPS can weight the thermal response more strongly toward the transducer, near-surface layers, and interfaces and are therefore commonly applied to measurements of cross-plane film thermal conductivity and interfacial thermal conductance. As modulation frequency decreases or the relevant observation time increases, the characteristic thermal-diffusion length increases, and the measured response becomes progressively more sensitive to buried layers, the substrate, and longer-range thermal transport. Low-frequency FDTR and SPS, together with TTR measurements over appropriate transient time windows, can therefore access larger heat-diffusion scales relevant to thicker films, multilayer structures, and bulk materials.

When the characteristic lateral thermal-diffusion length becomes comparable to the laser-spot size or pump–probe offset, lateral heat spreading contributes more strongly to the measured temperature field. By introducing lateral position or beam offset into the measurement geometry, SDTR and related beam-offset methods can enhance sensitivity to in-plane thermal transport and are therefore particularly useful for characterizing in-plane thermal conductivity and thermal anisotropy. At sufficiently long timescales or low modulation frequencies, the thermal response approaches the quasi-steady limit. SSTR directly exploits this regime, while sufficiently low-frequency periodic methods can approach quasi-steady behavior during portions of the excitation cycle, increasing sensitivity to the combined thermal resistance and longer-range heat spreading of the sample system.

The principal purpose of Fig. 4 is therefore to illustrate the continuity and substantial overlap among the characteristic heat-diffusion scales accessible to different thermoreflectance techniques, rather than to define absolute applicability boundaries. No single diffusion-length range uniquely corresponds to a particular technique or thermophysical property, and the appropriate experimental scheme should instead be selected according to film thickness, interface depth, thermal diffusivity, optical-spot size, sample architecture, and the target thermophysical parameters. For polymers, porous materials, and other systems with thermal diffusivities substantially lower than that of single-crystal silicon, the characteristic thermal-diffusion length at a given modulation frequency or observation time is correspondingly smaller. Accessing a comparable cross-plane diffusion length therefore generally requires a lower modulation frequency or a longer observation timescale, whereas adjustment of the laser-spot size can be used to control the relative importance of lateral and cross-plane heat flow.

## 4 Method Matching and Case Studies for Representative Complex Measurement Scenarios

The preceding sections compared thermoreflectance techniques in terms of heat-diffusion scale, parameter identifiability, and in-plane/cross-plane heat transport. This section examines four complex measurement problems and uses representative studies to show how the independent information required by a problem determines the experimental scheme. Several figures in this section are drawn from recent SPS and BO-SPS studies because they provide clear examples of how broadband frequency, periodic-waveform, and spatial-offset degrees of freedom can constrain parameters; their inclusion does not imply that SPS is generally superior to TDTR, FDTR, SDTR, or other methods in these scenarios. For a specific sample, the most appropriate scheme remains dependent on instrument bandwidth, sample dimensions, available prior information, and target uncertainty.

### 4.1 Isotropic Low-Thermal-Conductivity Materials: Joint Estimation of Thermal Conductivity and Volumetric Heat Capacity

Thermoreflectance signals are primarily sensitive to two combinations of sample properties: the in-plane thermal diffusivity ($\alpha = k_r/C$) and a cross-plane thermal-effusivity-related quantity ($e = \sqrt{k_z C}$). For a homogeneous isotropic material, the in-plane and cross-plane thermal conductivities satisfy $k_r = k_z = k$. If complementary experiments can determine $\alpha$ and $e$ independently, thermal conductivity and volumetric heat capacity can then be obtained jointly through the analytical relations $k = e\sqrt{\alpha}$ and $C = e/\sqrt{\alpha}$. Thus, simultaneous characterization of $k$ and $C$ in an isotropic material is not simply a matter of adding another unknown fitting parameter; it requires experimental conditions under which the signal exhibits distinguishable sensitivities to $\alpha$ and $e$ at different heat-diffusion scales, thereby breaking the parameter correlation.

TDTR and FDTR have established a mature technical foundation for such two-parameter estimation. Variable-frequency TDTR continuously adjusts the pump modulation frequency to change thermal penetration depth and has been used to measure $k$ and $C$ simultaneously in bulk and thin-film materials [71]. Dual-frequency TDTR further exploits differences in the thermal response at two modulation frequencies to perform volumetric-heat-capacity imaging with micrometer-scale spatial resolution [72]. In the frequency domain, the FDTR method introduced by Schmidt et al. acquires amplitude and phase as functions of modulation frequency, providing an independent frequency-control dimension for estimating thermal conductivity, thermal diffusivity, and volumetric-heat-capacity-related properties in homogeneous bulk and thin-film systems [10]. Subsequent broadband FDTR studies further demonstrated simultaneous extraction of thermal conductivity and volumetric heat capacity in isotropic materials [73]. More recently, variants including negative-delay TDTR, variable-spot-size TDTR, and beam-offset TDTR/FDTR [53, 69, 74, 75] have expanded the multiparameter-decoupling degrees of freedom of TDTR/FDTR through the delay-time window, beam geometry, and lateral spatial control, respectively. These studies show that conventional time- and frequency-domain thermoreflectance is not limited to single-parameter thermal-conductivity measurement; with coordinated design of multiple experimental variables, joint estimation of $k$, $C$, and interfacial thermal conductance is possible.

For an isotropic low-thermal-conductivity polymer such as poly(methyl methacrylate) (PMMA), however, the principal challenge in joint multiparameter estimation is whether the experimental system can reach sufficiently low modulation frequencies to establish a long heat-diffusion length scale. The thermal diffusivity of PMMA is on the order of $10^{-7}$ $\mathrm{m^2/s}$. With a laser spot on the order of tens of micrometers, the pump modulation frequency must be reduced to the hundreds-of-hertz range or below for the lateral diffusion length to become comparable to the spot radius and for in-plane heat transport to contribute appreciably to the signal. Conventional TDTR provides excellent temporal resolution and mature multilayer thermal models, but its commonly used modulation frequencies are in the MHz range. Constraints associated with the intrinsic laser repetition rate, long-time pulse accumulation, and the stability of low-frequency lock-in detection make it difficult for a conventional TDTR configuration to access directly the ultralow-frequency in-plane diffusion regime required for such polymers. FDTR can increase penetration depth by lowering modulation frequency, but low-frequency measurements generally require much longer integration times and become more susceptible to environmental thermal drift, phase-calibration error, and reduced signal-to-noise ratio. Consequently, when conventional TDTR or FDTR is used alone for low-thermal-conductivity isotropic materials, parameter coupling and low-frequency limitations often make it necessary to prescribe a literature value of volumetric heat capacity or determine heat capacity independently before thermal conductivity can be estimated stably. These are not absolute limits of TDTR or FDTR: lowering modulation frequency, changing spot size, introducing independent heat-capacity information, or jointly fitting multiple conditions can all provide workable solutions, depending on instrument performance and target uncertainty.

For this low-thermal-conductivity case, SPS provides one implementation route for building complementary constraints from a broad frequency range and complete periodic waveforms. A square-wave-modulated heat source is used and the full temperature waveform is recorded, allowing the diffusion length to be varied continuously from hertz to megahertz frequencies. Low-frequency waveforms can capture fully developed in-plane diffusion in a low-thermal-conductivity material, while high-frequency waveforms provide constraints related to near-surface cross-plane thermal effusivity. Compared with approaches such as negative-delay TDTR that mainly extend the use of existing time-domain information, this SPS implementation extends the accessible experimental conditions to lower modulation frequencies and thereby supplies complementary information for decoupling $\alpha$, $e$, and ultimately $k - C$.

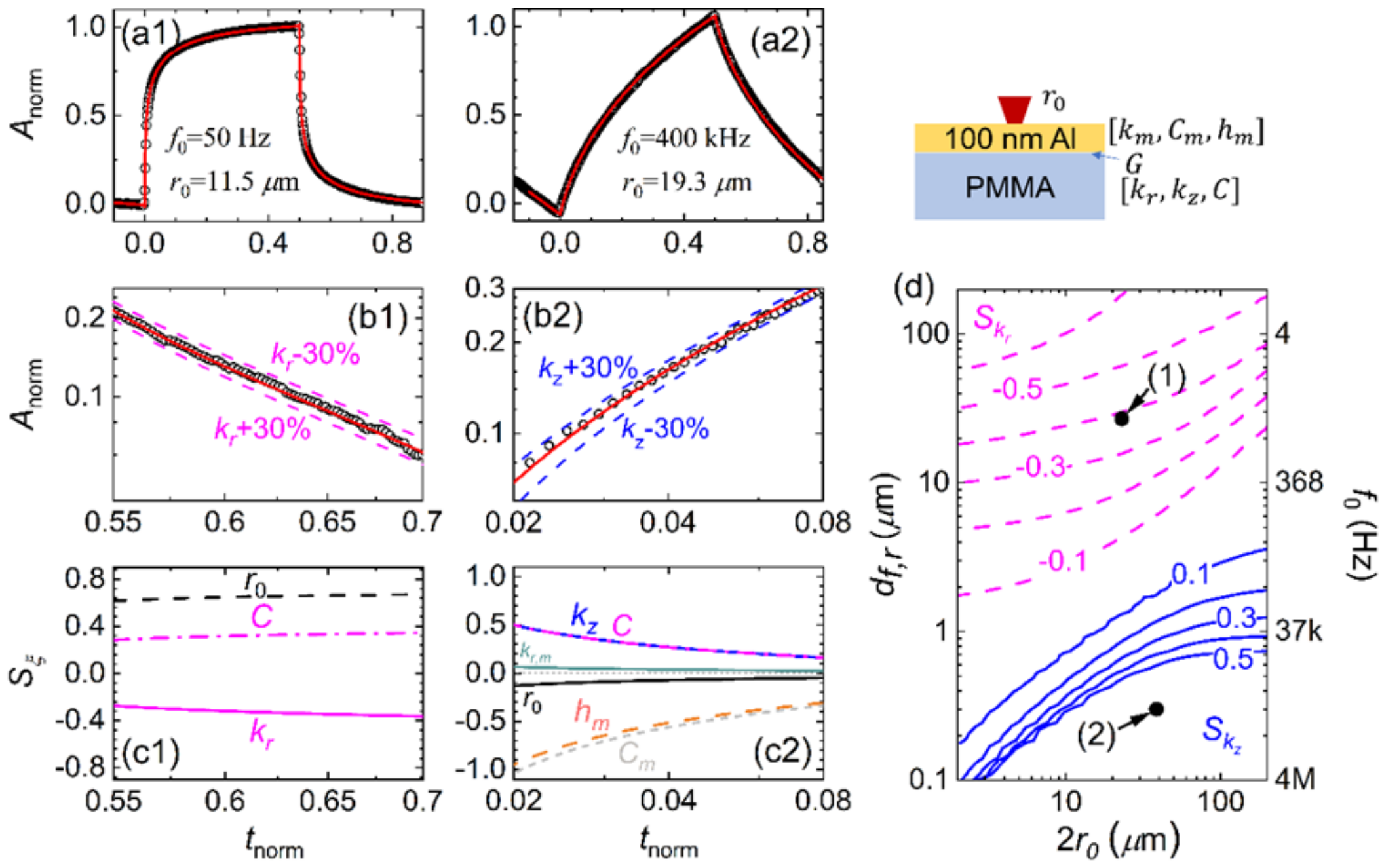

**Figure 5 Joint extraction of in-plane thermal diffusivity and cross-plane thermal effusivity of PMMA using SPS [40]. Copyright © 2024 Elsevier Ltd.**

Figure 5 shows the SPS measurements reported by Chen et al. [40] for joint characterization of the thermal conductivity and volumetric heat capacity of PMMA. Under the low-frequency condition, with a 23 µm laser-spot diameter and a modulation frequency of 50 Hz, the measured signal is primarily sensitive to PMMA $k_r$, $C$, and the spot radius $r_0$, while showing relatively weak sensitivity to the metal-transducer properties and interfacial thermal conductance. Once the spot radius is known, the combined parameter $k_r/C = 0.11557 \times 10^{-6}\ \mathrm{m^2/s}$ can therefore be fitted directly. Under the high-frequency condition, with a 38.6 µm spot diameter and a modulation frequency of 400 kHz, sensitivity shifts to the combined parameter $\sqrt{k_z C}/(h_m C_m)$. Given the metal-transducer areal heat capacity $h_m C_m$, $k_z C = 0.31236 \times 10^6\ \mathrm{W^2 \cdot s/(m^4 \cdot K^2)}$ can then be calculated.

Combining the two independent measurement conditions yields $k = 0.19 \pm 0.01\ \mathrm{W/(m \cdot K)}$ and a volumetric heat capacity of $C = 1.644 \pm 0.11\ \mathrm{MJ/(m^3 \cdot K)}$, with an overall uncertainty of approximately 7%. These results agree well with previously reported PMMA thermal conductivity $0.1934\ \mathrm{W/(m \cdot K)}$ and volumetric heat capacity $1.6104\ \mathrm{MJ/(m^3 \cdot K)}$ [76, 77], indicating that under these experimental conditions SPS can obtain a self-consistent estimate of $k$ and $C$ in an isotropic low-thermal-conductivity material by combining a low-frequency thermal-diffusivity measurement with a high-frequency thermal-effusivity measurement. As a comparison, TDTR applied to the same system using a literature heat-capacity value as an input gives a PMMA $k_z \approx 0.21\ \mathrm{W/(m \cdot K)}$; that comparison did not simultaneously solve for two independent properties, thermal conductivity and volumetric heat capacity [40].

### 4.2 Anisotropic Films: Joint Identification of In-Plane/Cross-Plane Thermal Conductivity and Volumetric Heat Capacity

Thermoreflectance characterization becomes substantially more challenging for anisotropic films than for isotropic materials. For an oriented polyimide (PI) polymer film, for example, the in-plane thermal conductivity $k_r$, cross-plane thermal conductivity $k_z$, and volumetric heat capacity $C$ jointly influence the surface thermoreflectance signal. With only a single modulation frequency, fixed spot size, or single measurement platform, the sensitivity curves of these three properties can readily overlap and become coupled, preventing independent determination of the in-plane conductivity, cross-plane conductivity, and heat capacity.

Early studies of heat transport in PI films commonly used separate platforms or sequential measurements.

Kurabayashi et al. [78] used micromachined cantilever and microbridge resistance methods to determine $k_z$ and the in-plane thermal diffusivity, then used a literature value of $C$ to calculate $k_r$. More recently, Chowdhury et al. [79] combined photothermal displacement phase spectroscopy (D-TOPS) with TDTR to extract $k_r$ and $k_z$, respectively, while still prescribing $C$. These approaches established an important foundation for studying anisotropic transport in PI films. However, because film volumetric heat capacity can be affected by spin coating, stretching, curing, and molecular orientation, direct use of a fixed literature heat-capacity value can introduce additional uncertainty into the extracted thermal conductivity.

The central strategy for characterizing anisotropic films is therefore not merely to maximize sensitivity to one parameter, but to acquire multiple thermal responses within an experimental framework that provide mutually independent and complementary constraints on $k_r$, $k_z$, and $C$. The thermoreflectance response of an anisotropic film can be decomposed into coupled contributions from several parameter combinations, including a cross-plane thermal-effusivity-related term $\sqrt{k_z C}$), an in-plane thermal-diffusivity-related term $k_r/C$, and the areal heat capacity $hC$. Only when experimental conditions produce sufficiently distinct sensitivity signatures for these parameter combinations across different modulation frequencies or characteristic heat-diffusion scales, and the film thickness $h$ is independently known, can $k_r$, $k_z$, and $C$ be separately identified with acceptable confidence.

For the PI-film example, SPS provides one way to construct multiscale constraints within a single platform. Broadband square-wave modulation and full-period waveform acquisition allow the heat-diffusion length to be tuned continuously while producing multiple independent sensitivity constraints. At high excitation frequencies, the thermal disturbance is concentrated mainly in the metal transducer and the near-surface portion of the film, and the signal is more sensitive to cross-plane thermal effusivity and transducer properties. As frequency decreases, the diffusion length increases and the contributions of through-thickness and lateral diffusion within the film grow. Under low-frequency, small-spot conditions, in-plane heat diffusion has a more pronounced effect on the periodic temperature waveform. Jointly fitting the periodic waveforms acquired at different frequencies and spot sizes can therefore yield $k_r$, $k_z$, and $C$ simultaneously without prescribing volumetric heat capacity.

Figure 6 shows the SPS measurements of suspended and spin-coated PI films reported by Zhang et al. [80]. For the suspended PI film (Kapton HN-100), the measured values are $k_r = 0.56\ \mathrm{W/(m \cdot K)}$ and $k_z = 0.175\ \mathrm{W/(m \cdot K)}$, corresponding to an in-plane-to-cross-plane thermal-conductivity anisotropy ratio of approximately 3.2. By contrast, the spin-coated PI film exhibits markedly weaker thermal anisotropy, with an anisotropy ratio of approximately $1.6 - 1.7$, while its cross-plane thermal conductivity increases to $0.29 - 0.34\ \mathrm{W/(m \cdot K)}$. These results indicate that film-processing routes can substantially alter chain orientation and through-thickness packing in PI films and thereby change both in-plane and cross-plane heat conduction.

Polarized Raman spectroscopy further supports this interpretation. The spin-coated PI film shows a higher depolarization ratio than the commercial film, indicating a more nearly isotropic molecular orientation; this is consistent with its lower thermal-conductivity anisotropy and higher measured cross-plane thermal conductivity. Importantly, volumetric heat capacity need not equal a bulk or literature value and can depend on film density, curing history, and molecular packing. Direct measurement of $C$ in such process-sensitive films therefore helps reduce systematic error introduced by an assumed heat-capacity input.

It should also be noted that simultaneous decoupling of $k_r$、$k_z$ and $C$ generally requires the film thickness and thermal penetration depth to fall within an appropriate relative range so that the film areal heat capacity $hC$ becomes an independently observable quantity. When the penetration depth is much smaller or much larger than the film thickness, the signal can degenerate toward a near-surface or semi-infinite response; under these conditions, only $\sqrt{k_z C}$ or $k_r/C$ may be accessible and $C$ becomes difficult to separate independently. The broad frequency range of SPS makes it possible to search a wide range of diffusion lengths for suitable sensitivity windows and thereby improve the stability of multiparameter estimation in anisotropic films.

Overall, the PI-film example illustrates a common difficulty in low-thermal-conductivity anisotropic films: in-plane thermal conductivity, cross-plane thermal conductivity, and volumetric heat capacity are strongly coupled in the thermoreflectance signal, and a single experimental condition rarely provides sufficient independent constraints on all three. TDTR, D-TOPS, and micromachined electrothermal techniques have provided an important basis for studying in-plane and cross-plane transport; the SPS example demonstrates a feasible route for joint estimation of $k_r$、 $k_z$ and $C$ on a single platform through broadband periodic-waveform measurements. This capability is relevant to thermal design of flexible electronics, heat dissipation in polymer packaging, and studies of thermal-management mechanisms in process-sensitive functional films. The same principle of adding independent constraints can also be implemented using multifrequency or multispot TDTR, BO-TDTR/FDTR, or cross-platform measurements; the essential requirement is that the observations contain enough information to distinguish the target parameters.

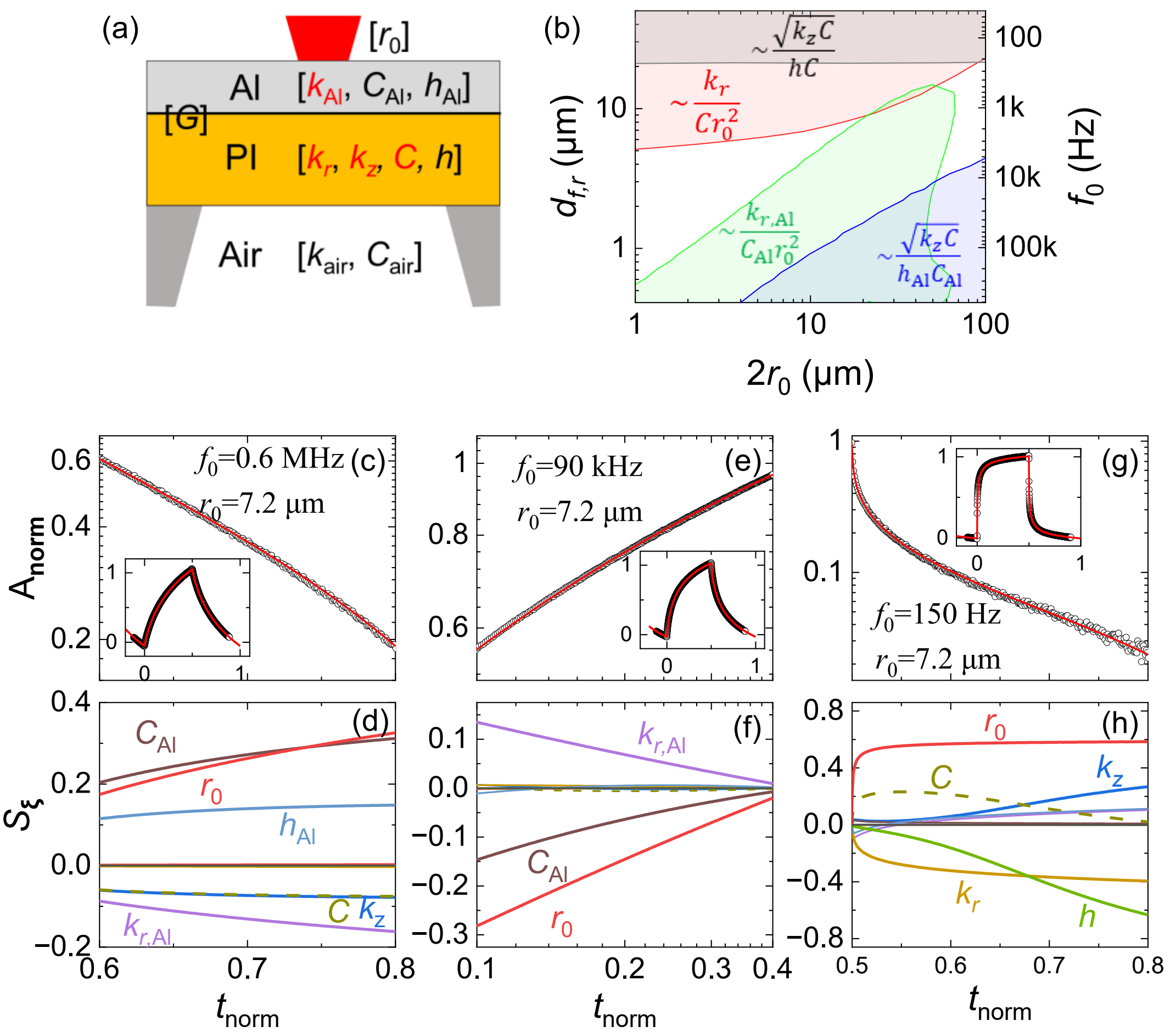

**Figure 6 Simultaneous extraction of in-plane thermal conductivity, cross-plane thermal conductivity, and volumetric heat capacity of suspended PI films using SPS [80]. Copyright © 2025 Elsevier Ltd.**

### 4.3 Thermoreflectance Measurement of the Three-Dimensional Anisotropic Thermal-Conductivity Tensor

For an anisotropic material, thermal conductivity is no longer a scalar but a second-rank tensor. In many layered materials and oriented films, the dominant anisotropy appears as a difference between in-plane and cross-plane thermal conductivity and can therefore often be approximated using $k_r$ and $k_z$. In low-symmetry crystals, off-cut crystals, or materials whose crystallographic axes are not aligned with the device coordinate system, however, heat flux and temperature gradient need not be parallel, and the thermal-conductivity tensor in the experimental coordinate system can contain off-diagonal components. In such cases, a thermoreflectance signal no longer reflects an effective conductivity along a single direction, but is jointly determined by multiple tensor components, making reconstruction of the full three-dimensional conductivity tensor a substantially more challenging measurement problem.

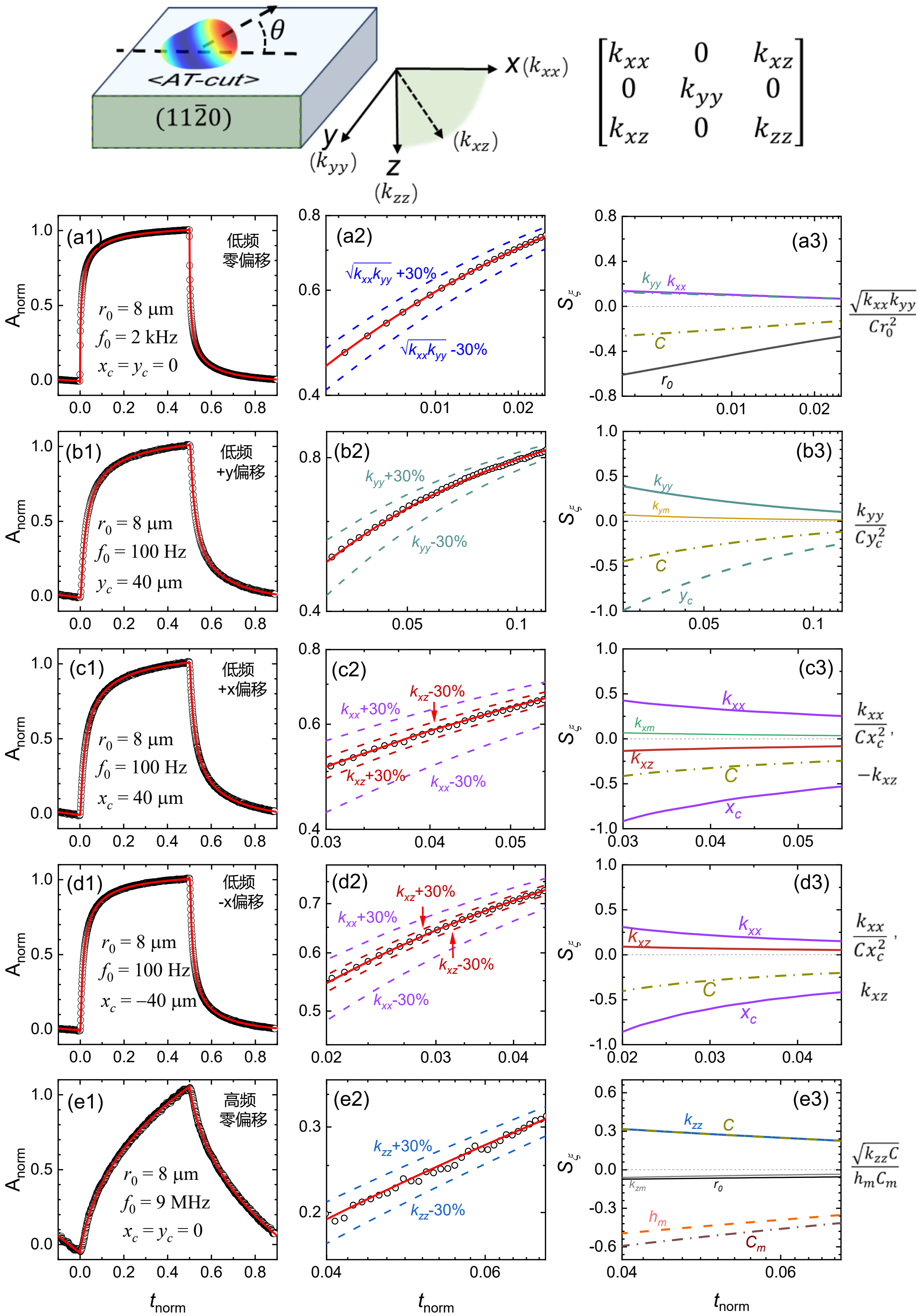

**Figure 7 Reconstruction of the full three-dimensional thermal-conductivity tensor of AT-cut quartz using beam-offset SPS [43]. Copyright © 2024 Elsevier Masson SAS.**

*Artwork label translations: (a1) low frequency, zero offset; (b1) low frequency, +y offset; (c1) low frequency, +x offset; (d1) low frequency, -x offset; (e1) high frequency, zero offset.*

Conventional thermoreflectance techniques including TDTR, FDTR, and SDTR have established a mature framework for quantitative characterization of anisotropic heat conduction. Variable-spot-size TDTR [53] and multifrequency TDTR [71] vary spot size or modulation frequency to change the relative contributions of in-plane and cross-plane diffusion and thereby estimate $k_r$ and $k_z$ in films or crystals. Elliptical-beam TDTR [75], beam-offset TDTR [67], and beam-offset FDTR [69] further introduce lateral spatial resolution so that differences in heat diffusion along different in-plane directions can be detected. SDTR directly scans the spatial distribution of the surface thermal response and provides an especially intuitive physical picture for measuring in-plane thermal diffusivity and in-plane anisotropy [39]. Together, these developments have expanded thermoreflectance from early measurements dominated by cross-plane transport to quantitative analysis of directional heat transport with angular resolution.

Even so, measuring only effective in-plane and normal-direction conductivities is insufficient to reconstruct a complete three-dimensional conductivity tensor. In particular, when the crystallographic principal axes are not aligned with the sample-surface coordinate system, the thermal-conductivity tensor in the experimental coordinates can be written as

$$\mathbf{k} = \begin{bmatrix} k_{xx} & k_{xy} & k_{xz} \\ k_{yx} & k_{yy} & k_{yz} \\ k_{zx} & k_{zy} & k_{zz} \end{bmatrix} \tag{5}$$

Both diagonal and off-diagonal terms can influence the surface temperature field. The contributions of different tensor components to the thermoreflectance signal vary with modulation frequency, beam-offset direction, and probe position, and the components can be strongly coupled. A single modulation frequency, one spatial scan direction, or a fit that prescribes part of the tensor is therefore generally insufficient for stable reconstruction of the complete tensor. Three-dimensional tensor estimation requires enough independent frequency, spatial, and directional constraints, together with sensitivity analysis to determine whether the individual components are separable.

AT-cut quartz provides a representative example. Its crystallographic c axis (the optical or Z axis of quartz) is inclined relative to the sample surface, with an angle of approximately $35.25°$ between the cut plane and the c axis. In the experimental coordinate system, the conductivity tensor contains not only the principal in-plane components ($k_{xx}$, $k_{yy}$) and cross-plane component ($k_{zz}$), but also, critically, an off-diagonal coupling term ($k_{xz}$) that directly governs coupling between in-plane and cross-plane heat flow. In other words, the surface thermal response cannot be attributed simply to one in-plane or one cross-plane conductivity; the complete three-dimensional conductivity tensor in the experimental coordinate system must be considered. Figure 7 illustrates the BO-SPS strategy used by Chen et al. [43] to measure the conductivity tensor of AT-cut quartz. The optical configuration combines low-frequency modulation (~100 Hz) with a large beam offset (greater than five times the spot radius), producing a pronounced anisotropic temperature gradient at the sample surface and enabling parameter decoupling through directional inversion symmetry. When the probe is displaced in the positive and negative directions along the $x$ axis, the signal is symmetric with respect to $k_{xx}$ but antisymmetric with respect to $k_{xz}$ (Fig. 7c-d). This even/odd difference arises directly from heat-flow asymmetry introduced by the inclined crystallographic axis. By combining positive and negative offsets along $x$ (Fig. 7c-d), zero offset (Fig. 7a), and offsets along $y$ (Fig. 7b), $k_{xx}$ and $k_{xz}$ can be constrained separately from $k_{yy}$, while high-frequency measurements at zero offset constrain the cross-plane transport component $k_{zz}$ (Fig. 7e).

The experimental results show that, in this case, a small set of offset directions and modulation-frequency configurations with complementary sensitivities provides sufficient constraints for reconstructing the three-dimensional conductivity tensor and jointly determining multiple tensor components in the experimental coordinate system. After diagonalization, the principal thermal conductivities $k_c = 10.76\,\mathrm{W/(m\cdot K)}$ and $k_a = 6.40\,\mathrm{W/(m\cdot K)}$ and the inclination angle $\theta = -35.39°$ of the $c$ axis are obtained. Agreement of the recovered angle with the known cut angle supports the feasibility of inferring the principal directions of three-dimensional heat transport from surface thermal responses.

Overall, the AT-cut quartz example shows that the essential requirement for three-dimensional anisotropic measurement is sufficient directional spatial information, together with multiple heat-diffusion scales that alter the relative responses of the tensor components. BO-SPS provides one implementation in this particular case; the same methodological principle can also be explored on other thermoreflectance platforms capable of multidirectional spatial scanning and multi-condition control.

### 4.4 Multilayer Heterostructures: Selective Estimation of Combined Parameters

In wide-bandgap semiconductor devices and heterogeneous integration structures, heat transport commonly involves multiple functional films, buffer layers, buried interfaces, and supporting substrates. In representative power-device structures such as GaN/Si, GaN/SiC, and GaN/diamond, heat generated during operation must be dissipated sequentially through metal contacts, the GaN epitaxial layer, transition or buffer layers, semiconductor/substrate interfaces, and a high-thermal-conductivity substrate [55]. The thermoreflectance signal is therefore influenced simultaneously by film thermal conductivity, volumetric heat capacity, interfacial thermal conductance, layer thickness, metal-transducer parameters, and substrate properties. If the in-plane conductivity, cross-plane conductivity, volumetric heat capacity, interfacial thermal conductance, and thickness of every layer are all treated as independent fitting parameters, the number of unknowns grows rapidly and the inverse problem becomes highly ill-posed. The key to multilayer-heterostructure measurement is therefore not simply to increase the number of fitting parameters, but to identify the combinations of parameters to which the signal is genuinely sensitive and determine whether those combinations can be separated using different frequencies, heat-diffusion scales, or portions of the waveform.

TDTR and FDTR have developed a substantial application base for thermophysical characterization of wide-bandgap semiconductor heterostructures. For epitaxial GaN/SiC, Cho et al. used TDTR to study samples with different GaN thicknesses and obtained the GaN film thermal conductivity and GaN-SiC interfacial thermal conductance, providing important experimental information for thermal-resistance analysis in HEMT devices [81]. TDTR was subsequently applied to bonded GaN/SiC, GaN/SiC with AlN transition layers, and epitaxial GaN/SiC, GaN/AlN, and AlN/SiC wafers to examine how epitaxial quality, transition layers, and interface structure affect interfacial thermal conductance [82-84]. For integration with high-thermal-conductivity substrates, Cheng et al. used TDTR to measure the thermal conductance of bonded GaN/diamond interfaces and combined the thermal measurements with structural analysis to show how bonding-layer thickness and interface quality affect heat dissipation [85]. TDTR has also been used for ultrawide-bandgap heterogeneously integrated systems such as β-$Ga_2O_3$/SiC to measure film thermal conductivity and buried-interface thermal conductance, and a dual-modulation-frequency TDTR imaging method has been developed to visualize spatial variations in the buried interfacial thermal conductance of β-$Ga_2O_3$/SiC [86].

FDTR continuously scans modulation frequency to vary thermal penetration depth, causing the relative contributions of different films, interfaces, and the substrate to change with frequency and providing another route for multilayer parameter estimation. Ziade et al. used FDTR to measure the temperature-dependent interfacial thermal conductance of epitaxial GaN/SiC [87]. Song et al. combined FDTR and SSTR to characterize GaN-on-SiC wafers and used the measured properties to validate device-level thermal models [88]. Delmas et al. used spatially resolved FDTR to map heat transport and stress across a compression-bonded GaN/diamond interface [89]. These studies show that changing frequency and spatial conditions can provide different sensitivity weights for multilayer and buried-interface problems, but this should not be interpreted as strict tomographic depth imaging.

For a multilayer stack such as GaN/Si, in which the cross-plane thermal-resistance distribution spans multiple characteristic scales (Fig. 8a1), however, one or a few fitted parameters are often insufficient to represent the actual heat-transport behavior. The near-surface GaN layer, AlGaN buffer layer, AlN transition layer, buried GaN/Si interface, and Si substrate can all contribute to the thermoreflectance signal, and their relative contributions change with modulation frequency and heat-diffusion length. If all individual layer properties are fitted directly without first determining which parameter combinations are actually measurable, the optimization may converge mathematically to a result that is not physically unique. Identifying the experimentally measurable parameters is therefore itself a central problem in thermoreflectance inversion for complex multilayer systems.

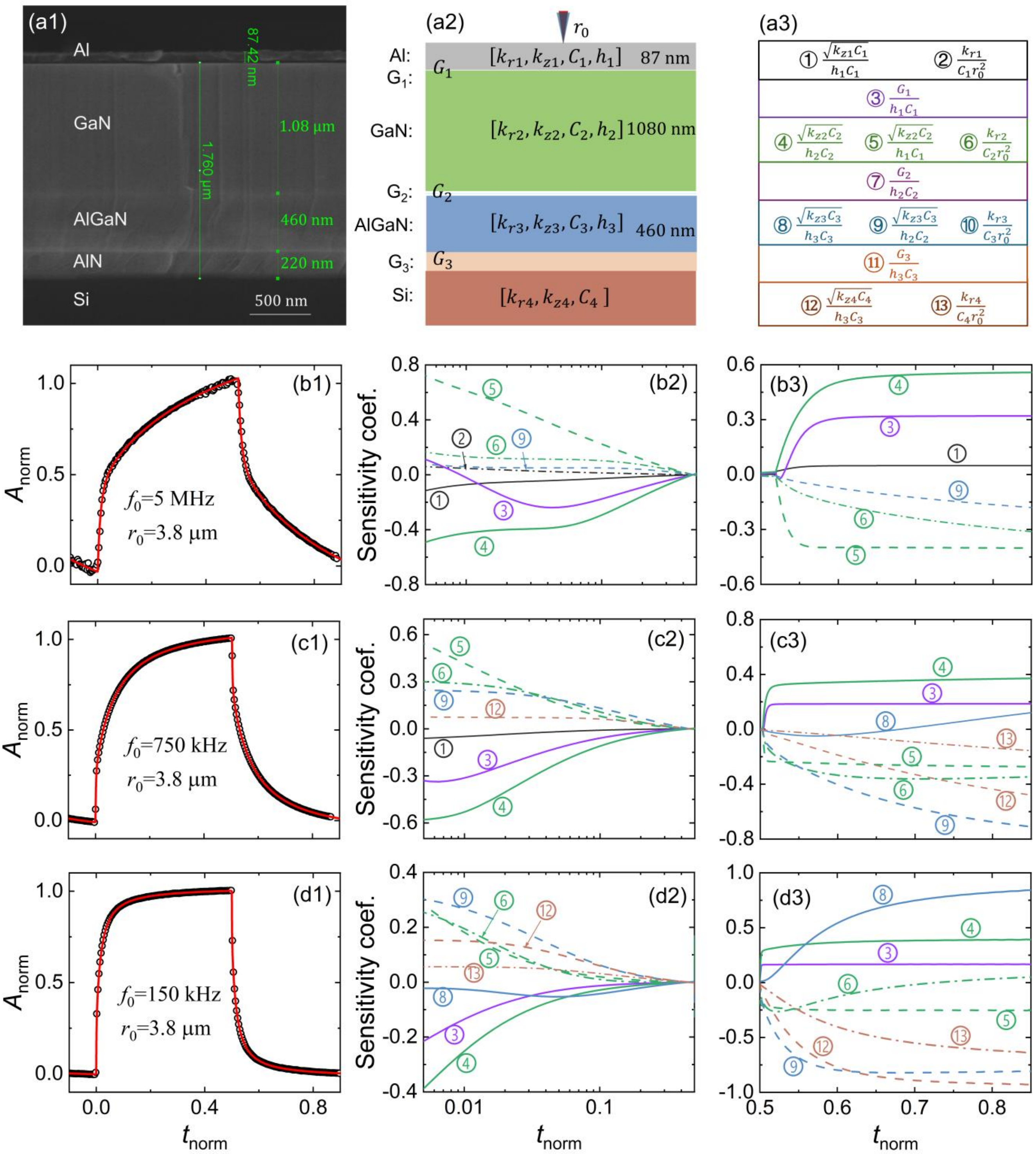

**Figure 8 Combination-parameter analysis and thermophysical-property extraction for a GaN/Si multilayer heterostructure using SPS [55]. Copyright © 2025 American Physical Society.**

Dimensionless-parameter analysis provides a systematic approach to this problem. By nondimensionalizing the heat-diffusion equations and combining the result with the normalized thermoreflectance signal and parameter-sensitivity analysis, the original physical parameters can be reorganized into a smaller set of combined parameters with clear physical meaning [55]. For a film layer, cross-plane diffusion can be characterized by $\tau_z = h^2C/k_z$, representing the characteristic time for heat to traverse a thickness $h$; in-plane diffusion can be characterized by $\tau_r = Cr_0^2/k_r$, representing the characteristic time for heat to diffuse to the spot boundary and thus reflecting radial heat dissipation; and interfacial thermal relaxation can be characterized by $\tau_G = hC/G$, representing the timescale over which the areal heat capacity of the film releases heat through the interface. Multiplication by the modulation frequency gives strictly dimensionless combinations such as $fh^2C/k_z$, $fCr_0^2/k_r$, and $fhC/G$. These combined parameters have direct physical interpretations corresponding to cross-plane diffusion, lateral diffusion, and interfacial thermal relaxation, rather than being merely mathematical regroupings of variables.

Figure 8 summarizes the study of a GaN/Si multilayer heterostructure by Chen et al. [55] using SPS and combined-parameter analysis. The sample consisted of an approximately 87 nm Al transducer, 1.08 μm GaN, 460 nm AlGaN, 220 nm AlN, and a Si substrate (Fig. 8a1). A direct full multilayer thermal model contains 19 physical parameters, including layer thermal conductivities, volumetric heat capacities, thicknesses, interfacial thermal conductances, and spot parameters (Fig. 8a2). Nondimensionalization and combined-parameter analysis reorganize these into 13 combined parameters (Fig. 8a3). Because the AlN layer is thin and has relatively low thermal resistance, its contribution can be combined with those of adjacent interfaces into an equivalent thermal resistance, further reducing model complexity. The value of combined-parameter analysis lies not only in reducing the dimensionality of the inverse problem, but more importantly in identifying which combinations are genuinely measurable in a complex multilayer system.

In this SPS example, different modulation frequencies produce clear changes in the relative sensitivities to the combined parameters. At 5 MHz (Fig. 8b1-b3), the thermal response is concentrated mainly in the Al transducer and the near-surface portion of the GaN. At 750 kHz (Fig. 8c1-c3), the diffusion region extends farther into the GaN and AlGaN buffer. At 150 kHz (Fig. 8d1-d3), substrate-related parameters carry greater weight. Thus, the "layer selectivity" in this case is more accurately understood as a frequency-dependent redistribution of sensitivity among layers and combined parameters, rather than independent layer-by-layer imaging or strict depth separation.

In addition to the sensitivity changes provided by frequency, this SPS example uses the heating and cooling portions of the periodic waveform to increase discrimination among combined parameters. Near-surface heat capacity and interfacial thermal conductance, cross-plane film diffusion, in-plane diffusion, and substrate properties do not influence every part of the waveform in exactly the same way. Jointly fitting full-period responses at multiple frequencies allows candidate combined parameters with sufficiently high sensitivity and relatively low correlation to be selected, after which known layer thicknesses, spot size, and transducer areal heat capacity can be used to recover specific thermophysical properties.

Using this combined-parameter estimation strategy, Chen et al. [55] obtained several key thermophysical properties of the GaN/Si multilayer structure: Al/GaN interfacial thermal conductance $G_1 = 125 \pm 5\ \mathrm{MW/(m^2 \cdot K)}$ (from parameter combination ③); cross-plane GaN thermal conductivity $k_{z2} = 147 \pm 10\ \mathrm{W/(m \cdot K)}$ (from combinations ④ and ⑤); in-plane GaN thermal conductivity $k_{r2} = 144 \pm 11\ \mathrm{W/(m \cdot K)}$ (from ⑥); GaN volumetric heat capacity $C_2 = 2.7 \pm 0.14\ \mathrm{MJ/(m^3 \cdot K)}$ (from ④, ⑤, and ⑥); and AlGaN thermal conductivity $k_{z3} = 16 \pm 2.5\ \mathrm{W/(m \cdot K)}$ (from ⑧). The low-frequency response also constrains properties such as the thermal conductivity and volumetric heat capacity of the Si substrate. Relative to directly fitting a large number of individual physical parameters, this strategy first determines the combinations to which the signal is genuinely sensitive and then converts the measurable combinations into physically meaningful layer thermal conductivities, volumetric heat capacities, or interfacial thermal conductances.

The broader significance of this example is the principle of determining what can be measured before estimating individual properties. In a multilayer heterostructure, the thermoreflectance signal often constrains combinations formed from characteristic diffusion times, thermal resistances, and areal heat capacities before it constrains every layer property independently. In this particular study, SPS uses multiple frequencies and periodic waveforms to generate complementary conditions, but the principle of starting from combined parameters and reducing correlations through multiple conditions is not platform-specific; it can also be applied to TDTR, FDTR, and other methods that vary temporal, frequency, or spatial observation conditions.

### 4.5 From Case Studies to Method Selection: Observation Information Before Method Labels

Taken together, these examples show that the key to a complex thermoreflectance measurement is not to choose a particular technique first, but to identify which thermophysical properties must be distinguished and then determine what mutually independent temporal, frequency, spatial, or waveform information the available experiment can provide. The frequent use of SPS and BO-SPS examples in this section reflects the availability of representative recent studies and figures, not a ranking of method performance. For the same scientific problem, multifrequency TDTR/FDTR, variable-spot or beam-offset measurements, SPS, SDTR, and cross-platform combinations can all be viable. The final choice should be evaluated on parameter identifiability, measurement uncertainty, experimental complexity, and reproducibility.

# 5 Challenges, Trends, and Outlook

## 5.1 Ill-Posed Inversion and Uncertainty Quantification

Sections 1.4 and 3.2 discussed identifiability in thermoreflectance inversion from the perspectives of local sensitivity and parameter correlation. For multilayer heterostructures, anisotropic materials, and complex devices, however, local sensitivity analysis alone is not sufficient to establish the uniqueness or credibility of the inferred parameters. Even when the experimental signal is reproduced well by a heat-diffusion model, the objective function may contain multiple parameter combinations that are statistically or practically indistinguishable within the experimental uncertainty. Strong parameter correlations, dependence on initial guesses, and uncertainties in nominally fixed inputs can further affect the inferred results. The central challenge in complex inverse problems is therefore not simply to obtain a satisfactory fit, but to determine which parameters or parameter combinations are actually constrained by the available data and to quantify the uncertainty associated with those estimates.

Error propagation and Monte Carlo sampling are widely used to propagate experimental and input uncertainties into inferred thermophysical properties, while Bayesian inversion provides a probabilistic framework for multiparameter inference. Bayesian approaches combine prior information, experimental observations, and an assumed measurement-error model to construct posterior distributions that describe parameter uncertainty and correlation conditioned on the adopted physical and statistical models [91]. Compared with reporting only a best-fit value and an associated standard uncertainty, posterior distributions can more directly reveal poorly constrained parameters, non-Gaussian uncertainty, multimodality, and correlations among inferred quantities. Such approaches shift thermoreflectance analysis from point estimation toward uncertainty-aware parameter inference, although the resulting posterior remains dependent on the assumed priors, likelihood, and forward-model fidelity.

Parameter uncertainty alone, however, does not fully characterize the credibility of a thermoreflectance measurement. Most inversion procedures assume that the selected heat-diffusion model adequately represents the underlying transport physics. When the relevant thermal length and time scales approach regimes in which Fourier diffusion becomes inadequate, or when complex interfaces, nonequilibrium processes, or other unmodeled physics become important, the forward model itself can introduce systematic bias. Under such conditions, a small fitting residual may simply indicate that adjustable parameters compensate for model inadequacy rather than that the inferred properties accurately represent the physical system. Uncertainty analysis for complex systems should therefore distinguish uncertainty in fitted parameters and experimental inputs from uncertainty or discrepancy associated with the forward model. The applicable regime of the model should also be assessed explicitly so that fitted thermophysical parameters are not inadvertently used to absorb deficiencies in the assumed transport description.

Uncertainty quantification can further be incorporated directly into experimental design rather than being applied only after parameter fitting. Conventional measurements often select modulation frequencies, delay-time windows, laser-spot sizes, or spatial offsets in advance and subsequently invert the resulting data. In strongly coupled systems, however, collecting additional observations under conditions with similar sensitivity signatures may provide little new information about the unknown parameters. Sensitivity matrices, parameter-correlation measures, and the Fisher information matrix can be used before measurement to evaluate the expected information content of candidate experimental conditions and identify configurations that improve parameter separability [92]. After initial measurements are acquired, subsequent conditions can in principle be selected according to the updated parameter uncertainty or expected information gain, establishing a closed loop among experimental design, data acquisition, parameter inference, and uncertainty updating. Such adaptive strategies seek not merely to increase the quantity of data, but to acquire measurements that most effectively reduce uncertainty or distinguish among competing parameter combinations. This framework also provides a natural foundation for incorporating fast surrogate models and machine-learning-assisted optimization into thermoreflectance experiments.

Future thermoreflectance data analysis is therefore likely to place increasing emphasis on credible inference, model adequacy, and information-efficient experimental design, rather than on fitting speed or residual minimization alone. Combining probabilistic inversion, model-discrepancy assessment, computationally efficient surrogate models, and adaptive experimental design could enable measurement conditions to be selected dynamically according to the information already acquired. Such a closed-loop framework can target experimental conditions that reduce parameter correlation, constrain the most uncertain quantities, and avoid redundant measurements. For complex materials and devices, these developments could move thermoreflectance beyond conventional offline parameter fitting toward adaptive, uncertainty-aware characterization in which measurement and inference are jointly optimized according to parameter identifiability and information gain.

### 5.2 Scale Effects and Model Boundaries in Extending Spatiotemporal Resolution

The effective spatiotemporal resolution of thermoreflectance measurements is jointly constrained by multiple characteristic scales, including thermal penetration depth, phonon mean free path, laser-spot size, and metal-transducer thickness. Changes in modulation frequency, spot size, and detection scheme affect not only the spatial and temporal scales sampled by the experiment but also the relative importance of different heat-transport processes in the measured response. Further improvement in spatiotemporal resolution therefore cannot be understood simply as extending measurements to higher frequencies or smaller optical spots. Instrumental resolution, the physical origin of the measured signal, characteristic transport length and time scales, and the validity of the adopted forward model must instead be considered together.

At high modulation frequencies, the thermal penetration depth decreases and can become comparable to the thickness of an Al or Au transducer. Under such conditions, an increasing fraction of the temperature response is localized within the transducer and its immediate thermal environment, and the measured signal can become more sensitive to transducer heat capacity and metal/sample interfacial thermal conductance while exhibiting reduced sensitivity to the intrinsic thermophysical properties of the underlying material [94]. Reducing the transducer thickness can partially alleviate this limitation, but may also alter optical absorption and thermoreflectance sensitivity and increase the relative importance of uncertainties in film thickness, continuity, morphology, and interface quality. Transducerless thermoreflectance provides an alternative by exploiting the material's intrinsic optical or electronic response for direct excitation and/or detection [95–98], thereby avoiding thermal and interfacial contributions introduced by a deposited metal film. This approach, however, places more stringent requirements on the optical response, surface condition, and signal stability of the material itself. The transducer should therefore be regarded not merely as an optical component but as part of the thermal measurement system whose dimensions and properties can ultimately constrain the accessible heat-diffusion scale.

As the characteristic thermal length scales—such as the thermal penetration depth or laser-spot size—become comparable to relevant phonon mean free paths, Fourier's local diffusive description can progressively lose validity, and quasiballistic or more generally nondiffusive transport can contribute appreciably to the measured response. Because heat in real materials is carried by phonons spanning broad distributions of mean free paths and relaxation times, this transition generally occurs over a range of experimental length and time scales rather than at a single sharply defined boundary. Under such conditions, an effective or apparent thermal conductivity inferred using a purely diffusive model can exhibit systematic dependence on modulation frequency, spot size, or other experimental length scales [99–101]. TDTR and FDTR have been used to investigate phonon mean-free-path-dependent transport and to infer cumulative contributions to thermal conductivity [8, 102, 103]. Such inverse problems, however, are intrinsically model-dependent and can require assumptions regarding phonon dispersion, relaxation-time distributions, or the form of the suppression function relating quasiballistic transport to the measured response. In multilayer heterostructures, spectral mismatch, mode-dependent transmission, and interfacial scattering can occur simultaneously, making intrinsic bulk phonon transport and interface-specific transmission difficult to disentangle from thermoreflectance data alone [104]. Further investigation of nonequilibrium interfacial transport will therefore require closer integration of experimental inversion with spectrally resolved or nongray Boltzmann transport models and microscopic calculations of mode-dependent interfacial transport [105], rather than representing complex interface effects solely through an effective bulk mean free path or a single phenomenological parameter.

At the low-frequency end, the dominant limitations increasingly arise from measurement stability, acquisition time, and low-frequency noise. As the modulation period becomes longer, ambient-temperature drift, laser-power fluctuations, sample-stage instability, and other low-frequency disturbances can become increasingly important relative to the thermal signal. For low-frequency or steady-state absolute thermal-conductivity measurements of bulk materials, established approaches such as the (3\omega) method and steady-state calorimetry can offer complementary advantages in measurement stability and absolute-property determination [106, 107]. Thermoreflectance measurements at long timescales are therefore particularly valuable when their intrinsic advantages—noncontact optical interrogation, localized heating and detection, microscale spatial selectivity, and tunable heat-flow geometry—are important to the measurement problem. Rather than requiring a single thermoreflectance platform to span all relevant timescales, complementary use of optical, electrical, and calorimetric techniques can provide more robust coverage across different heat-diffusion regimes.

Further improvement in spatial resolution is ultimately constrained by the far-field optical diffraction limit. Near-field optical approaches provide a possible route to nanoscale optical interrogation beyond conventional diffraction-limited spot sizes [108]. Once excitation or detection occurs in the optical near field, however, localized electromagnetic fields, evanescent coupling, and the local electromagnetic density of states can influence energy deposition and signal formation [109]. The measured thermoreflectance response may then depend not only on heat diffusion but also on probe geometry, probe–sample separation, material dielectric functions, and electromagnetic boundary conditions. Consequently, interpreting a near-field optical signal as a local temperature measurement requires careful separation of thermal effects from changes in the local optical interaction itself. Reliable extraction of temperature and thermophysical properties in this regime will therefore require coupled electromagnetic–thermal forward models, together with quantitative assessment of probe perturbation, spatial averaging, and local optical-response effects on the inverse problem.

As measurement dimensions continue to shrink and thermoreflectance is increasingly applied to operating devices and geometrically complex structures, sample geometry itself becomes an important constraint on model validity. Heat-flow paths in curved structures, nonuniform films, micro- and nanodevices, and flexible electronics are governed jointly by local geometry, heterogeneous material distributions, heat-source dimensions, and boundary conditions and may no longer be represented adequately by conventional one-dimensional or axisymmetric multilayer models. Coupling finite-element or other numerical forward models with parameter inversion allows experiment-specific geometry, spatially varying material properties, localized heat sources, and realistic boundary conditions to be incorporated directly into the forward problem [110]. Such numerical inversion frameworks can extend thermoreflectance beyond idealized planar structures toward complex engineering devices, although their increased number of geometric and material inputs also makes uncertainty propagation, model validation, and computational efficiency increasingly important.

Overall, further improvement in thermoreflectance spatiotemporal resolution will increasingly depend on co-designing experimental resolution with transport modeling and inversion fidelity. The objective should therefore not be higher modulation frequency, shorter timescale, or smaller optical spots alone, but the development of multiscale forward and inverse frameworks whose physical description remains appropriate as the relevant transport regime changes. Such frameworks may need to incorporate transducer effects, nondiffusive phonon transport, near-field electromagnetic coupling, and complex device geometry, while quantifying the uncertainty and model discrepancy introduced at each scale. Only when increased experimental resolution is accompanied by commensurate model fidelity and parameter identifiability can thermoreflectance measurements be interpreted reliably in regimes involving nonequilibrium interfacial transport and highly localized heat transfer in real micro- and nanodevices.

### 5.3 Multiphysics Measurements Under Device Operating Conditions and Engineering Applications

As thermoreflectance moves from standard thin-film samples toward practical devices, measurements are increasingly extending from ex situ or near-equilibrium characterization to in situ and operando measurements under realistic operating conditions. Electrical excitation, transient self-heating, mechanical deformation, and other physical processes can occur simultaneously during device operation. Coupling among these processes can modify the underlying heat-transport behavior while also introducing nonthermal or indirectly thermal contributions to the measured optical response. The central challenge for device-oriented thermoreflectance is therefore not simply to achieve higher spatiotemporal resolution, but to separate coupled physical contributions quantitatively while retaining the electrical, thermal, and mechanical conditions relevant to actual device operation.

In wide-bandgap power devices, transient self-heating is strongly coupled to carrier transport, and the local temperature field evolves dynamically with switching conditions, current distribution, and power dissipation. Measurements performed after interrupting or substantially perturbing the electrical bias may therefore fail to reproduce the thermal state present during active operation. Recently developed on-the-fly thermoreflectance synchronizes electrical excitation with optical probing to capture in situ temperature evolution within selected temporal windows during device transients [111]. The measured reflectance can also contain electrically modulated contributions associated with electrothermal and thermoelectric phenomena, including the Seebeck effect [112]. Distinguishing temperature-induced reflectance changes from reflectance variations associated with electrical excitation or other coupled mechanisms may provide access to complementary information on electrothermal transport, but requires appropriate modulation schemes, calibration, and physical modeling. Thermoreflectance is thus evolving from conventional temperature sensing toward operando characterization of coupled electrical and thermal behavior, with reliable separation and interpretation of the different signal contributions remaining a central challenge.

In flexible electronics, polymer-based devices, and advanced packaging, heat transport is also frequently coupled to mechanical deformation and evolving interface conditions. Thermal cycling, mechanical bending, or applied strain can alter interfacial contact, local microstructure, and material properties, thereby modifying interfacial thermal conductance and lateral heat spreading. At the same time, thermal expansion and thermoelastic deformation can modify the optical response or provide an additional measurable observable. Photothermal deformation techniques exploit surface displacement or curvature changes to infer thermophysical or thermomechanical properties such as thermal conductivity and thermal-expansion coefficient [113–115]. Combining thermoreflectance-based temperature sensing with photothermal deformation measurements and controlled in situ mechanical loading could therefore provide complementary observables for tracking the coupled evolution of temperature, deformation, and interface state during strain or thermal cycling. Such measurements could provide a multiphysics framework for investigating thermomechanical degradation, interface evolution, and heat-transfer failure mechanisms in flexible devices and advanced packaging.

From an engineering perspective, extending thermoreflectance toward wafer-scale mapping, high-throughput screening, and in-line metrology will require simultaneous improvements in measurement throughput, process compatibility, automated data analysis, and uncertainty-controlled parameter extraction. Conventional TDTR commonly relies on mechanical delay-line scanning, whereas asynchronous optical sampling and other electronically or optically encoded time-domain acquisition schemes can reduce or eliminate mechanical delay scanning and thereby increase acquisition speed. For frequency- and waveform-domain approaches such as FDTR and SPS, multifrequency excitation, broadband acquisition, and digital demodulation provide potential routes to more efficient collection of information-rich datasets. Metal transducers can introduce additional interfaces and may be incompatible with semiconductor process flows or finished-device inspection; transducerless measurements based on intrinsic optical responses are therefore particularly attractive for engineering applications, although variations in surface condition, optical properties, and thermoreflectance coefficients can complicate quantitative calibration and large-area measurement consistency. Physics-informed machine learning, surrogate forward models, and precomputed model libraries may further accelerate inversion for complex structures, but physical constraints, uncertainty quantification, and validation against independent or reference measurements remain necessary for reliable deployment.

Overall, device-oriented thermoreflectance is progressing from idealized samples toward operando device characterization, from predominantly thermal measurements toward coupled multiphysics interrogation, and from serial laboratory experiments toward higher-throughput engineering metrology. Future progress will depend not only on improved spatiotemporal resolution, but also on the ability to distinguish thermal signals from electrical, mechanical, and optical contributions without substantially perturbing the operating state of the device. At the same time, increased acquisition speed, process-compatible measurement configurations, automated inversion, and rigorous uncertainty assessment will be required to translate laboratory techniques into robust engineering tools. Together, these developments could broaden the role of thermoreflectance from thermophysical-property measurement to spatially and temporally resolved thermal diagnostics and metrology for power electronics, flexible devices, advanced packaging, and semiconductor manufacturing.

### 5.4 Natural-Language-Driven Data Analysis and Intelligent Experimentation

Beyond measurement resolution and inversion methodology, broader adoption of thermoreflectance is also constrained by the complexity of its analysis workflow and its dependence on specialized domain knowledge and user-defined analysis choices. Methods such as TDTR, FDTR, and SPS typically require data organization and preprocessing, definition of the multilayer structure, specification of known and unknown parameters, selection of fitting ranges and experimental conditions, sensitivity analysis, parameter estimation, and uncertainty propagation. Even when mature computational codes or dedicated software packages are available, analysis of complex multilayer structures can remain strongly dependent on user experience, particularly in model construction, parameter selection, identifiability assessment, and interpretation of fitting results. This dependence complicates workflow standardization, reproducible high-throughput analysis, and broader interdisciplinary use.

AI agents based on large language models (LLMs) provide a potential route to workflow-level automation that is conceptually distinct from machine-learning models trained to predict thermophysical properties directly from experimental signals. Recent studies have begun to explore tool-augmented LLMs for scientific user facilities, automated experimentation, and interaction with complex laboratory instrumentation [116–119]. For thermoreflectance, a scientifically appropriate role for an LLM-based agent is to translate user intent into structured analysis tasks, coordinate validated computational tools, and interpret their outputs, rather than to generate thermophysical-property values independently. Numerical calculations can remain within deterministic and validated heat-diffusion solvers, fitting algorithms, sensitivity-analysis routines, and uncertainty-quantification modules, while the LLM provides natural-language interaction, workflow orchestration, and contextual reasoning across these components.

A layered architecture could therefore consist of a natural-language interaction layer, an AI-agent orchestration layer, and a validated physics-computation layer. Researchers could specify the sample structure, known parameters, experimental data, and analysis objectives in natural language, after which the agent would translate these inputs into a machine-readable configuration, check required parameters and units, and sequentially invoke data preprocessing, forward-model evaluation, sensitivity analysis, parameter fitting, and uncertainty-quantification modules. The outputs of each step—including model assumptions, parameter bounds, fitting settings, sensitivity results, uncertainties, and software versions—could be recorded automatically to provide a traceable provenance chain. Such an architecture could shift thermoreflectance analysis from manual operation of specialized software toward reproducible, auditable workflows in which scientific intent is translated systematically into validated computational operations, while retaining the underlying physics model as the basis of quantitative inference.

When such an agent is further connected to data-acquisition and instrument-control interfaces, it could form a closed loop with the adaptive experimental-design framework discussed in Section 5.1. Based on the measurements already acquired, the system could update parameter sensitivities and uncertainties and evaluate candidate experimental conditions according to expected information gain or another predefined design criterion. It could then propose—or, where appropriate, execute after user confirmation—new modulation frequencies, delay-time windows, spot sizes, or spatial offsets that are expected to improve parameter identifiability or reduce uncertainty. This creates a potential progression from natural-language-assisted data analysis to adaptive experimentation, in which measurement selection and parameter inference are iteratively coupled rather than performed as separate stages.

Because thermophysical metrology requires quantitative reliability and reproducibility, however, LLM-based automation should not be treated as an unconstrained decision layer. Intelligent thermoreflectance systems should instead be built around validated domain-specific models and numerical codes, rigorous checks of parameter ranges, dimensions, and units, machine-readable records of model assumptions and data transformations, complete provenance of computational and experimental operations, and explicit human confirmation for critical actions. Numerical outputs should remain traceable to the deterministic tools and experimental data that generated them, allowing calculations to be independently reproduced and audited rather than relying on the language model as the source of quantitative results. The objective of intelligent thermoreflectance experimentation is therefore not automation for its own sake, but the development of a human-supervised, physics-grounded, and traceable analysis–experiment loop that combines natural-language accessibility with the rigor required for quantitative thermophysical measurement.

## 6 Summary

This review has systematically examined thermoreflectance-based micro- and nanoscale thermophysical-property measurement, covering the physical foundations, experimental configurations, signal-acquisition strategies, and application regimes of representative techniques including TTR, TDTR, FDTR, SSTR, SDTR, and SPS. Although these methods differ substantially in experimental implementation, they can be understood within a common framework of photothermal excitation, heat transport, thermoreflectance detection, and model-based parameter inference. Their principal distinction lies in the measurement coordinate and observable through which the thermal response is interrogated: TTR records single-pulse transient decay; TDTR resolves ultrafast thermal evolution through pump–probe delay; FDTR measures frequency-dependent amplitude and phase; SSTR probes the quasi-steady temperature response; SDTR and beam-offset approaches introduce lateral spatial information; and SPS captures the full temperature waveform under periodic square-wave excitation. These techniques should therefore be viewed not as competing variants of a single measurement, but as complementary experimental projections of the same underlying heat-transport problem, providing different sensitivity signatures across characteristic length and time scales.

In practical applications, method selection should account jointly for sample geometry, target thermophysical properties, available prior information, instrument bandwidth, and the identifiability of the desired parameter set under the accessible experimental conditions. TDTR and FDTR are widely applied to cross-plane film thermal conductivity and interfacial thermal conductance, whereas SDTR and beam-offset configurations can introduce enhanced spatial sensitivity to in-plane transport and thermal anisotropy. For low-thermal-conductivity materials or strongly coupled multilayer systems, low-frequency FDTR, multi-condition TDTR, SPS, extended transient measurements, or joint analysis of complementary measurement configurations can be employed according to the relevant heat-diffusion scales and required parameter sensitivities. Regardless of technique, strong sensitivity to an individual parameter does not by itself guarantee that the parameter can be uniquely determined when other unknowns produce correlated responses. Measurement credibility therefore depends on an appropriate forward model, accurate structural and optical inputs, sufficiently distinct parameter sensitivities, and rigorous uncertainty assessment. No single technique or experimental condition can consequently be regarded as universally optimal for complex multiparameter characterization.

Future development of thermoreflectance can be considered along four interconnected directions. First, advances in temporal and spatial resolution will extend measurements toward shorter timescales and smaller transport dimensions, enabling further investigation of ultrafast thermal relaxation, quasiballistic phonon transport, nonequilibrium interfaces, and localized heat transfer in micro- and nanostructures. Such gains in resolution must be accompanied by corresponding advances in model fidelity as experiments approach the boundaries of conventional Fourier heat diffusion. Second, integration with electrical excitation, mechanical loading, near-field optical techniques, and in situ or operando device measurements will expand thermoreflectance from isolated thermal characterization toward multiphysics measurements under realistic operating conditions. Third, data analysis is expected to progress from single-optimum fitting toward uncertainty-aware inference that combines sensitivity and identifiability analysis, probabilistic parameter estimation, model-credibility assessment, and physics-based constraints. Fourth, advances in scientific AI may enable natural-language interfaces and agent-based workflow orchestration, in which experimental intent is translated into structured analysis and tool execution while quantitative inference remains grounded in validated physical models and traceable numerical calculations. Coupling these capabilities with uncertainty-aware experimental design could further establish closed loops among measurement, inference, and condition selection, enabling more efficient and adaptive acquisition of informative data.

Overall, thermoreflectance has developed into a versatile platform for micro- and nanoscale thermophysical characterization. Its strength lies not only in noncontact optical interrogation, high temporal resolution, and localized measurement capability, but also in the ability to manipulate heat-flow geometry and parameter sensitivity through experimental variables such as modulation frequency, pump–probe delay, spot size, and spatial position or offset. The central opportunity for future thermoreflectance is therefore not simply to extend measurement ranges or increase data-acquisition speed, but to integrate experimental design, multiscale physical modeling, parameter identifiability, uncertainty quantification, and reproducible workflow control. As these elements become increasingly integrated, thermoreflectance can evolve from a collection of specialized laboratory techniques toward a more unified, quantitative, traceable, and adaptive thermal-metrology framework, supporting increasingly complex applications in functional thin films, heterogeneous interfaces, power electronics, advanced packaging, and semiconductor-process monitoring.

**Chinese Version (Original) / 中文原文**

# 基于热反射的微纳尺度热物性测量技术：原理、方法与前沿进展

江普庆[1*]，陈珊珊[2]，李小波[1*]，杨荣贵[1,3*]

1. 华中科技大学能源与动力工程学院， 武汉 430074；

2. 中国人民大学物理学院，北京 100872；

3. 北京大学工学院能源与资源工程系， 北京 100871；

* 联系人, E-mail: jpq2021@hust.edu.cn; xbli35@hust.edu.cn; ronggui@pku.edu.cn

**摘要** 随着半导体器件特征尺寸持续缩小至纳米尺度，材料与界面的热输运及热储能行为呈现出日益显著的尺度、结构和界面依赖性，对热导率、界面热导和体积热容等关键热物性参数的准确表征，已成为功能材料设计与先进电子器件热管理中的重要基础问题。基于光热效应的热反射测量技术具有非接触、高时空分辨率和宽测量范围等优势，已成为微纳尺度热物性表征的重要方法。本文系统梳理了瞬态热反射(TTR)、时域热反射(TDTR)、频域热反射(FDTR)、稳态热反射(SSTR)、空间域热反射(SDTR)及方脉冲热源法(SPS)的物理原理与技术特征，建立了统一的热扩散分析框架，并比较了不同方法在参数敏感性、测量不确定度及适用范围方面的差异。结合典型应用案例，本文进一步分析了热反射技术在低导热材料、各向异性薄膜、晶体材料及多层异质结构热物性测量中的适用条件，阐明了不同方法在薄膜热导率、界面热导和面内/面外热输运表征中的互补作用。最后，本文展望了热反射测量在超高时空分辨、多物理场耦合、工业在线检测及智能化数据处理与自然语言驱动分析等方向的发展趋势，以期为微纳尺度热输运研究和先进器件热管理提供参考。

随着功能材料与高性能电子器件的快速发展，微纳尺度热输运行为的准确表征已成为揭示材料热输运机制、优化器件热管理及评估服役可靠性的重要基础[1, 2]。在材料层面，低维材料、各向异性晶体、聚合物薄膜、多层异质结构及热功能材料中的热输运行为通常呈现明显的尺度、界面与微结构依赖性。其中，热导率($k$)和界面热导($G$)等输运参数易受边界与界面散射、晶体取向及加工工艺等因素影响；体积热容($C$)也可能随材料组成、相态和结晶度等发生变化。因此，传统宏观热物性参数往往难以直接反映微纳结构中的实际热响应[3-5]。在器件层面，随着 5G 通信和氮化镓(GaN)高电子迁移率晶体管(HEMT)等高功率、高频技术的发展，器件局部功率密度与热流密度不断升高，沟道热点、薄膜热阻及异质界面热阻已成为制约器件性能与寿命的重要因素[6]。尤其当材料或器件的特征尺寸与声子平均自由程相当时，边界散射、界面热阻及各向异性热扩散等效应更加突出[7, 8]，传统宏观测量方法与均匀介质假设面临局限，从而对微纳尺度热物性参数的准确测量提出了更高要求[1, 9]。

与宏观尺度测量相比，微纳尺度热物性参数的实验获取并非测量尺度的简单缩小，而是同时受到空间分辨率、时间分辨率及多参数耦合反演等因素的制约。首先，微纳材料与器件中的温升和热流分布往往具有显著的空间局域性，热点可集中于薄膜、异质界面、晶体管沟道或微区缺陷附近，因此，测量技术需要具备微米级乃至亚微米级空间分辨率，才能有效分辨目标区域的局部热响应。其次，声子弛豫、界面能量传递和薄膜瞬态热扩散等过程可跨越皮秒至毫秒的宽时间尺度，要求探测手段兼具足够的时间分辨率和动态响应范围。更为关键的是，实验信号通常由热导率、体积热容和界面热导等待测参数，以及薄膜厚度、光斑尺寸等结构与实验参数共同决定。不同参数之间的敏感性相关可能导致反演问题不适定，使实验噪声、模型简化及先验参数误差被显著放大，进而降低测量结果的准确性与可靠性[10, 11]。因此，亟需发展兼具高时空分辨率、非接触探测能力和可靠参数反演方法的微纳尺度热物性测量技术，以满足复杂材料热输运研究和先进电子器件热设计的需求。

面对上述测量挑战，现有热物性表征方法在微纳尺度应用中仍存在不同程度的局限[12]。在稳态测量方法中，稳态平板法原理成熟、结果直观，但受样品尺寸和形貌限制，难以用于微区热物性测量；微热桥法虽可实现微纳结构的热输运表征，却易受到接触热阻、边缘热损失及传感器引线寄生热流等因素影响[13]。瞬态平面热源法(TPS) 在块体材料测量中具有较高的可靠性，但其热穿透深度与传感器尺寸相互耦合，空间分辨率通常处于毫米量级，难以满足微米乃至亚微米尺度薄膜、界面和局部热点的表征需求[14]。此外，对于低维材料、各向异性薄膜和多层异质结构，上述接触式或依赖传感器集成的方法往往难以有效区分面内与面外热输运贡献，也难以在低扰动条件下准确解析层间界面热导。因此，微纳尺度热物性的精准测量亟需发展兼具局部激励、非接触探测和高灵敏温度响应能力的新型表征技术。

在此背景下，基于光热激励与光学探测的热反射测量技术(Thermoreflectance, TR)逐渐发展成为微纳尺度热物性表征的重要手段[15]。该技术利用泵浦光在样品表面或金属传感层中产生局部温升，并通过探测光检测温度变化引起的微弱反射率变化，结合热传导模型解析材料的热扩散行为并提取相关热物性参数。通过调节光斑尺寸、调制频率或泵浦—探测延迟时间等实验条件，不同热反射技术可实现微米级乃至更高的空间分辨率，并在整体上覆盖皮秒至秒级的宽时间尺度，从而适用于薄膜、异质界面、多层结构及各向异性材料的热输运研究。经过数十年的发展与技术迭代[16-18]，热反射测量已形成多种各具特点的技术路线（图 1），主要包括瞬态热反射(transient thermoreflectance, TTR)、时域热反射(time-domain thermoreflectance, TDTR)、频域热反射(frequency-domain thermoreflectance, FDTR)、稳态热反射(steady-state thermoreflectance, SSTR)、空间域热反射(spatial-domain thermoreflectance, SDTR)及方脉冲热源法(square-pulsed source, SPS)。这些方法在热源激励方式、信号采集维度和参数反演策略等方面各有侧重，可面向不同材料构型、时空尺度及参数测量需求，提供相互补充的热物性表征方案。

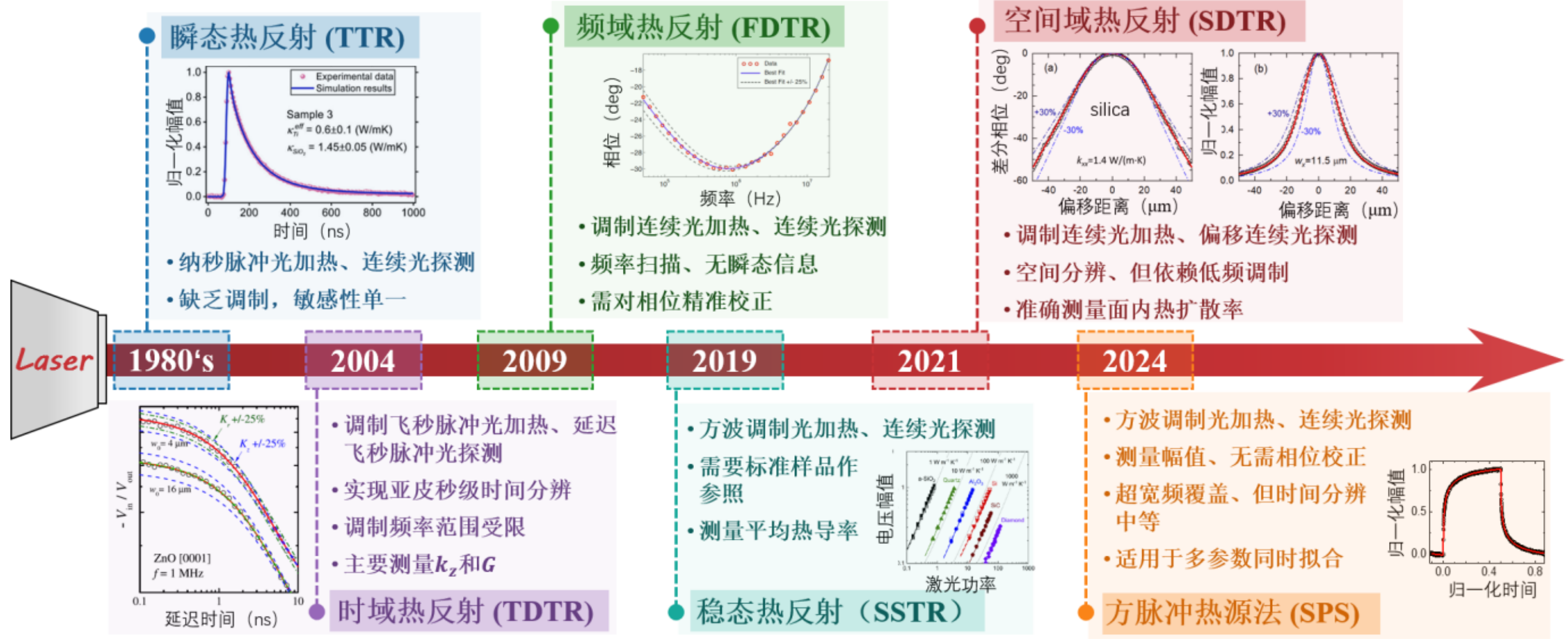

**图 1** 热反射测量技术的发展脉络及主流技术分支

**Figure 1** Development timeline and mainstream branches of thermoreflectance measurement

现代泵浦—探测式热反射测量技术的发展可追溯至 20 世纪 80 年代。1986 年，Paddock 与 Eesley 采用皮秒脉冲激光泵浦—探测系统及机械延迟装置，首次利用瞬态热反射方法定量测量金属薄膜的热扩散率，奠定了瞬态热反射（TTR）用于薄膜热物性表征的实验基础[19]。 随后， Maris 团队围绕皮秒超声、相干声子及超快泵浦—探测过程开展了一系列研究，推动了该技术在薄膜声学与热输运表征中的发展[20-26]。Kading 等进一步将瞬态光热测量拓展至介质薄膜及多层结构的热输运研究[27]。在此基础上，Cahill 等系统建立了高斯光束加热条件下多层结构的轴对称热扩散模型，并将其与飞秒激光泵浦—探测、泵浦调制和锁相检测相结合，完善了时域热反射（TDTR）的实验与数据分析框架[28]，推动薄膜热导率和界面热导的定量测量逐步走向成熟[29-37]。

随着研究对象由相对简单的单层薄膜逐步拓展至低热导材料、各向异性晶体及复杂多层异质结构，传统单一测量变量在参数灵敏度、可测范围和多参数解耦等方面逐渐显现局限，推动热反射技术持续向频域、稳态域和空间域拓展，并形成多测量域协同发展

的趋势。频域热反射（FDTR）通过扫描泵浦光调制频率调节热穿透深度及不同参数的相对灵敏度，为多层结构热输运参数的识别提供了频率维度的信息[10]。稳态热反射（SSTR）通过测量连续或低频加热条件下的稳态温升响应，实现对宽热导率范围材料的热物性表征[38]。空间扫描热反射及光束偏移类方法则引入泵浦光与探测光之间的相对空间位置作为测量变量，增强了信号对面内热扩散和热导率各向异性的敏感性[39]。近年来发展的方脉冲热源法（SPS）采用方波调制和全周期温升波形分析，同时利用加热与冷却阶段的热响应信息，为热导率、体积热容及相关界面参数的协同反演提供了新的测量维度，并逐步拓展至低热导材料和多层异质结构的热物性表征[40-43]。

尽管热反射技术已广泛应用于微纳尺度热物性的定量测量，现有综述与研究大多聚焦于单一技术路线或特定材料体系[44-46]，仍缺乏对不同方法内在物理关联、性能边界及适用场景的系统比较。由于各类热反射方法在激励方式、实验构型和信号采集形式上存在明显差异，它们往往被视为彼此独立的测量技术。然而，这些方法均遵循相同的基本物理过程，即通过光热激励在样品中产生温度场，并利用反射率对温度变化的响应获取热学信号，再结合热扩散模型反演相关热物性参数。不同技术路线的主要差异在于对时间延迟、调制频率、稳态温升或空间位置等观测变量的选择，以及相应的信号处理和参数反演策略。若缺乏统一的热扩散分析框架，研究人员在面对多层薄膜、异质界面、低热导率材料和各向异性晶体等复杂体系时，往往难以合理选择测量方法与实验参数，也难以系统评估参数敏感性、反演稳定性及测量不确定度。

基于上述研究现状，本文从热反射效应与热扩散基本方程出发，系统梳理 TTR、TDTR、FDTR、SSTR、SDTR 及 SPS 等典型热反射方法的物理原理、实验构型和参数敏感特性，并在统一的热扩散分析框架下厘清不同技术路线之间的内在联系与核心差异。进一步地，本文从热扩散特征尺度、观测变量与调控方式、面内和面外热输运的辨识能力以及多参数联合反演的稳定性四个维度，比较各类方法的性能特点与适用范围；同时，以低热导率聚合物、各向异性薄膜、晶体材料及 GaN/Si 异质结构为典型案例，分析热反射技术在复杂材料与器件热物性表征中的应用特征。最后，本文展望热反射测量技术在更高时空分辨率、原位多物理场表征、快速参数反演、工业在线无损检测及自然语言驱动的智能数据分析与自适应实验等方向的发展趋势，以期为微纳尺度热输运研究和新一代微电子器件热管理提供理论与实验参考。

# 1　物理基础与理论框架

## 1.1　热反射效应与光学测温

热反射测量的基本物理依据是材料表面反射率对温度变化的响应。从光学本质上看，温度变化会引起材料复折射率及相关光学参数的改变，进而导致表面反射率发生变化。在参考温度（$T_0$）附近，将反射率对温度作泰勒展开；当二阶及更高阶项相对于一阶项可以忽略时，反射率的相对变化量可表示为[47]：

$$\frac{\Delta R}{R} = C_{\mathrm{TR}}\,\Delta T \qquad (1)$$

其中，$R$ 为材料反射率，$\Delta R$为温度变化引起的反射率变化量，$\Delta T$ 为被探测区域的温度变化量，$C_{\mathrm{TR}}$为参考温度$T_0$下的热反射系数，单位为 $K^{-1}$。$C_{\mathrm{TR}}$的数值和正负不仅取决于材料种类，还与探测波长、入射角、偏振状态以及薄膜厚度和微结构等因素有关。对于 Al、Au 等常用金属传感层，$|C_{\mathrm{TR}}|$通常处于 $10^{-5}$~$10^{-4}\,K^{-1}$ 量级，即 1K 的温度变化仅对应约万分之一至十万分之一的相对反射率变化[48-50]。由于热反射信号十分微弱，实验中通常采用锁相放大、平衡探测及周期累加平均等方法提高信噪比；在优化的探测带宽和积分条件下，系统可获得毫开尔文乃至更高的等效温度分辨率[51]。

实际测试过程中，通常在待测样品表面沉积一层数十至约 100 nm 厚的金属薄膜作为传感层(Transducer)。该金属层兼具光热转换与温度探测功能：一方面吸收泵浦光并在样品表面形成可控的局部热源，另一方面利用自身较强的热反射响应将温度变化转化为可测的光学信号。同时，光学性质相对明确的金属层有助于降低样品本征光学响应对测量信号的干扰，并便于建立多层热扩散模型。

传感层材料、厚度及探测波长需要根据具体实验条件协同选择。探测波长通常应使传感层具有较大的热反射系数和适当的反射率，例如 532 nm 探测光与 Au 传感层、785 nm 探测光与 Al 传感层均属于常见组合，但并非固定配置。传感层厚度一般需要达到足够的光学不透明性，以减弱来自下层材料的寄生反射信号；与此同时，厚度又不宜过大，以避免其热容和横向热扩散对待测结构热响应造成过强扰动。因此，常用金属传感层厚度多为 50～100 nm，但具体取值仍需结合光学吸收深度、界面性质、待测参数敏感性及实验信噪比进行优化。

需要指出的是，带有金属传感层的热反射测量并非直接读取裸露待测材料的表面温度，而是探测金属传感层所表征的局部温度响应。实验信号反映的是由金属传感层、待测薄膜、层间界面及衬底共同构成的多层体系的综合热响应。金属传感层提高了热反射测

量对不同材料体系的适用性和实验可重复性，但也向测量体系中引入了额外的热容、热导率及界面热导。因此，在热物性参数反演过程中，需要将传感层厚度、热物性及其与待测材料之间的界面热导纳入热扩散模型，并评估这些参数的不确定度对反演结果的影响。

### 1.2 热扩散控制方程

热反射测量通常利用调制或脉冲激光对样品进行局部光热激励。对于覆盖金属传感层的样品，泵浦光能量主要被传感层吸收并转化为热能，在样品近表面形成随时间和空间变化的温度场。在多数微纳尺度热反射实验中，当特征尺度大于主要载热声子的非平衡输运尺度时，样品内部的热传导可采用经典傅里叶热扩散模型描述[28]。对于具有轴对称泵浦光斑的各向异性层状体系，在柱坐标系下，各材料层内部的控制方程可写为：

$$C\frac{\partial T}{\partial t}=\frac{k_r}{r}\frac{\partial}{\partial r}\left(r\frac{\partial T}{\partial r}\right)+k_z\frac{\partial^2 T}{\partial z^2} \tag{2}$$

式中，$T$ 为相对于初始温度的温升，$C$ 为体积热容，$k_r$和$k_z$分别为面内(径向)和面外(纵向)热导率。对于各向同性材料，有$k_r=k_z=k$。

在多数热反射模型中，金属传感层内的光学吸收可等效为表面热流边界条件或有限吸收深度内的体热源，其具体形式取决于传感层厚度与光学吸收深度。对于尺寸远大于热扩散区域的样品，通常将衬底视为半无限体，并采用远场温升趋于零的边界条件：$T(r,z\to\infty,t)\to 0$，$T(r\to\infty,z,t)\to 0$。

对于理想接触界面，界面两侧满足温度和热流连续条件；当存在有限界面热阻时，热流连续而界面允许出现温度跃变，其关系可表示为：

$$q=G(T_1-T_2) \tag{3}$$

其中，$G$ 为界面热导，$T_1$和$T_2$分别为界面两侧温度。对于由金属传感层、待测薄膜和衬底组成的多层结构，各层热扩散方程与相应界面条件共同构成热反射信号分析的基本物理模型。

由上述控制方程可见，热反射信号并非由某一单独热物性参数决定，而是热导率、体积热容、界面热导以及膜厚、光斑尺寸等结构与实验参数共同作用下的综合热响应。不同热反射技术本质上均对这一受控热扩散过程进行激励和观测，其主要差异在于热源形式、观测变量及信号获取方式。因此，各类热反射方法可在统一热扩散框架下理解为对时间、频率、空间及其组合等不同测量维度的选择。

### 1.3 热穿透深度与特征尺度

为定量描述各类热反射测量中的有效热扩散范围，并分析实验信号对不同热物性参数的敏感区域，需引入热穿透深度(thermal penetration depth)作为表征热扰动传播范围的核心特征尺度[52]。对于周期性热源，在傅里叶热扩散近似下，热扰动在材料内部的特征传播深度可表示为：

$$d_p=\sqrt{\frac{k}{\pi f C}} \tag{4}$$

其中，$d_p$为热穿透深度，$k$为材料热导率，$C$为体积热容，$f$为泵浦光调制频率。对于层状各向异性体系，在讨论厚度方向的热穿透时，$k$对应相应方向上的热导率。由式（4）可见，热穿透深度同时取决于材料本征热物性和实验调制频率。其中，$k$和$C$为材料热物性参数，而$f$是可调实验变量；$d_p$与调制频率呈$f^{-1/2}$关系，因此提高调制频率会减小热穿透深度，使热扰动更加局限于样品近表面区域。

需要强调的是，式（4）给出的$d_p$表征的是热扰动自身的特征传播尺度，但实际热反射信号所涉及的材料区域及热流路径并不只由$d_p$的绝对大小决定，还取决于其与样品厚度和光斑尺寸之间的相对关系。薄膜厚度和光斑尺寸虽然不直接出现在式（4）中，却分别构成约束纵向和径向热扩散的重要几何尺度。因此，在热反射实验中，更具实际意义的是比较$d_p$与待测薄膜厚度 $h$ 以及泵浦光斑直径$2r_0$的相对大小：前者决定热扰动在厚度方向能够覆盖的材料层及界面，后者则决定热扩散更接近一维纵向传热还是具有显著的径向扩散分量。

首先，对于多层堆叠薄膜体系，将热穿透深度$d_p$与待测薄膜厚度 $h$ 比较，可以判断热扰动在厚度方向上的有效覆盖范围。当$d_p \ll h$时，热扰动主要局限于待测层的近表面区域，难以到达下层界面，测量信号主要反映表层材料的热响应；当$d_p \sim h$时，热扰动能够传播至薄膜/衬底界面，界面热导以及下层材料的热输运特性开始对信号产生明显影响；若$d_p \gg h$时，热扰动将进一步深入衬底，测量信号逐渐受到衬底热物性以及多层结构整体热阻的共同控制。因此，$d_p/h$的相对大小决定了测量信号在纵向上的主要贡献区域，也决定了实验对薄膜、界面和衬底热物性的相对敏感程度。

其次，将热穿透深度$d_p$与泵浦光斑的横向尺度$2r_0$比较，可以判断热扩散过程中纵向与径向传热的相对重要性。如图 2 所示，当$d_p \ll 2r_0$时，热扰动的纵向传播尺度远小于光斑横向尺度，热量主要沿样品厚度方向扩散，体系可近似表现为一维传热，此时测量信号对面内热导率的敏感性较弱。随着$d_p$逐步趋近

乃至超过 $2r_0$，径向热扩散所占比例持续增加，温度场表现出更明显的二维或三维扩散特征，测量信号对面内热导率的敏感性也随之增强[53]。因此，$d_p/(2r_0)$的相对大小可作为判断径向热扩散是否显著的重要尺度依据。

由此可见，热穿透深度实际上建立了材料热物性、实验调控参数与样品几何尺度之间的联系。通过改变调制频率 $f$，可以直接调节 $d_p$，进而改变热扰动在厚度方向上的有效传播范围；通过改变光斑尺寸 $r_0$，虽然并不改变式（4）所定义的 $d_p$，却可以调节 $d_p/(2r_0)$的相对大小，从而改变纵向与径向热扩散的相对贡献。实验设计中因而需要综合考虑 $d_p/h$和 $d_p/(2r_0)$等尺度关系，以调控测量信号的主要贡献区域，并增强对目标材料层或特定热物性参数的敏感性。

需要指出的是，$d_p$并不是一个仅由实验条件决定的独立可控量。由式（4）可知，其同时包含可调实验变量$f$和待求热物性参数$k$与$C$。因此，在未知材料的实际测量中，热穿透深度通常只能依据先验热物性参数进行初步估算，并据此选择合适的调制频率和光斑尺寸；最终实验条件及有效敏感区间仍需结合敏感性分析和参数拟合进行迭代修正。

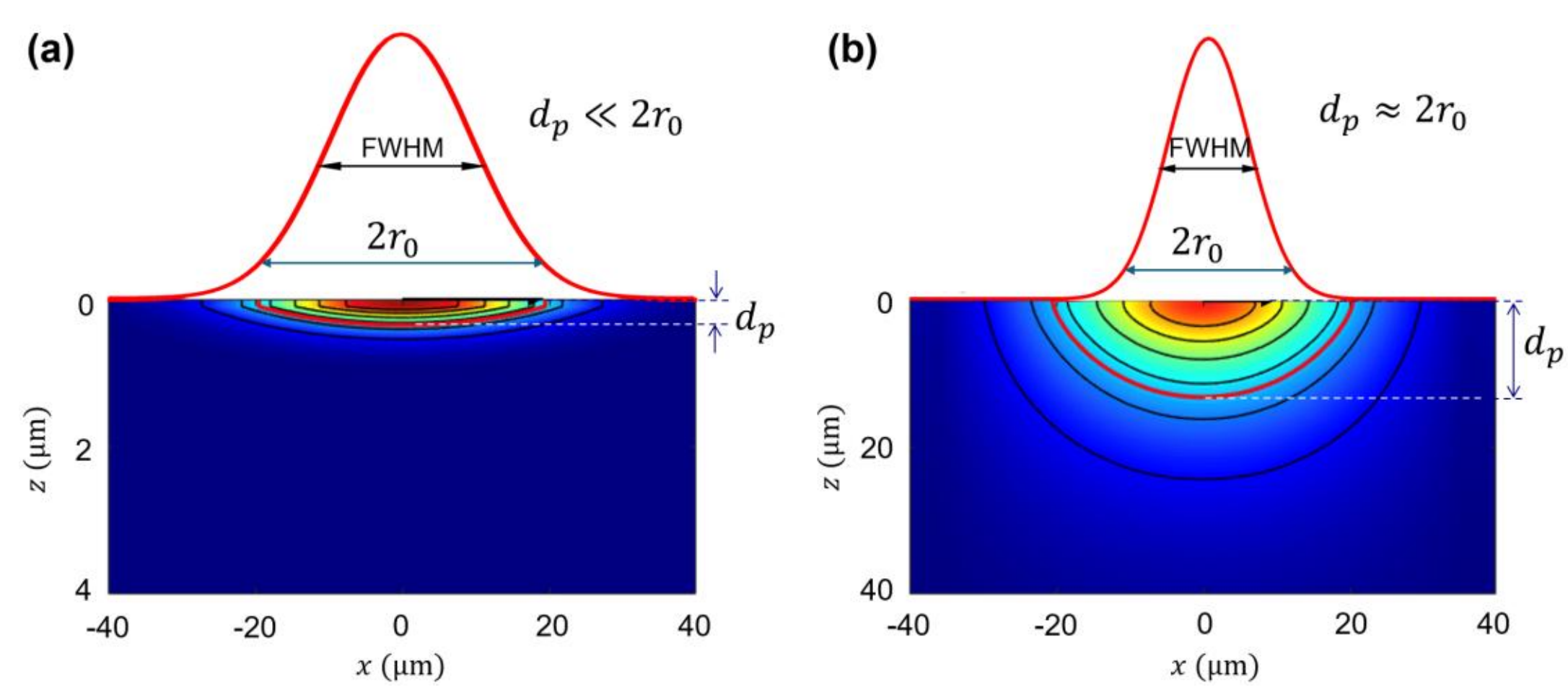

**图 2** 热穿透深度与光斑尺寸对热扩散路径的影响

**Figure 2** Effects of thermal penetration depth and laser spot size on heat diffusion pathways

## 1.4 参数敏感性与反演策略

热反射测量本质上属于导热反问题：通过测量样品表面的热响应，并结合热扩散模型反向求解未知热物性参数。为定量描述测量信号对参数$x$的响应强弱，引入无量纲敏感性系数[54]：

$$S_x = \frac{\partial \ln A}{\partial \ln x} \tag{5}$$

其中，$A$ 表示具体测量方法所采用的实验观测量，例如幅值、相位、锁相信号比值、归一化温升响应等。$S_x$表征参数$x$发生微小相对变化时测量信号产生的相对变化量。一般而言，$|S_x|$越大，说明实验信号对该参数的响应越明显，参数具有更高的潜在可辨识性；若$|S_x|$接近于零，则表明信号对该参数变化不敏感，仅依靠当前实验条件难以获取可靠估值。

然而，较大的单参数敏感性并不必然意味着该参数能够被独立反演。对于多参数问题，更关键的是不同参数对实验信号的影响是否能够相互区分。当两个或多个参数的敏感性曲线在时间域、频率域或空间域中具有高度相似的变化趋势时，它们对测量信号产生的影响近似相关，参数之间将出现较强耦合，反演问题也随之表现出明显的不适定性。此时，即使原始实验信号具有较高的信噪比，微小的测量误差、模型误差或先验参数不确定度仍可能在反演过程中被显著放大，导致待测热物性的不确定度增加。因此，敏感性分析需要同时回答两个问题：一是目标参数对实验信号是否具有足够大的响应，即参数是否“可测”；二是目标参数与其他未知参数的响应特征能否有效区分，即参数是否“可分离”。

这种参数耦合与热反射信号对组合热物性参量的响应密切相关。在许多典型实验条件下，测量信号并非分别独立响应于 $k_r$、$k_z$ 和 $C$，而主要受到某些组合参数控制，例如面外热逸散率$\sqrt{k_z C}$、面内热扩散率$k_r/C$以及薄膜面积热容$hC$等[55]，其中$h$为薄膜厚度。这意味着，在单一时间窗口、单一调制频率或单一光斑尺寸条件下，仅凭一组测量信号往往难以同时独立确定多个强耦合热物性参数。因此，热反射参数反演的关键并不只是选择合适的数值拟合算法，更在于通过实验变量设计改变不同参数的敏感性特征，使目标参数与其他耦合物性参数在不同测量条件下表现出足够不同的响应，从而提高多参数反演的可辨识性。

敏感性分析因此也是实验设计和参数解耦的核心工具。通过计算$S_x$随延迟时间、调制频率、光斑尺寸或空间偏移量等实验变量的变化，可以识别目标参数的高敏感区间，并判断不同参数之间的耦合程度。具体而言，敏感性分析主要可用于以下几个方面：(1)选择适当的时间窗口、频率区间或空间测量范围，提高信号对目标参数的敏感性；(2)比较不同参数敏感性曲线的幅值和变化趋势，判断多参数之间的可分离程度；(3)结合实验噪声和先验参数不确定度，评估待测参数可能达到的测量不确定度；(4)指导实验条件与样品结构设计，例如优化金属传感层厚度、光斑尺寸、调制频率及衬底选择等，从而提高参数反演的稳定性。由此，前述热穿透深度与几何特征尺度的调控可以进一步转化为对参数敏感性的调控：改变调制频率和光斑尺寸，本质上是在改变温度场及热流路径，从而重新分配实验信号对不同热物性参数的响应权重。

在确定实验构型和进行敏感窗口优化后，待测热物性参数通常通过实验信号与热扩散模型之间的拟合获得。非线性最小二乘法是热反射数据分析中常用的反演方法，其中 Levenberg–Marquardt 算法通过最小化实验测量值与模型预测值之间的残差，迭代获得未知参数的最优估计[56]。对于参数数量较少、初值合理且参数相关性较弱的问题，该类局部优化方法通常具有较高的计算效率。然而，当参数耦合较强、目标函数存在多个局部极值或初始参数范围较宽时，反演结果可能对初始值、实验噪声及先验参数误差较为敏感。因此，实际分析中通常需要将数值优化与物理约束、敏感性分析及不确定度量化结合，以提高反演结果的稳定性和可靠性。

对于参数空间较宽或局部优化容易受到初值影响的问题，粒子群优化、遗传算法等全局搜索方法，以及全局搜索与局部优化相结合的混合算法，也逐渐应用于热反射数据反演[57]。这类方法能够在更大的参数空间内搜索可能的最优解，但通常需要更多的模型计算次数，因此计算效率与搜索稳健性之间需要进行权衡。近年来，机器学习方法也开始应用于热物性参数快速估计和高通量热反射数据处理[58-60]。在训练数据充分覆盖目标参数空间的条件下，此类方法可以显著降低重复反演的计算成本；但其预测可靠性仍取决于训练样本质量、参数空间覆盖程度、模型泛化能力以及训练数据与实际热输运物理模型之间的一致性。

## 2 热反射测量技术的分类与演进

在前述热反射效应、热扩散方程及参数敏感性分析的基础上，本节进一步讨论 TTR、TDTR、FDTR、SSTR、SDTR 和 SPS 等典型热反射测量方法的分类依据及技术演进。尽管这些方法在实验光路、光热激励形式和信号采集形式上存在差异，但其基本物理过程是一致的：通过激光光热激励在样品中产生温度响应，利用温度变化引起的反射率变化获取热学信号，并结合热扩散模型反演相关热物性参数。因此，各类方法之间的主要差异并非热传导控制方程本身，而在于对同一热扩散过程所采用的激励方式、观测变量及信号获取维度不同。

图 3 按照技术发展时序概括了几类主流热反射方法的实验构型。早期 TTR 主要通过记录脉冲激光加热后的瞬态温度衰减过程获取热扩散信息；TDTR 在超快泵浦-探测光路基础上引入调制检测和延迟时间扫描，显著提升了薄膜及界面热输运的定量测量能力；FDTR 则以调制频率扫描为主要观测手段，通过改变热穿透深度，为多层结构热物性的辨识提供了独立的频率调控维度。此后发展的 SSTR、SDTR 和 SPS 等方法进一步拓展了准稳态温升、空间温度场以及周期全波形等信息获取方式，使热反射测量由单一时间域或频率域观测逐步向多测量维度协同表征发展。

需要指出的是，图 3 所示的技术发展时序并不完全等同于本文后续的论述顺序。若单纯按照历史时间线展开，容易弱化不同方法在热扩散尺度、观测变量和参数敏感性上的本质差异。因此，以下内容将从热扩散响应特征出发，按照“准稳态响应→单脉冲瞬态响应→频率域响应→超快泵浦－探测时域响应→周期全波形响应→空间域响应”的逻辑展开，依次讨论 SSTR、TTR、FDTR、TDTR、SPS 和 SDTR。其中，SSTR 首先讨论，并非因为其出现最早，而是因为稳态响应可视为热扩散方程在长时间极限下的特殊情形，可作为理解其他非稳态方法的物理参照。在此基础上，本文进一步比较各类技术在有效热扩散尺度、参数敏感性、多参数反演能力及适用范围方面的差异，并结合图 4 分析其测量能力边界与典型适用场景。

(a) TTR

反射镜
滤光片
光电探测器
连续波激光器
探测光
偏振分光镜
示波器
$\lambda/4$
纳秒脉冲激光器
泵浦光
分色镜
物镜
样品

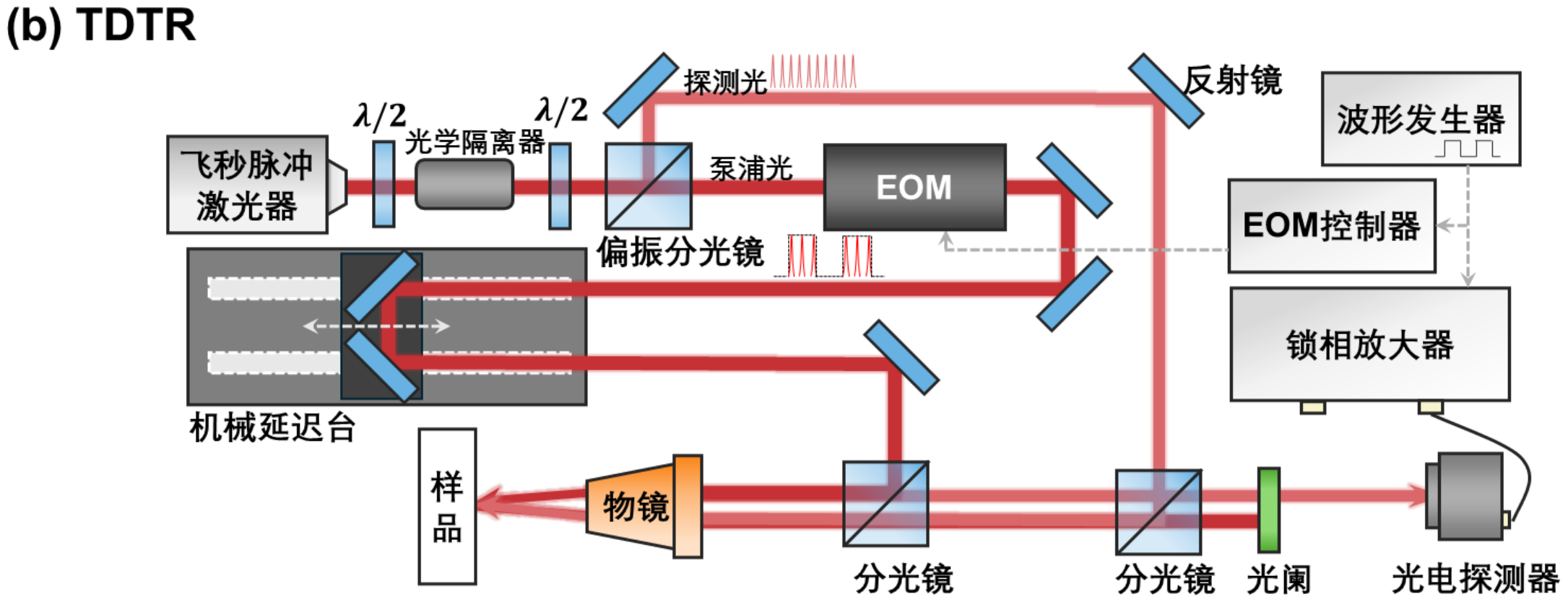

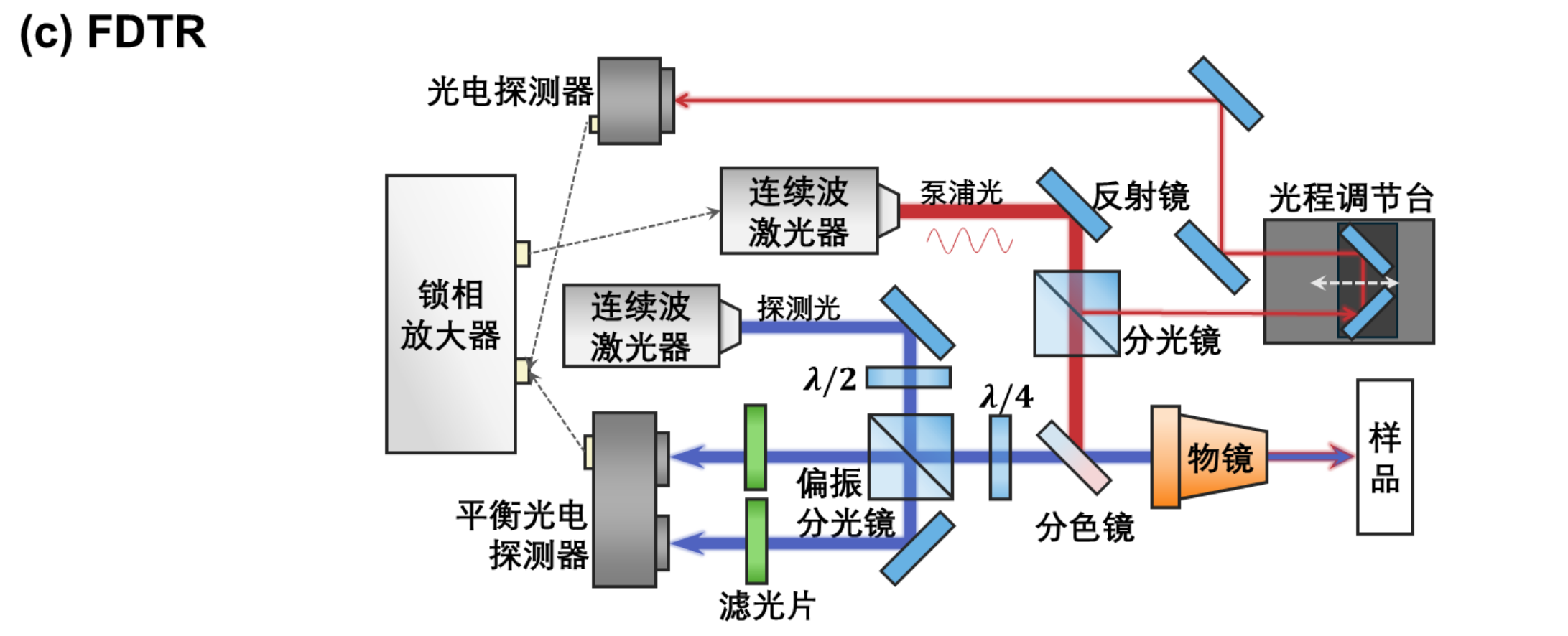

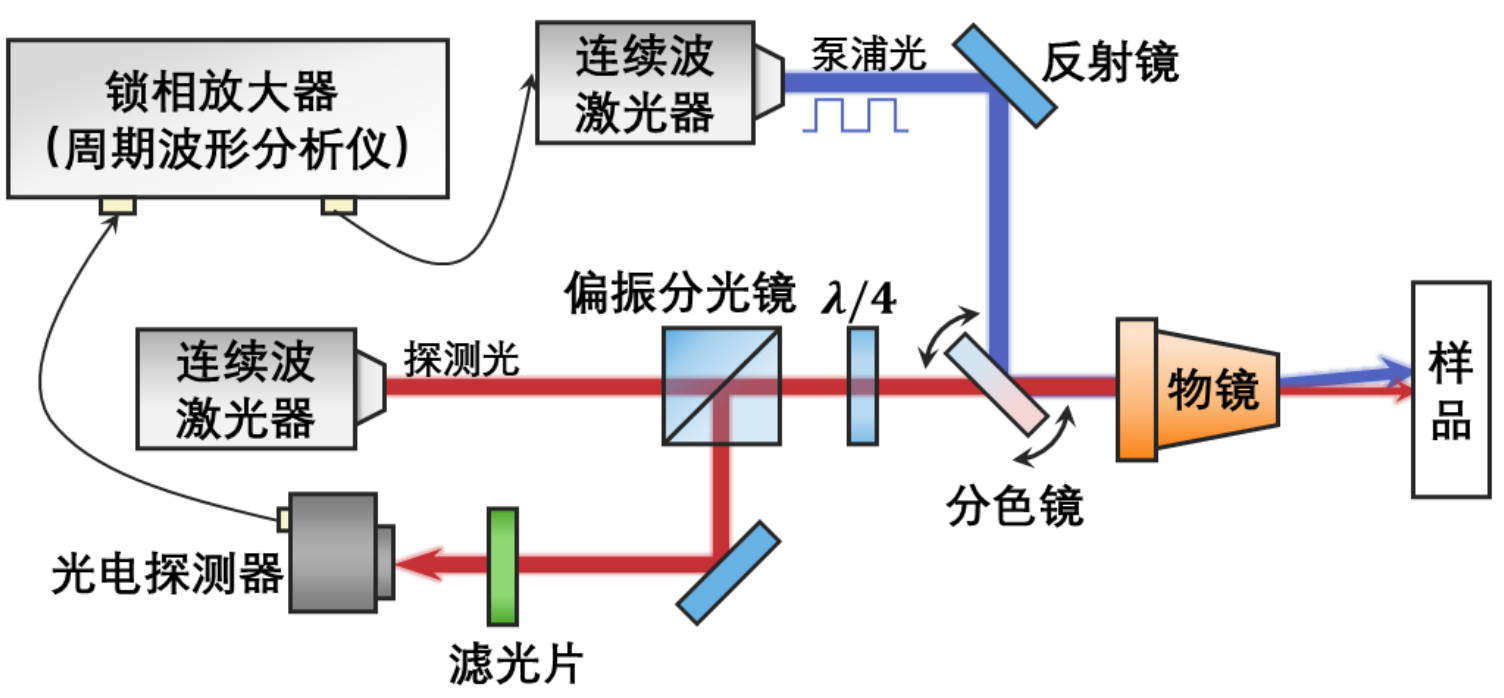

**图 3** 不同热反射测量技术的实验系统示意图

**Figure 3** Schematic diagrams of experimental systems for different thermoreflectance techniques

### 2.1 稳态极限：SSTR 方法

稳态热反射法(SSTR)[38]通常采用低频调制的连续波激光作为泵浦热源加热样品。当加热变化的时间尺度远大于体系的特征热弛豫时间时，瞬态热响应逐渐衰减，样品表面温升趋于准稳态，因此可将 SSTR 视为热扩散过程在长时间尺度下的极限情形。该方法关注的核心并非脉冲激励后的瞬态温度衰减，而是在给定吸收功率和光斑尺寸条件下形成的准稳态温升。其基本物理图像较为直观：在相同加热条件下，热导率较高的材料能够更快地将热量输运至周围区域，因而表面温升较低；热导率较低的材料散热能力较弱，热量更易在加热区域积聚，对应较高的准稳态温升。

如图 3d 所示，SSTR 实验系统通常由低频调制的连续波泵浦光、连续探测光以及反射信号采集模块组成[38]。泵浦光在样品表面或金属传感层中产生局部温升，探测光通过检测温度变化引起的反射率变化获取热学响应。数据分析时，可通过比较待测样品与参考样品在相同加热条件下的温升差异，或结合吸收功率、光斑尺寸及稳态热扩散模型，反演待测材料的有效热导率或体系等效热阻。

SSTR 的优势在于物理图像直观、实验构型相对简单，并且在稳态极限下温度场不再显式依赖体积热容，因此适用于块体或较厚材料有效热导率以及体系等效热阻的快速表征。然而，对于多层薄膜、埋藏界面或传热路径复杂的异质结构，准稳态温升通常包含薄膜热阻、界面热阻及衬底散热等多种贡献，仅依靠单一稳态温升幅值往往难以独立区分不同层和界面的热物性参数。因此，当研究目标涉及薄膜/界面分辨、多参数解耦或瞬态热扩散动力学时，需要进一步引入时间、频率或空间等额外观测维度。其中，最直接的扩展是记录脉冲激励后温度随时间的瞬态演化，即下面讨论的 TTR 方法。

### 2.2 单脉冲瞬态响应：TTR 方法

瞬态热反射(TTR)[61]利用脉冲激光在样品表面产生瞬时热扰动，并记录随后表面温度随时间的衰减过程，是最直接的时域热反射测量方法之一。与 SSTR 主要利用准稳态温升幅值不同，TTR 的热物性信息主要包含在瞬态温度衰减曲线的时间尺度和形状之中：热量扩散越快，表面温升衰减越迅速；材料热输运能力越弱，热扰动维持的时间越长。因此，瞬态温度衰减曲线能够为材料热扩散率及相关热物性参数提供直接的时域约束。

如图 3a 所示，TTR 通常采用纳秒或皮秒脉冲激光作为泵浦光[27, 62]，在样品表面或金属传感层中诱发快速温升，同时采用连续波激光作为探测光照射同一区域。温度变化引起的探测光反射率变化由高速光电探测器采集，从而获得表面热响应随时间的演化曲线。在数据解析过程中，常对瞬态温升信号进行峰值归一化，以弱化泵浦吸收能量、热反射系数及系统增益等幅值因素的影响，使分析更多依赖温度衰减曲线本身的时间尺度与形状特征。

TTR 的优势在于物理图像清晰，能够直接观测脉冲加热后热扰动的弛豫与扩散过程，尤其适用于较厚薄膜、块体材料以及特征热弛豫时间处于仪器可分辨时间范围内的样品。然而，对于多层异质结构、低热导率材料以及界面热阻显著的体系，瞬态温度衰减曲线通常同时受到薄膜热导率、体积热容、界面热导和衬底热物性等多个参数影响，不同参数的敏感性容易发生耦合，仅依靠单一衰减曲线往往难以实现多参数独立反演。此外，受探测带宽、信噪比及温升幅值衰减等因素限制，脉冲激励后期的长时间尺度信号较弱，容易受到实验噪声影响，从而限制对较慢热扩散过程的有效表征。

总体而言，TTR 建立了以瞬态温度衰减为核心的基本时域观测框架，但单一时间响应在热穿透尺度主动调控和多参数解耦方面仍存在局限。为获得额外的参数敏感性调控维度，后续热反射方法进一步引入调制频率、泵浦–探测延迟时间以及空间位置等实验变量。其中，调制频率可以主动改变热穿透深度及不同热物性参数的相对敏感性，由此形成了下面讨论的频域热反射方法 FDTR。

### 2.3 频率域响应：FDTR 方法

频域热反射(FDTR) 通过对泵浦光进行周期性强度调制，在样品中产生周期温度响应，并测量热反射信号相对于调制热源的幅值和相位随频率的变化。不同于 TTR 直接记录脉冲加热后的瞬态温度衰减过程，FDTR 将调制频率作为主要观测变量，在频域框架下表征样品的复热响应。其核心特点在于利用调制频率主动调节热穿透深度：调制频率升高时，热扰动更加局限于样品近表面区域；调制频率降低时，热穿透深度增大，较深层薄膜、界面及衬底对测量信号的贡献随之增强。

如图 3c 所示，典型 FDTR 系统采用连续波泵浦光和连续波探测光。泵浦光可通过电光调制器、声光调制器或直接强度调制形成周期性热源，探测光聚焦于相同区域并检测温度振荡引起的反射率变化，再由锁相放大器提取调制频率对应的幅值和相位信号[10,

63]。通过扫描一定范围内的调制频率，可获得频域热响应曲线，并结合多层热扩散模型反演薄膜热导率、界面热导、衬底热物性以及传感层面积热容等参数。

从物理机制来看，不同调制频率对应不同的有效热扩散尺度，使 FDTR 能够通过频率扫描调节不同深度区域对测量信号的相对贡献，从而具备一定的深度选择性。在较高调制频率下，热扰动主要集中于传感层及近表面区域，信号对传感层热容及近表层热输运参数的响应通常更加明显；随着调制频率降低、热穿透深度增加，下层薄膜、界面和衬底热物性对测量信号的影响逐渐增强。需要强调的是，这种深度选择性并非对不同材料层进行严格的空间分离，而主要体现为不同频率条件下各层及各参数敏感性权重的变化。高、中、低频所对应的实际敏感区域也不存在统一的固定边界，而是受到材料热扩散率、薄膜厚度、光斑尺寸、界面热导及多层结构构型等因素共同影响。因此，在 FDTR 实验设计中，需要结合目标参数的敏感性分析选择合适的调制频率范围，以增强信号对目标热物性的响应并降低参数间的相关性。

FDTR 的优势在于实验光路相对简洁、频率调控灵活，无需机械延迟线即可获取多频点热响应信息，适用于薄膜、界面及多层结构热输运特性的表征。通过改变调制频率，可在不改变样品结构的条件下主动调节热扩散尺度，为实验条件优化和参数敏感性调控提供较大的自由度。然而，FDTR 获得的仍是周期稳态条件下的频域幅值和相位响应，当多个热物性参数在所选频率区间内具有相似的敏感性变化趋势时，复杂多层体系中的联合反演仍可能存在较强参数耦合。此外，低频测量容易受到环境温漂和低频噪声影响，而高频测量则对调制器带宽、探测器响应速度、系统相位校准及信噪比提出更高要求。

总体而言，FDTR 将热反射测量由瞬态时间响应拓展至以调制频率为主要扫描变量的频率域表征，并利用频率扫描主动调控热扩散尺度及参数敏感性。从测量维度来看，若进一步引入独立的时间延迟变量，则可以同时利用时间尺度和频率尺度调控测量信号对不同热物性参数的响应，这种时频协同特征在 TDTR 方法中得到典型体现。

### 2.4 超快泵浦-探测时域响应：TDTR 方法

时域热反射(TDTR) 是基于超快泵浦－探测光路发展起来的代表性时域热反射技术。与 TTR 直接记录单次脉冲加热后的连续温度衰减不同，TDTR 通常采用高重复频率飞秒激光产生泵浦光和探测光，对泵浦光施加周期性强度调制，并通过机械延迟线连续改变两束脉冲之间的相对时间延迟，以获得锁相信号随延迟时间的变化。泵浦光的周期调制主要用于锁相检测，同时也会影响脉冲间热积累和有效热扩散尺度；因此调制频率是重要实验控制参数，但 TDTR 的主要扫描与拟合维度仍是泵浦–探测延迟时间。

如图 3b 所示，典型 TDTR 测试系统主要由飞秒脉冲激光器、泵浦光调制模块、机械延迟台、泵浦-探测光路及锁相检测模块组成[64]。高重复频率激光脉冲被分为泵浦光和探测光，其中泵浦光经电光调制器等器件进行 MHz 量级的强度调制后聚焦于样品表面，在金属传感层中产生周期性脉冲加热；探测光通过机械延迟线改变光程，并与泵浦光聚焦于相同测试区域，其反射率变化用于表征不同延迟时间下的表面温度响应。锁相放大器提取调制频率对应的同相分量($V_{\mathrm{in}}$)和正交分量($V_{\mathrm{out}}$)，实验中通常采用二者比值$-V_{\mathrm{in}}/V_{\mathrm{out}}$作为拟合信号[28]。这种比值形式可在一定程度上弱化热反射系数、光学增益以及激光功率等绝对幅值因素对参数反演的影响，从而提高测量和拟合的稳定性。

从物理图像来看，TDTR 信号是高重复频率脉冲加热在周期调制包络下形成的累积热响应。泵浦–探测延迟时间扫描能够解析皮秒至纳秒尺度内的温度演化，使不同延迟时间区间对金属传感层热容、薄膜热输运和界面热导等参数呈现不同的敏感性。与此同时，泵浦调制频率会影响周期热扩散尺度和脉冲间热积累程度，从而改变薄膜、界面及衬底等不同区域对测量信号的相对贡献。因此，通过联合选择延迟时间窗口和调制频率，可以调节不同热物性参数的敏感性分布，在一定程度上降低参数相关性并提高反演稳定性。这也是 TDTR 区别于单一时域或单一频域观测的重要特征。

凭借皮秒量级时间分辨能力、较成熟的多层热扩散模型以及锁相检测带来的较高信噪比，TDTR 已成为薄膜面外热导率和界面热导测量中应用广泛的代表性方法，尤其适用于金属传感层/功能薄膜/衬底等典型多层结构。针对纳米至微米尺度薄膜，可利用不同延迟时间区间的信号敏感性，并结合多层热扩散模型反演薄膜热导率、界面热导等关键参数[65]。然而，TDTR 信号仍可能同时对多个热物性和结构参数产生响应，当参数敏感性高度相关时，单组实验条件下仍难以实现多个未知参数的独立确定。因此，实际拟合中通常需要预先给定或独立测量部分参数，如体积热容、薄膜厚度、金属传感层热物性及部分界面参数等，以降低参数相关性并提高反演稳定性[46]。此外，激光重复频率、脉冲热积累、调制器带宽及系统信噪比等因素也会限制可用调制频率和延迟时间范围。

总体而言，TDTR 以泵浦–探测延迟时间扫描为

核心，结合周期调制和锁相检测实现微弱超快热响应的提取，在薄膜热导率、界面热导及多层结构热输运研究中具有成熟应用。通过改变调制频率、光斑尺寸和拟合时间窗口，可进一步获得互补敏感性信息；但对于低热导材料、深埋界面或强参数耦合体系，单一TDTR 构型仍可能需要独立先验参数或与其他测量条件联合使用。

### 2.5 周期全波形响应：SPS 方法

方脉冲热源法(SPS)[40]是一种基于周期方波光热激励和全周期温度响应采集的热反射测量方法。与TDTR 主要通过泵浦–探测延迟扫描获取超快瞬态热响应、FDTR 在不同调制频率下提取幅值和相位信息不同，SPS 采用方波调制的连续波泵浦光对样品进行周期性加热，并通过高速信号采集系统记录单个调制周期内完整的温度演化波形。因此，SPS 不仅利用调制频率改变热扩散尺度，还保留每个周期内加热、冷却及热恢复过程的时序信息，从而形成“频率维度 + 全周期波形维度”共同约束热物性参数的测量方式。

如图 3d 所示，SPS 的实验构型与 SSTR 具有一定相似性，二者均可采用连续波泵浦–探测光路，但在激励方式、信号采集及信息利用方面存在明显差异。SSTR 主要利用低频条件下形成的准稳态温升信息反演材料有效热导率或体系等效热阻；SPS 则采用周期方波作为热源，并在较宽调制频率范围内记录完整的温度响应波形。对于每一个调制频率，SPS 获得的不再只是单一温升幅值或一组幅相参数，而是一整条周期温升曲线，由此增加了可用于参数反演的时序约束信息。已有实验系统报道的 SPS 调制频率可覆盖约 1 Hz～20 MHz 范围[40]，但实际可用频率区间仍取决于调制器、探测器及信号采集系统的带宽。

调制频率的变化首先使 SPS 具备较宽范围的热扩散尺度调控能力。根据前述热穿透深度关系，较高调制频率对应较小的热扩散尺度，热扰动主要集中于传感层及样品近表面区域；随着调制频率降低，热穿透深度逐渐增加，下层薄膜、界面和衬底对测量信号的贡献随之增强；在足够低的频率下，体系响应可逐渐趋近准稳态，此时测量信号更多反映多层结构的综合热阻特征。因此，通过改变调制频率，SPS 能够在同一实验构型下连续调节不同深度区域及不同热输运过程对测量信号的相对贡献。需要指出的是，不同频率并不与特定材料层或热物性参数形成严格的一一对应关系，各参数的实际敏感区间仍受到材料热扩散率、薄膜厚度、光斑尺寸、界面热导及多层结构构型等因素共同影响，需要结合敏感性分析确定合适的测量频率范围。

SPS 除利用调制频率改变热扩散尺度外，还保留每个周期内加热、冷却及恢复阶段的温度波形。不同时间阶段可能对传感层、界面、薄膜和衬底参数呈现不同的响应权重，因而能够为参数反演增加额外约束。相较于仅使用单个频率点的幅值或相位信息，全周期波形在合适的实验条件下有助于提高部分参数的可区分性。该特性使 SPS 尤其适用于低热导材料、多层异质结构、深埋藏界面及参数强耦合体系的热物性反演。

尽管 SPS 能够在相对统一的实验构型中同时拓展热扩散尺度和时序信息维度，但其测量能力同样受到实验条件和参数耦合的限制。低频测量容易受到环境温漂、激光长期功率漂移及低频噪声影响；在高频条件下，调制器的上升和下降时间、探测器及采集系统带宽会影响方波及温度响应波形的保真度，实际热源也可能偏离理想方波形式。此外，多频全周期数据虽然增加了反演约束，但并不能从根本上消除参数相关性。当多个参数在所选频率和时间区间内仍表现出相似的敏感性特征时，仍需结合先验参数、敏感性分析和不确定度评估确定可以可靠反演的参数组合。

总体而言，SPS 通过宽频方波调制和全周期温度波形采集，在频率调控热扩散尺度的基础上进一步引入周期内时序信息，为复杂材料体系的多参数热物性反演提供了更多独立约束。与 TDTR 相比，SPS 更侧重较宽时间尺度下的周期热响应，而 TDTR 在皮秒至纳秒尺度超快热过程及纳米薄膜界面输运表征方面仍具有明显优势，两者在适用时间尺度和参数敏感性方面具有互补性。需要指出的是，SSTR、TTR、FDTR、TDTR 及 SPS 主要通过时间或频率变量改变热扩散响应；对于面内热输运及热导率各向异性的辨识，还需要进一步引入空间位置这一独立观测维度，由此形成下一节讨论的空间域热反射方法。

### 2.6 空间域响应：SDTR 及光斑偏移方法

空间域热反射(SDTR)[39]通过改变探测光相对于泵浦光的横向位置，测量热反射信号随空间偏移距离的变化，从而表征样品表面温度场的横向扩散特征。不同于前述主要以时间、频率或周期波形作为观测变量的热反射方法，SDTR 将横向空间位置作为独立观测维度，使面内热扩散直接参与热物性参数反演。由于空间温度场对横向热输运具有较强响应，该方法特别适用于各向异性晶体、取向薄膜以及其他具有显著面内热输运特征材料的表征。

在典型 SDTR 实验中，调制泵浦光在样品表面形成局部周期性热源，探测光相对于泵浦中心沿横向方向逐点扫描，并记录不同偏移位置处的热反射响应。

随着探测位置逐渐远离加热中心，温度振荡幅值通常逐渐减小，同时相对于周期热源的相位滞后发生变化。通过分析信号幅值或相位随横向偏移距离的空间演化，并结合相应热扩散模型，可以反演材料的面内热导率或热扩散率。为增强信号对面内热扩散的敏感性，需要合理匹配调制频率、光斑尺寸和空间扫描范围，使横向热扩散尺度与光斑尺度具有适当的相对关系。当热扩散尺度远小于光斑尺寸时，空间响应更容易受到局部光斑分布影响，对面内热物性的敏感性较弱；适当降低调制频率、增大热扩散尺度，则有利于增强面内热输运对空间响应的贡献。

光斑偏移时域热反射法（Beam-Offset TDTR，BO-TDTR)[66, 67]是将横向空间信息引入TDTR平台的代表性方法。该方法通常选择对面内热输运具有较高敏感性的延迟时间，并系统改变泵浦光与探测光之间的横向偏移量，获得锁相信号随空间位置的分布。通过分析正交分量($V_{\mathrm{out}}$)等信号的空间展宽特征，例如半高全宽，并结合各向异性热扩散模型，可以反演材料的面内热导率或热扩散率。与 SDTR 常采用幅值或相位随偏移距离的空间衰减进行分析相比，BO-TDTR 更多利用温度场空间分布的展宽特征提取面内热输运信息。二者虽然在具体光路和信号形式上有所不同，但共同特点都是将横向空间变量纳入测量与反演过程，从而主动增强信号对面内热输运参数的敏感性。

空间维度的引入对于各向异性热输运表征尤其重要。对于热导率具有明显方向依赖的晶体或取向薄膜，可沿不同面内方向进行空间扫描，比较温度场在不同方向上的扩散范围和衰减特征，并结合各向异性热扩散模型获得不同方向的面内热导率；在具有足够方向信息和适当模型约束的条件下，还可进一步辨识面内热导率的主轴方向。与主要依赖时间或频率变化的测量相比，空间扫描提供了直接反映横向热扩散的信息维度，能够改变面内热导率与面外热导率、界面热导等参数的敏感性分布，从而为降低不同热输运参数之间的相关性提供额外约束。

空间域方法的测量能力同样受到热扩散尺度与实验几何条件的限制。为增强温度场的横向展宽，通常需要适当降低调制频率以增大热扩散尺度，但与此同时，纵向热穿透深度也会增加，使衬底及深层结构对测量信号的贡献增强，并可能削弱对目标薄膜面内热输运的辨识能力。此外，较低频率下的测量更易受到环境温漂、激光功率漂移及低频噪声影响；若调制频率过高，横向热扩散尺度相对于光斑尺寸过小，则测得的空间响应更容易受到光斑卷积和局域纵向传热的影响，导致信号对材料本征面内热物性的敏感性降低[39]。因此，空间域实验同样需要结合特征尺度和参数敏感性分析，对调制频率、光斑尺寸及扫描范围进行协同优化。

需要进一步区分的是，并非所有采用泵浦 – 探测光斑偏移的热反射方法都属于本文所定义的空间域测量。为避免不同光斑偏移技术在分类上的混淆，本文依据“横向空间偏移量是否作为独立扫描变量参与参数反演”进行区分。当泵浦 – 探测偏移量作为独立自变量进行系统扫描，并通过幅值、相位或其他热响应随空间位置的变化规律反演面内热输运参数时，本文将其归入空间域热反射方法。相反，若横向偏移量仅作为固定的实验几何条件，用于改变热流路径或增强某一目标参数的敏感性，而最终反演仍主要依赖频率扫描或周期波形，例如BO-FDTR[68, 69]和BO-SPS[43]，则更适合归入具有空间约束的频域或全周期响应方法。这一区分的关键不在于实验中是否存在光斑偏移，而在于空间位置本身是否构成独立的观测和拟合维度。

总体而言，SDTR 及相关空间扫描方法通过引入横向位置这一独立观测变量，将热反射测量由时间和频率维度进一步拓展至空间维度，为面内热导率及热输运各向异性的辨识提供了重要手段。然而，空间响应仍受到面外热扩散、界面热导、衬底热物性及光斑几何等因素的共同影响，增加空间维度并不意味着参数耦合可以被完全消除。对于复杂多层结构和多参数反演问题，仍需结合敏感性分析及先验参数约束，并与 TDTR、FDTR 或 SPS 等时间域、频率域和全周期响应方法形成互补，从而获得更加可靠的多维热物性表征。

### 2.7 小结

综上，各类热反射测量方法均建立在相同的热反射效应与热扩散理论框架之上，其主要差异体现在热激励方式、主要观测变量和信号获取形式。SSTR 主要利用准稳态温升；TTR 记录单脉冲加热后的瞬态温度衰减；FDTR 以调制频率为主要扫描变量获取幅值和相位响应；TDTR 以泵浦–探测延迟时间为主要扫描变量，调制频率则作为锁相检测和敏感性调控的重要实验参数；SPS 记录周期方波激励下的完整温度波形，并可结合不同调制频率；SDTR 及相关空间扫描方法则以横向空间位置为主要观测变量。

由此可见，不同热反射技术并非彼此替代的线性演进关系，而是围绕准稳态、瞬态时间、调制频率、超快延迟时间、周期波形和空间位置等不同观测方式形成的互补测量范式。这种分类强调“主要观测变量”，同时允许调制频率、光斑尺寸等实验控制参数跨方法

使用。基于这一框架，下一节将进一步比较不同技术的热扩散尺度、参数可辨识性、面内/面外热输运能力和适用场景。

## 3 不同热反射方法的测量范围与适用条件

前文从观测维度出发，分别介绍了 SSTR、TTR、FDTR、TDTR、SPS 和 SDTR 等典型热反射方法。尽管这些技术共享相同的热反射效应与热扩散物理基础，但由于激励方式、观测变量及实验条件不同，其有效热扩散尺度、参数敏感性、面内/面外热输运辨识能力及适用样品体系存在明显差异。因此，本节不再按方法逐一展开，而是围绕实际测量中的关键问题，对不同技术的测量能力和适用条件进行横向比较，重点讨论有效热扩散尺度及其调控方式、热物性参数敏感性与可辨识性、各向异性热输运表征能力以及不同材料体系下的方法选择。上述比较可为后续典型应用中的实验方案设计、参数选择和热物性反演提供依据。

### 3.1 有效热扩散尺度与实验调控

热反射测量的有效探测范围首先受到热扩散特征尺度的制约。对于周期性调制热源，可采用前述热穿透深度$d_p = \sqrt{\alpha/\pi f}$表征周期热扰动的特征传播尺度，其中$\alpha$为材料热扩散率，$f$为泵浦光调制频率。对于单脉冲瞬态过程，则可采用$d_p = \sqrt{\alpha\tau}$估算观测时间$\tau$内的特征热扩散长度。需要指出的是，这些特征尺度并不是严格的探测边界，其实际意义在于与薄膜厚度、界面深度和光斑尺寸等几何尺度进行比较。实验中的调制频率、延迟时间、脉冲重复周期及信号采集窗口共同影响热扰动的有效传播范围，进而改变表层薄膜、埋藏界面和衬底对测量信号的相对贡献。

对于准稳态或长时间尺度测量，热扰动可以传播至较大的空间范围。SSTR 利用准稳态温升信息，测量结果主要受到样品整体热阻网络控制，因此更适合热输运路径相对简单、无需进行纵向分层辨识的材料体系。其优势在于能够表征有效热导率或体系等效热阻，但当多层薄膜、界面及衬底热阻同时参与热输运时，单一稳态温升难以区分各部分的独立贡献。TTR 则通过记录脉冲激励后的温度衰减过程，利用观测时间窗口对应的瞬态热扩散尺度获取热输运信息，适用于较厚薄膜、块体材料以及特征热弛豫时间落在仪器有效时间范围内的体系。对于超薄薄膜或界面主导结构，若热弛豫时间接近仪器时间分辨极限，或长时间衰减阶段的信噪比较低，则不同热物性参数的响应更难区分，从而增加反演不确定度[62, 70]。

FDTR 主要利用调制频率连续改变周期热穿透深度，因此具有较直接的热扩散尺度调控能力。提高调制频率可使热扰动更加集中于样品近表面区域，降低频率则可增大热穿透深度，使下层薄膜、界面及衬底对测量信号的贡献逐渐增强。由此，频率扫描不仅改变有效探测尺度，也会重新分配不同热物性参数的敏感性权重。对于多层结构，不同功能层或界面的参数敏感频区可能发生重叠，因此单纯扩大频率扫描范围并不必然实现多参数独立辨识，仍需结合敏感性分析确定有效频率窗口[68]。

TDTR 主要通过泵浦－探测延迟时间扫描获取皮秒至纳秒尺度的瞬态热演化信息；泵浦调制频率则影响脉冲间热积累和周期热扩散背景，因而可作为改变参数敏感性的实验控制量。通过在不同调制频率、光斑尺寸或延迟窗口下重复测量，可以构建多组互补的时域约束。其可用范围仍受到激光重复频率、机械延迟范围、热积累效应以及调制和探测带宽等条件限制。

SPS 进一步拓展了可利用的频率范围和周期内时间信息。通过方波调制和全周期温度波形采集，SPS 不仅利用调制频率改变热扩散尺度，还同时保留不同加热与冷却时刻的热响应。在较宽频率范围内改变调制条件，可以使近表面区域、深层结构及整体热阻网络在不同程度上参与测量响应；周期内部的不同时序阶段又提供额外的参数敏感性差异。因此，SPS 能够在同一实验构型下利用多频率、多时间尺度热响应约束参数反演。但这种尺度扩展并不意味着不同频率或时间阶段与特定材料层之间存在严格的一一对应关系，实际有效敏感区间仍需结合样品结构和敏感性分析确定。

与上述主要沿时间或频率方向调节热扩散尺度的方法不同，SDTR 及空间扫描类方法进一步引入了横向空间尺度。通过协同选择调制频率、光斑尺寸及空间扫描范围，可以使横向热扩散尺度与泵浦–探测几何尺寸相匹配，从而增强测量信号对面内热输运的敏感性。当横向热扩散尺度远小于光斑尺寸时，空间响应更容易受到局部光斑分布和纵向热扩散的影响；当二者处于可比尺度时，横向温度场的展宽和衰减能够提供更明显的面内热输运信息。因此，空间域方法的主要特点并不是简单扩大热扩散长度，而是通过引入横向位置这一独立观测维度，使尺度分析由纵向热穿透进一步扩展到面内热扩散。

总体而言，不同热反射技术对有效热扩散尺度的调控方式各有侧重：SSTR 利用准稳态热场；TTR 通过观测时间窗口选择瞬态扩散尺度；FDTR 通过调制频率扫描改变周期热穿透深度；TDTR 以延迟时间解析超快瞬态过程，并可借助调制频率和光斑尺寸改变

实验敏感性；SPS 利用调制频率与周期波形共同提供多尺度信息；SDTR 则通过横向空间位置、光斑尺寸和调制频率调控面内扩散响应。实际实验设计仍需根据样品特征尺度和目标参数选择合适的观测变量与控制参数。

### 3.2 参数敏感性与可辨识性

参数敏感性的定义、参数相关性以及“可测”与“可分离”的区别已在 1.4 节给出。本节不再重复这些基本概念，而侧重比较不同热反射方法能够提供何种独立观测信息，以及这些信息如何改变参数可辨识性。

从方法比较的角度看，SSTR 和 TTR 主要依赖单一准稳态或瞬态时间响应，适合参数数目较少或先验信息较充分的体系；FDTR 通过调制频率扫描获得一组频率依赖响应，可利用不同频率下敏感性权重的变化约束薄膜、界面和衬底参数。其可靠性取决于所选频率窗口内不同参数响应是否具有足够差异。

TDTR 以延迟时间为主要观测变量，并可通过改变调制频率、光斑尺寸和拟合窗口获得多组互补条件；SPS 则在不同调制频率下保留周期内完整波形。二者都可以通过增加具有不同物理敏感性的实验条件来改善参数辨识，但“更多数据”本身并不保证强耦合参数能够被独立反演。

SDTR 及其他空间扫描/光斑偏移方法进一步引入横向空间信息，对面内热扩散和方向性热输运尤其有价值。对于各向异性或多层体系，空间信息可与频率或时间信息联合使用，从而改变面内热导率、面外热导率和界面热导之间的相关结构。

因此，多参数问题的核心并非寻找某一种“最强”的热反射方法，而是判断现有观测条件是否为目标参数提供了线性独立或近似独立的信息。实际分析中可结合敏感性矩阵、Fisher 信息矩阵、奇异值分解或后验相关性评估，必要时增加新的频率、延迟时间、光斑尺寸、空间偏移或独立先验参数。

对于复杂体系，跨测量维度的互补信息通常比在单一条件下简单增加数据点更有价值。方法选择应围绕目标参数、样品尺度、可获得先验信息和预期不确定度展开，并在实验前后通过敏感性与不确定度分析验证参数是否真正可辨识。

### 3.3 面内与面外热输运的辨识能力

对于薄膜、层状结构和各向异性材料，面内与面外热导率可能存在显著差异，因此不同方向热输运参数的辨识能力是方法选择的重要依据。从特征尺度来看，当热穿透深度相对于光斑尺寸较小、纵向热扩散占主导时，测量信号通常对面外热导率、界面热导及纵向热阻更为敏感；随着横向热扩散尺度增大并与光斑尺寸或空间偏移尺度接近，面内热输运对信号的贡献增强，从而有利于面内热导率的辨识。

在常规同轴泵浦－探测构型下，TTR、FDTR 和 TDTR 通常更侧重面外热输运表征。其中，FDTR 可通过调制频率改变热穿透深度，TDTR 进一步利用延迟时间和调制频率调节面外热导率、界面热导等参数的敏感性。通过减小光斑尺寸或降低调制频率也可以增强径向热扩散贡献，但轴对称同轴构型通常难以独立辨识不同面内方向的热导率，尤其对于强各向异性材料更为明显。

当研究目标转向面内热导率或面内各向异性时，引入横向空间位置或光斑偏移可以显著改变参数敏感性分布。SDTR、BO-TDTR 以及具有空间约束的 BO-FDTR、BO-SPS 等方法[66, 68, 69]均可通过增强横向热扩散贡献，提高信号对面内热输运参数的敏感性。对于各向异性材料，沿不同面内方向进行空间测量并结合各向异性热扩散模型，可获得方向相关的面内热导率信息。需要指出的是，空间维度的引入并不会自动消除面外热导率、界面热导和衬底参数的影响，因此仍需通过光斑尺寸、调制频率及空间尺度的合理匹配降低参数相关性。

总体而言，若研究重点为薄膜面外热导率、界面热导或纵向热阻网络，同轴 TDTR、FDTR 及 SPS 等方法通常更为适合；若目标为面内热导率及其各向异性，则引入空间扫描或光斑偏移约束更为有利。对于需要同时获得$k_r$、$k_z$和$G$等多个参数的复杂各向异性体系，单一构型往往难以提供充分的独立约束，通常需要联合不同光斑尺寸、调制频率或空间测量构型，利用互补敏感性信息提高参数可辨识性。

### 3.5 材料体系与应用场景匹配

在实际实验测量中，热反射测试方案的选取主要由样品结构、目标待测热物性参数和可接受的实验复杂度共同决定。各类热反射表征技术不存在绝对的优劣层级划分，其适用边界由热扩散特征尺度和多参数反演约束条件共同决定。表 1 总结了主流热反射技术在探测维度、有效热扩散尺度与典型应用场景上的差异。

针对块体材料或热物性标准参考样品，若目标是获得等效整体热导率与整体热阻，SSTR 可实现简洁直观的定量测量。该类样品内部结构简单，界面热阻、薄膜厚度等结构参数对测温信号的影响较小，准稳态温升信号能够直接表征体系主导热阻特征，十分适用

于材料快速热物性筛查和测试系统基准标定工作。

对于纳米至微米尺度的薄膜及界面结构，TDTR 和 FDTR 是面外热导率、界面热导定量表征的常用手段。TDTR 通过延迟时间扫描结合调制锁相检测解析快速热响应；FDTR 则利用调制频率扫描改变热穿透深度，使不同层和界面对信号的相对敏感性随频率变化。需要强调的是，这种频率依赖性提供的是“敏感性权重变化”，并不等同于严格的纵向深度剖析。对于参数较多的多层结构，两种方法均通常需要结合敏感性分析和独立先验参数，以提高反演稳定性。

针对低热导率材料、多孔介质和多层异质结构，往往需要更长的热扩散时间尺度或多组具有互补敏感性的测量条件。可根据具体仪器带宽和样品尺度采用低频 FDTR、扩展时间窗口的瞬态方法、多频/多光斑 TDTR、SPS 或跨平台联合测量。SPS 利用较宽频率范围和周期全波形提供了一种可行方案，但其优势是否成立仍取决于低频稳定性、波形保真度、先验参数和目标参数的实际可辨识性。

面向二维材料、取向薄膜和各向异性晶体，测量技术通常需要主动引入横向热扩散信息。SDTR、BO-TDTR、BO-FDTR 和 BO-SPS 可通过空间扫描或光斑偏移增强对面内热输运的响应，并用于比较不同方向的热扩散行为。与此同时，面外热导率、界面热导和衬底参数仍可能影响空间信号，因此完整的三维或界面表征通常需要联合频率、时间或其他先验信息。

因此，表 1 的核心价值并非对各类表征技术进行优劣排序，而是为实验方案设计提供定性参考。实际选择测量方法时，需综合考虑膜厚、热导率范围、界面深度、热各向异性、仪器带宽及可获得的先验参数，并通过敏感性与不确定度分析判断目标物性是否能够在可接受的不确定度范围内辨识。表中给出的尺度和适用对象均应理解为典型条件下的参考，而非严格边界。

**表 1** 不同热反射测量方法的主要测量信息、适用对象与使用限制

**Table 1** Main measurement information, applicable objects, and limitations of different thermoreflectance techniques

| 方法 | 主要观测量 | 热扩散尺度调控方式与范围 | 适用对象 | 使用限制 |
| --- | --- | --- | --- | --- |
| **SSTR** | 准稳态温升幅值 | $d_p \to 0$(准稳态) | 块体/较厚样品的有效热导率、整体热阻 | 多层体系中各热阻贡献易耦合；对吸收功率和稳态边界条件较敏感 |
| **TTR** | 温度衰减曲线 | $d_p \sim \sqrt{\alpha \cdot t_{cool}}$(脉冲间距决定) | 较厚薄膜(>1 μm)、块体材料 | 受探测带宽、长时间信噪比和参数相关性限制 |
| **FDTR** | 调制频率响应 | $d_p \propto 1/\sqrt{f}$(频率可调) | 薄膜、界面及多层结构的频率依赖热响应与参数反演 | 不同参数的敏感性频区可能重叠；频率变化不等同于严格深度剖析 |
| **TDTR** | 延迟时间响应及锁相信号比值 | $d_p$中等可调(通常 1-10 μm) | 薄膜面外热导率、界面热导及超快热过程 | 常需已知膜厚、热容或部分界面参数，频率范围有限 |
| **SDTR** | 空间分布扫描 | $d_p \sim \sqrt{\alpha/\pi f}$(横向扩散主导) | 面内各向异性、晶体取向 | 对面外热导率和界面热导不敏感 |
| **SPS** | 不同调制频率下的完整周期波形 | $d_p$较宽(已有系统报道 1 Hz~20 MHz) | 低热导材料、多层异质结构、参数强耦合体系 | 低频漂移与高频波形保真会限制有效范围；不用于皮秒级超快过程 |

图 4 以单晶硅作为典型参照材料，直观展示各类热反射表征技术可实现的等效热扩散长度范围。需要注意的是，图中标注的尺度区间并非刚性临界边界，而是基于各类技术典型调制频率、时域采集窗口和激光光斑尺寸计算得到的特征参考范围。实际热扩散长度会随材料本征热扩散率、样品层状构型和实验工况参数的调整发生动态变化。

在亚微米至十微米的深度探测区间，高频工况下的 TDTR、FDTR 和 SPS 可将热扰动约束于金属传感层、近表面薄膜及界面区域，因而常用于薄膜面外热导率、界面热导定量测量。随着调制频率下调或观测时间延长，热扩散长度持续拓展，测量信号将叠加更深层薄膜、衬底以及整体热阻的传热贡献，此时低频 FDTR、SPS 和 TTR 可用于厚薄膜与块体材料的整体热输运特性解析。

当热扩散长度与激光光斑尺寸、横向光斑偏移距离处于同一量级时，横向热扩散行为对测温信号的贡献显著提升，SDTR 与光斑偏移类方法凭借横向空间调控能力，更适配样品面内热扩散系数、各向异性导热系数的测量。对于低频 SPS 和 SSTR，热场演化趋近准稳态条件，测量信号主要反映样品较大尺度范围内的等效总热阻或整体平均热导率。

因此，图 4 的主要作用是厘清不同热反射技术在特征热扩散长度上的衔接与覆盖关系，而非界定绝对

化的适用边界。各类技术的有效量程存在大范围重叠，实际测试方案的具体选择仍需结合待测薄膜厚度、材料热导率范围、界面深度、光斑尺寸以及目标热物性参数类型综合判断。对于热扩散率系数远低于单晶硅的高分子、多孔材料，在相同调制频率下热扩散长度会相应减小，需通过降低调制频率、放大激光光斑尺寸等方式，实现目标探测深度的匹配。

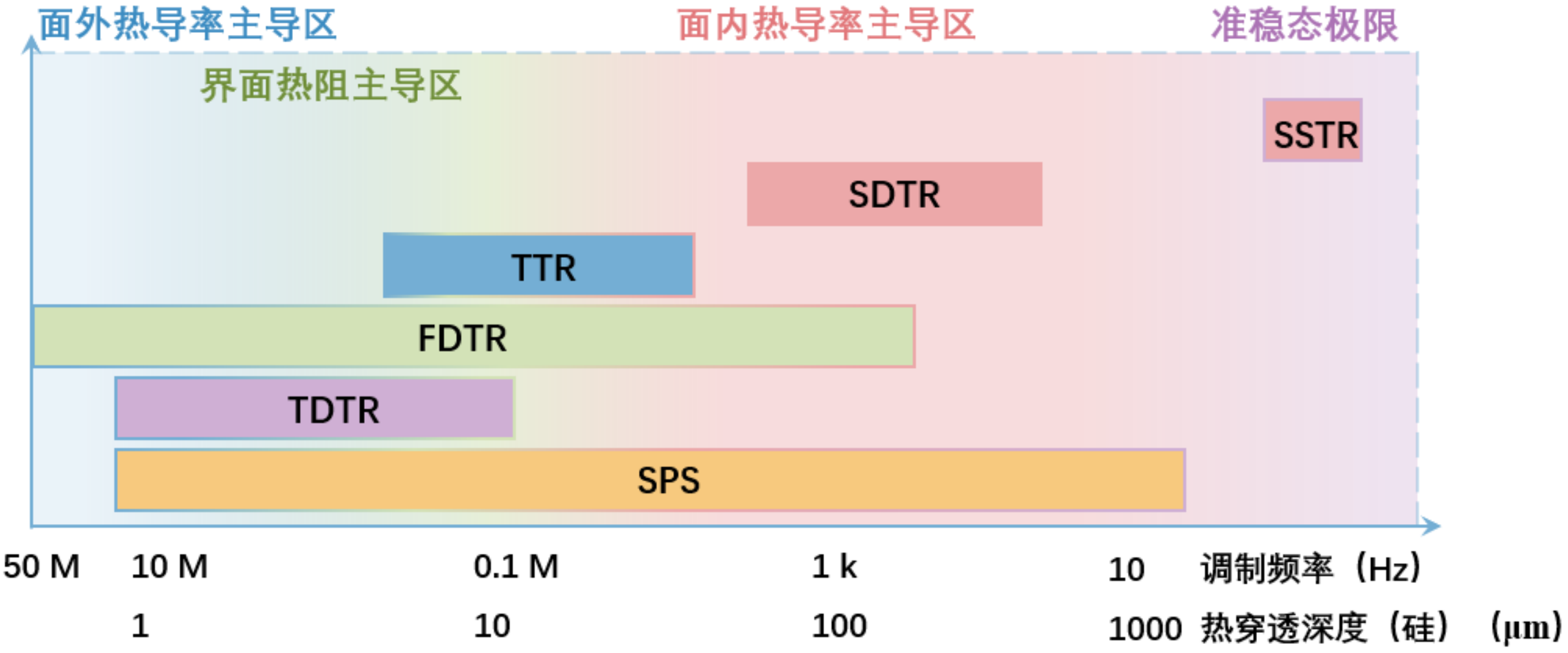

**图 4**　不同热反射方法在单晶硅中的等效热扩散长度范围

**Figure 4**　Accessible effective thermal diffusion lengths of different thermoreflectance methods in single-crystal silicon

## 4　典型复杂测量场景中的方法匹配与案例分析

前述章节已经从热扩散尺度、参数可辨识性和面内/面外热输运等角度比较了不同热反射技术。本节进一步选取四类复杂测量问题，通过代表性文献说明“需要什么独立信息”如何决定实验方案。需要说明的是，本节若干图例来自近年来 SPS/BO-SPS 工作，主要用于展示宽频、周期波形和空间偏移等观测自由度如何参与参数约束，并不意味着 SPS 在这些场景中普遍优于 TDTR、FDTR、SDTR 或其他方法。对于具体样品，最合适的方案仍取决于仪器带宽、样品尺度、可获得的先验参数和目标不确定度。

### 4.1　各向同性低热导材料：热导率与体积热容联合反演

热反射测温信号主要对样品面内热扩散率($\alpha = k_r/C$)、面外热逸散率($e = \sqrt{k_z C}$)两类组合物性敏感响应。对于各向同性均匀材料，面内、面外导热系数满足$k_r = k_z = k$。基于该关系，若依托差异化实验能够分别获得$\alpha$和$e$，便可进一步通过解析公式$k = e\sqrt{\alpha}$和$C = e/\sqrt{\alpha}$联合求解热导率和体积热容。由此可见，各向同性材料中$k$与$C$的协同表征并非单纯增加一个拟合未知数，而是依托检测信号在不同热扩散尺度下分别对$\alpha$和$e$具有可区分的敏感性特征，破除参数关联性。

TDTR 和 FDTR 技术已为该类双参数联合反演奠定了成熟的技术基础。变频 TDTR 通过连续调节泵浦调制频率改变热穿透深度，已成功实现块体、薄膜材料$k$与$C$的同步定量测量[71]。双频 TDTR 进一步利用不同调制频率下的热响应差值，实现了微米尺度空间分辨的体积热容成像表征[72]。在频域表征方面，Schmidt 等人提出的 FDTR 技术通过扫描调制频率采集幅值与相位演化曲线，为均质块体及薄膜结构的热导率、热扩散率和体积热容相关物性参数的反演构建了独立的频域调控维度[10]；后续发展的宽频 FDTR 进一步验证了其在各向同性材料热导率和体积热容双物性同步提取的能力[73]。近年来，负延迟时间 TDTR、变光斑尺寸 TDTR 以及光斑偏移 TDTR/FDTR 等衍生方法[53, 69, 74, 75]又分别从延迟时间采集窗口、光斑几何尺寸和横向空间调控等方面拓展了 TDTR/FDTR 的多参数解耦自由度。上述研究结果说明，经典频域/时域热反射技术不再局限于单一热导率测量，依托多维实验变量协同设计，可实现$k$、$C$及界面热导的多参数联合反演。

然而，对于聚甲基丙烯酸甲酯（PMMA）这类各向同性低热导率聚合物，多参数联合反演的核心瓶颈在于实验系统能否实现超低频调制、构建长尺度热扩散测量实验区间。PMMA 的热扩散率处于$10^{-7}$ m$^2$/s

量级，在十微米量级光斑条件下，若要使得横向热扩散长度与光斑半径达到可比量级，从而使面内热输运行为在探测信号中形成显著贡献，泵浦光调制频率需要降至百赫兹甚至更低频段。常规 TDTR 虽然具备超高时序分辨率和成熟完备的多层结构热解析模型，但其常用调制频率多处于 MHz 频段。受激光器固有脉冲重复频率约束、长时脉冲累积效应和低频锁相检测稳定性等因素影响，常规 TDTR 难以直接覆盖低热导聚合物所需的超低频面内热扩散区间。FDTR 虽可通过降低调制频率增加热穿透深度，但低频测量往往需要大幅延长信号积分时长，并更容易受到环境热漂移、相位校准误差和信噪比下降的影响。因此，仅依靠常规 TDTR 或 FDTR 开展低热导各向同性材料测试时，受参数耦合与低频测试短板制约，通常需要引入文献给定体积热容值，或是通过独立热容标定实验固定热容参数，才能实现热导率的稳定反演。需要指出的是，这些限制并非 TDTR 或 FDTR 的绝对边界；通过降低调制频率、改变光斑尺寸、增加独立热容信息或采用多工况联合拟合，仍可形成有效方案，具体取决于仪器性能和目标不确定度。

在该低热导材料案例中，SPS 提供了一种利用较宽频率范围和完整周期波形建立互补约束的实现路径。该技术采用方波调制热源并完整采集周期温升波形，可在赫兹至兆赫兹超宽频段内连续调控热扩散长度：低频段的温升波能够充分捕捉低热导材料内部充分发展的面内热扩散信息；高频波形则提供与近表面热逸散率相关的约束。与负延迟时间 TDTR 等主要扩展既有时域信息利用方式相比，该 SPS 实现将可用测量条件进一步延伸至较低调制频率，从而为$\alpha$、$e$以及进一步的$k-C$解耦提供互补测试信息。

图 5 展示了 Chen 等人[40]采用 SPS 方法对 PMMA 开展热导率与体积热容协同表征的实验结果。在低频测量条件下，激光光斑直径 23 μm、调制频率 50 Hz，测量信号主要对 PMMA 的$k_r$、$C$以及光斑半径$r_0$敏感，对金属传感层物性、界面热导的敏感性较弱；因此在光斑半径已知的条件下，可直接拟合得到组合参数$k_r/C = 0.11557\times10^{-6}\ \mathrm{m^2/s}$。在高频测量条件下，设置光斑直径 38.6 μm、调制频率 400 kHz，信号敏感性转向组合参数$\sqrt{k_zC}/(h_mC_m)$。在金属传感层面积热容$h_mC_m$已知的条件下，计算得到$k_zC = 0.31236\times10^6\ \mathrm{W^2\cdot s/(m^4\cdot K^2)}$。

联立上述两组独立工况的测量结果，可同时求解得$k = 0.19\pm0.01\ \mathrm{W/(m\cdot K)}$，体积热容$C = 1.644\pm0.11\ \mathrm{MJ/(m^3\cdot K)}$，总体不确定度约 7%。该测试结果与已有文献报道的 PMMA 热导率$0.1934\ \mathrm{W/(m\cdot K)}$与体积热容$1.6104\ \mathrm{MJ/(m^3\cdot K)}$吻合良好[76, 77]，说明在该实验条件下，SPS 通过低频热扩散率测量和高频热逸散率测量的组合，可实现各向同性低热导材料中$k$与$C$的自洽反演。作为对照，在相同体系中，TDTR 在采用文献热容值作为输入时可得到 PMMA 的$k_z\approx0.21\ \mathrm{W/(m\cdot K)}$，该对照方案未同时求解热导率和体积热容两个独立物性参数[40]。

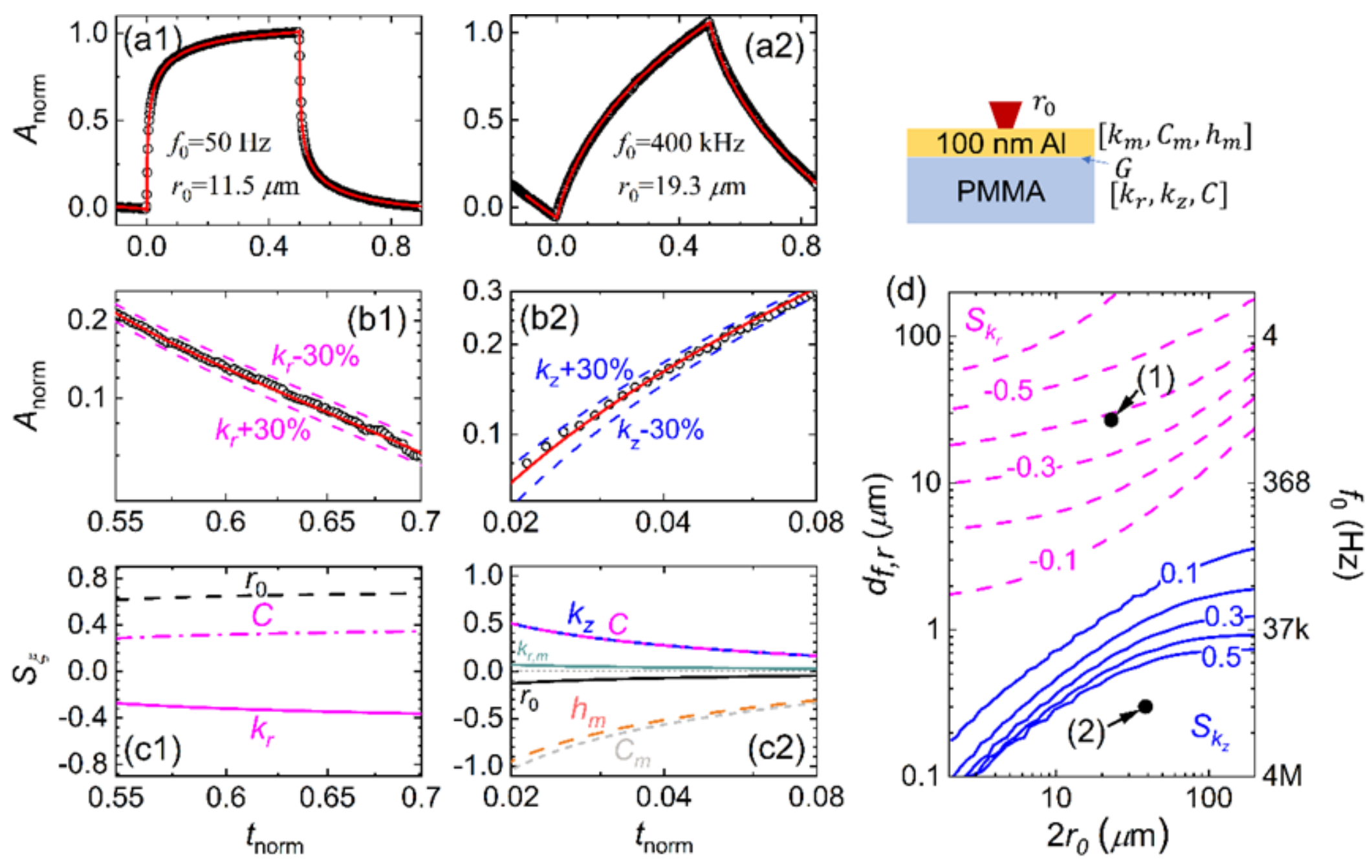

**图 5** 基于 SPS 的 PMMA 面内热扩散率与面外热逸散率联合反演[40]，Copyright © 2024 Elsevier Ltd.

**Figure 5** Joint extraction of in-plane thermal diffusivity and cross-plane thermal effusivity of PMMA using SPS[40], Copyright © 2024 Elsevier Ltd.

## 4.2 各向异性薄膜：面内/面外热导率与体积热容联合辨识

相较于各向同性材料，各向异性薄膜的热反射表征难度显著提升。以取向聚酰亚胺(PI)聚合物薄膜为例，面内热导率$k_r$、面外热导率$k_z$和体积热容$C$会共同耦合调制表面测温信号。若仅采用单一调制频率、固定光斑尺寸或单一测量平台，三类物性参数的敏感性曲线极易发生重叠耦合，导致面内/面外热导率和体积热容无法实现独立解耦求解。

早期针对 PI 薄膜热输运研究多采用分平台或分步骤测试方案。Kurabayashi 等人[78]利用微加工悬臂和微桥电阻法分别标定$k_z$与面内热扩散率，再结合文献$C$值换算$k_r$。近期 Chowdhury 等人[79]结合光热位移相位谱(D-TOPS)[1]与 TDTR 分别提取$k_r$与$k_z$，但仍需预设$C$。上述测试手段为 PI 薄膜各向异性热输运研究提供了重要基础；但当薄膜体积热容受到旋涂、拉伸、固化和分子取向等加工过程影响时，直接引用文献恒定热容值，可能将额外不确定度传递到热导率结果中。

由此可见，各向异性薄膜热物性表征的核心思路并非片面提高单一参数的敏感性，而是依托同一套实验系统采集多组热响应信号，为$k_r$、$k_z$和$C$三者提供相互独立、互补的约束条件。各向异性薄膜的热反射响应可拆解为若干组合参数的耦合贡献，包括面外热逸散率($\sqrt{k_z C}$)相关项、面内热扩散率($k_r/C$)相关项以及面积热容$hC$。只有当实验条件能够使这些组合参数在不同频率或不同热扩散尺度下形成差异化敏感性特征并且膜厚$h$已知时，才能进一步反演$k_r$、$k_z$和$C$。

在该 PI 薄膜案例中，SPS 提供了一种在同一平台内构建多尺度约束的实现方式：利用宽频方波调制和完整周期波形采集，在同一平台内实现热扩散长度的连续调谐，并构建多维度独立敏感性约束。高频激励条件下，热扰动主要局限于金属传感层和薄膜近表面区域，探测信号对面外热逸散率及金属传感层参数更为敏感；随着频率降低，热扩散长度增大，薄膜厚度方向和横向扩散贡献占比逐步提升；在低频和小光斑测试条件下，面内热扩散对周期温度波形的影响更加明显。通过联合拟合不同频率和不同光斑条件下的周期温升波形，可在不预设体积热容的情况下同时获得$k_r$、$k_z$与$C$。

图 6 展示了 Zhang 等人[80]采用 SPS 方法对悬空 PI 薄膜和旋涂 PI 薄膜进行热物性测量的结果。对于悬空 PI 薄膜 (Kapton HN-100)，测得$k_r = 0.56\ \mathrm{W/(m \cdot K)}$，$k_z = 0.175\ \mathrm{W/(m \cdot K)}$，面内-面外热导率各向异性比约为3.2。相比之下，旋涂 PI 薄膜的热输运各向异性显著弱化，各向异性比约为$1.6-1.7$，且面外热导率提升至$0.29-0.34\ \mathrm{W/(m \cdot K)}$。这一结果说明，不同薄膜制备工艺会显著改变 PI 薄膜的链取向和厚度方向堆积结构，从而影响面内和面外热传导能力。

拉曼偏振光谱结果进一步支持了这一解释。旋涂 PI 膜的退偏比高于商业膜，表明其分子取向更趋各向同性；这与其较低的热导率各向异性和较高的面外热导率测试结果相一致。值得强调的是，体积热容并不一定等同于块体或文献值，而可能受到薄膜致密度、固化过程和分子堆积状态的影响。因此，在该类工艺敏感薄膜中，直接测量$C$有助于减少由先验热容输入引入的系统误差。

还需指出的是，$k_r$、$k_z$和$C$的同步解耦通常要求薄膜厚度与热穿透深度处于合适的相对范围，使薄膜的面积热容$hC$ 成为独立可观测量。当热穿透深度远小于或远大于薄膜厚度时，信号可能退化为近表面响应或半无限体响应，此时只能获得$\sqrt{k_z C}$或$k_r/C$，而难以独立分离$C$。SPS 的宽频调制能力使实验可以在较宽的热扩散长度范围内寻找合适的敏感性窗口，从而提高各向异性薄膜多参数反演的稳定性。

总体而言，PI 薄膜案例体现了各向异性低热导薄膜测量中的一个共性难点：面内热导率、面外热导率和体积热容三者深度耦合于热反射探测信号中，单一实验条件往往难以为三个物性参数提供充分的独立约束。TDTR、D-TOPS、微加工电热测试等方法为面内/面外热输运研究提供了重要基础，该 SPS 案例通过宽频周期波形测量展示了在同一平台内联合反演$k_r$、$k_z$和$C$的可行路径。这一能力对于柔性电子器件热设计、聚合物封装散热优化和工艺敏感功能薄膜的热管理机理研究具有重要意义。类似的“增加独立约束”思路也可通过多频/多光斑 TDTR、BO-TDTR/FDTR 或跨平台联合测量实现，关键在于观测信息是否足以区分目标参数。

[1] D-TOPS（displacement thermo-optic phase spectroscopy，位移型热光相位谱；早期称 frequency-domain probe-beam deflection）是一种频域泵浦–探测光热方法。与 FDTR 检测热反射信号不同，D-TOPS 利用周期加热产生的热弹性表面形变，使偏移探测光发生角偏转，并由位置敏感探测器测量其频率响应。该方法对横向热扩散具有较高敏感性，因而适合面内热导率/热扩散率表征；聚合物的典型测量频率多处于 $10^2-10^4$ Hz量级。

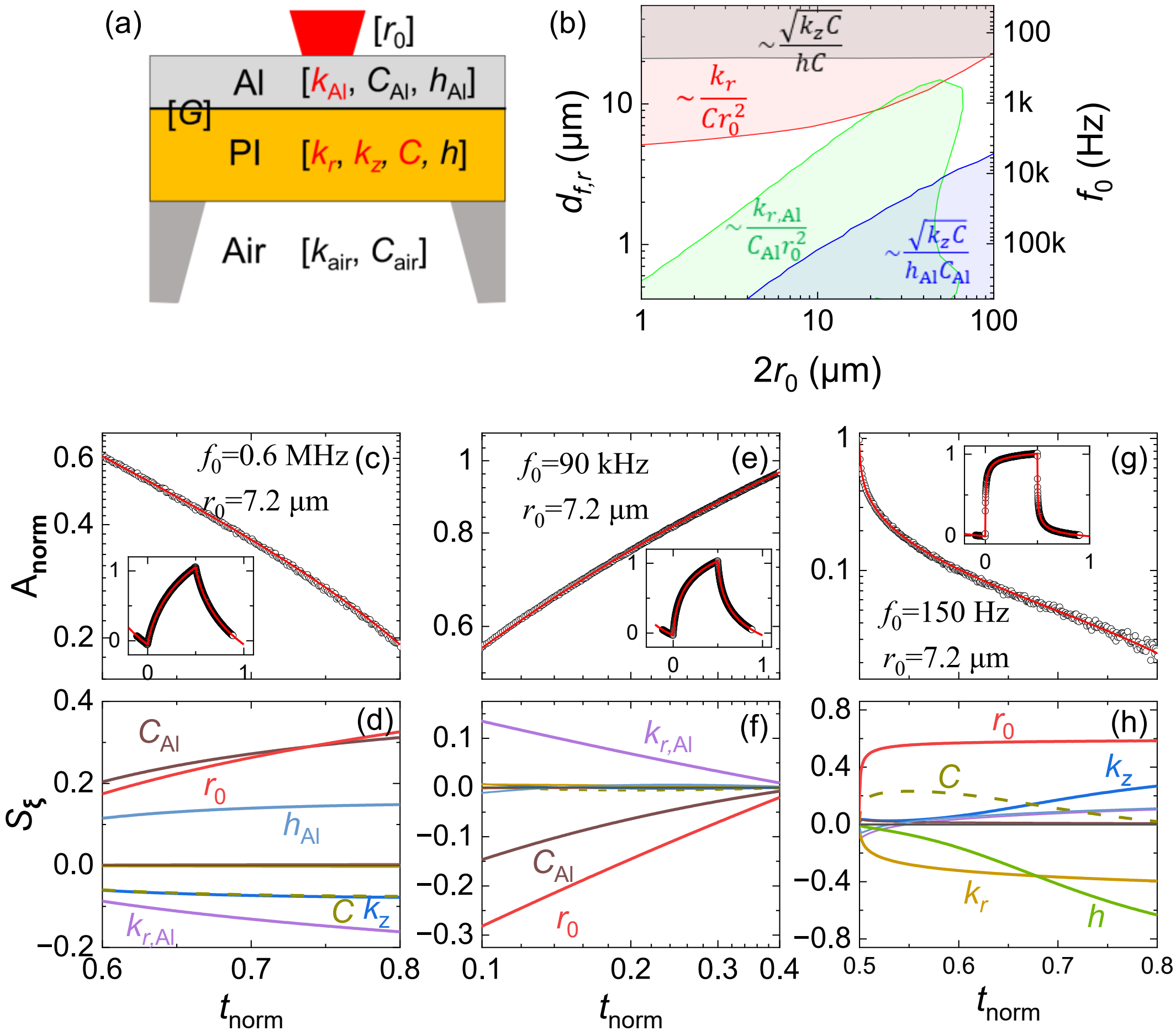

**图 6** 基于 SPS 的悬空 PI 薄膜面内/面外热导率与体积热容同步反演[80], Copyright © 2025 Elsevier Ltd.

**Figure 6** Simultaneous extraction of in-plane thermal conductivity, cross-plane thermal conductivity, and volumetric heat capacity of suspended PI films using SPS[80], Copyright © 2025 Elsevier Ltd.

### 4.3 三维各向异性热导率张量的热反射测量

对于各向异性材料，热导率不再是单一标量，而应表示为二阶张量。对于许多层状材料和取向薄膜，热输运差异主要表现为面内热导率与面外热导率的不同，因此常可用$k_r$和$k_z$近似描述。然而，在低对称晶体、斜切晶体或晶轴与器件坐标系不重合的材料中，热流方向与温度梯度方向可能并不平行，实验坐标系下的热导率张量还可能包含非对角分量。此时，热反射信号不再只反映某一方向上的等效热导率，而是由多个张量分量共同决定，三维热导率张量的反演成为更具挑战性的测量问题。

TDTR、FDTR 和 SDTR 等经典热反射技术，为各向异性导热特性的定量测试搭建了成熟的技术框架。变光斑 TDTR[53]和多频 TDTR[71]可通过改变光斑尺寸或调制频率，调节面内与面外热扩散在总信号中的占比，从而反演薄膜或晶体的$k_r$与$k_z$。椭圆光斑 TDTR[75]、光斑偏移 TDTR[67]和光斑偏移 FDTR[69]进一步引入横向空间分辨维度，使不同面内方向上的热扩散差异能够被探测。SDTR 则通过直接扫描表面温度响应的空间分布，在面内热扩散率和面内各向异性测量中具有较直观的物理图像[39]。这些工作推动了热反射测量从早期单一面外热输运表征拓展至具备角度分辨能力的方向性热输运定量解析。

即便如此，仅依靠面内、法向等效热导率的二元测量，依旧无法重构完整三维热导率张量。特别是当晶体主轴与样品表面坐标系不重合时，实验坐标系中的热导率张量可写为

$$\mathbf{k}=\begin{bmatrix}k_{xx} & k_{xy} & k_{xz}\\ k_{yx} & k_{yy} & k_{yz}\\ k_{zx} & k_{zy} & k_{zz}\end{bmatrix} \tag{5}$$

其中对角项和非对角项均可能影响表面温度场。不同张量分量对热反射信号的贡献会随调制频率、光斑偏移方向和探测位置而变化，且彼此之间存在较强耦合。因此，若只依赖单一调制频率、单一空间扫描方向或预设部分张量分量，往往难以稳定获得完整张量信息。三维张量反演要求实验同时提供足够多的频率、空间和方向约束，并结合敏感性分析判断各分量是否可解耦。

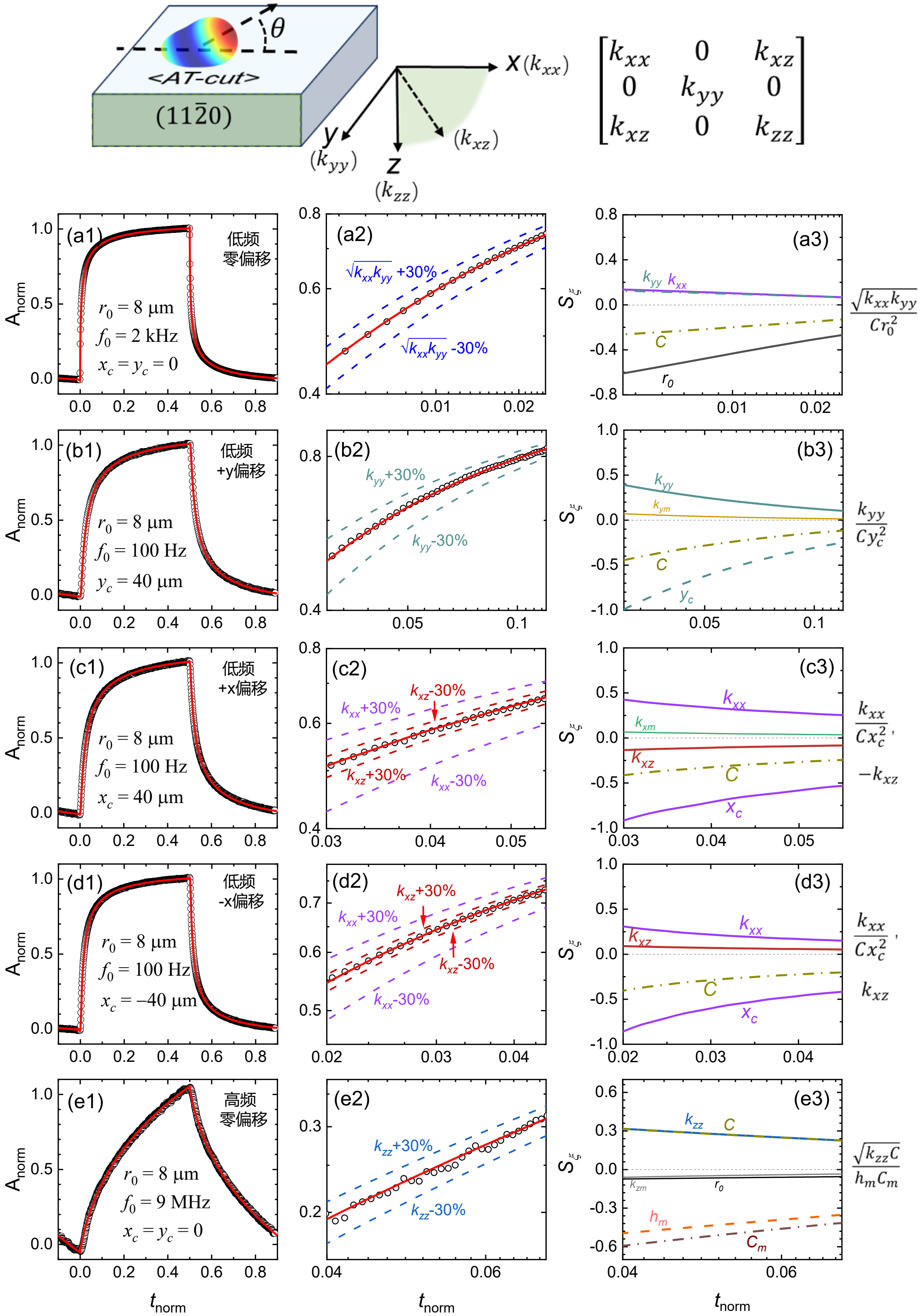

**图 7** 基于光斑偏移 SPS 的 AT 切石英完整三维热导率张量反演[43]. Copyright © 2024 Elsevier Masson SAS.

**Figure 7** Reconstruction of the full three-dimensional thermal-conductivity tensor of AT-cut quartz using beam-offset SPS[43]. Copyright © 2024 Elsevier Masson SAS

以 AT 切石英晶体为例，其晶体$c$轴(即石英的光轴或$Z$轴)相对于样品表面倾斜，切割面与$c$轴的夹角约为35.25°。在实验坐标系下，其热导率张量不仅包含面内($k_{xx}$，$k_{yy}$)与面外($k_{zz}$)主值，更关键的是存在非对角耦合项($k_{xz}$)，直接决定面内与面外热流的耦合程度。换言之，表面温度响应不再能简单归因于某一个面内热导率或面外热导率，而需要在实验坐标系下考虑三维热导率张量的共同作用。图 7 展示了 Chen 等人[43]应用 BO-SPS 方法测量 AT 切石英热导率张量的实验思路。该光路采用低频调制(~100 Hz) 搭配大光束偏移(偏移量>5 倍光斑半径)的联合设计，在样品表面建立显著的各向异性温度梯度，从而利用方向反演对称性实现参数解耦：当探测光斑沿$x$轴正负偏移时，信号对$k_{xx}$呈对称响应，对$k_{xz}$呈反对称响应(图 7c-d)。这种奇偶性差异直接源于倾斜晶轴引入的热流耦合不对称性。联合$x$方向正负偏移(图 7c-d)、零偏移(图 7a)及$y$方向偏移测量(图 7b)，即可分别锁定$k_{xx}$与$k_{xz}$、$k_{yy}$，而零偏移下的高频测量则约束面外热输运分量$k_{zz}$(图 7e)。

实验结果表明，通过少数组具有互补敏感性的偏移方向和调制频率配置，在该案例中可为三维热导率张量反演提供足够约束，从而实现实验坐标系下多个张量分量的联合求解。对角化后得到 AT 切石英的主热导率$k_c = 10.76\ \mathrm{W/(m\cdot K)}$ 、$k_a = 6.40\ \mathrm{W/(m\cdot K)}$及$c$轴倾角$\theta = -35.39°$，其与已知切割角的一致性支持了通过表面热响应反演三维热输运主轴方向的可行性。

总体而言，AT 切石英案例说明，三维各向异性测量的核心在于获得足够的方向性空间信息，并通过多组热扩散尺度改变不同张量分量的相对响应。BO-SPS 在该案例中提供了一种可行实现；同样的方法学原则也可在具备多方向空间扫描和多工况控制能力的其他热反射平台中探索。

### 4.4 多层异质结构：组合参数的选择性反演

在宽禁带半导体器件和异质集成结构中，热输运路径往往由多层功能薄膜、缓冲层、埋藏界面和支撑衬底共同决定。以 GaN/Si、GaN/SiC 和 GaN/diamond 等典型功率器件结构为例，器件工作时产生的热量需要依次通过金属接触层、GaN 外延层、过渡层或缓冲层、半导体/衬底界面以及高热导衬底实现纵向散热导出[55]。此时，热反射信号同时受到薄膜热导率、体积热容、界面热导、层厚、金属传感层参数和衬底热物性的影响。若将每一层的面内热导率、面外热导率、体积热容、界面热导和层厚均作为独立参数处理，待反演参数数目会迅速增加，导致反问题高度不适定。因此，多层异质结构测量的关键并不是简单增加拟合参数，而是识别热反射信号真正敏感的组合参数，并判断这些组合参数能否在不同频率、不同热扩散尺度或不同波形阶段被有效分离辨识。

TDTR 和 FDTR 技术在宽禁带半导体异质结构热物性表征领域已形成了较成熟的应用基础。对于 GaN/SiC 外延体系，Cho 等人采用 TDTR 研究不同厚度 GaN/SiC 样品，获得 GaN 薄膜热导率和 GaN–SiC 界面热导，为 HEMT 器件热阻分析提供了重要实验依据[81]。随后，TDTR 进一步被用于键合 GaN/SiC、带 AlN 过渡层的 GaN/SiC 以及 GaN/SiC、GaN/AlN 和 AlN/SiC 等外延晶圆，以分析外延质量、过渡层和界面结构对界面热导的影响[82-84]。在高导热衬底集成方面，Cheng 等人利用 TDTR 测量 GaN/diamond 键合界面的热导，并结合界面结构分析说明键合层厚度和界面质量对散热性能的影响[85]。对于超宽禁带半导体，TDTR 也被用于 β-$Ga_2O_3$/SiC 等异质集成体系中薄膜热导率和埋藏界面热导的测量，并发展出双调制频率 TDTR 成像方法，用于可视化 β-$Ga_2O_3$/SiC 埋藏界面热导分布[86]。

FDTR 通过连续调制频率扫描改变热穿透深度，从而使不同薄膜层、界面和衬底对信号的相对贡献随频率发生变化，形成多层结构参数反演的另一类方案。Ziade 等人利用 FDTR 测量外延 GaN/SiC 界面的温度依赖界面热导[87]；Song 等人结合 FDTR 和 SSTR 对 GaN-on-SiC 晶圆进行热表征，并将测得的热物性用于器件级热模型验证[88]；Delmas 等人采用空间分辨 FDTR 对 GaN/diamond 压缩键合界面开展热输运和应力映射[89]。这些工作表明，频率与空间条件的改变可以为多层和埋藏界面问题提供不同的敏感性权重，但并不等同于严格的层析深度成像。

然而，对于 GaN/Si 这类堆叠层数较多、纵向热阻分布跨越多个特征尺度的结构(如图 8a1 所示)，单纯依靠一个或少数几个拟合参数往往不足以描述真实热输运行为。近表面 GaN 层、AlGaN 缓冲层、AlN 过渡层、GaN/Si 埋藏界面和 Si 衬底均可能对热反射信号产生贡献，且这些贡献会随调制频率和热扩散长度变化而改变。若不先判断哪些参数组合真正可测，而是直接拟合所有单层物性参数，容易得到数学上可收敛但物理上不唯一的结果。因此，在复杂多层体系中，确定“可测参数”本身就是热反射反演的核心问题。

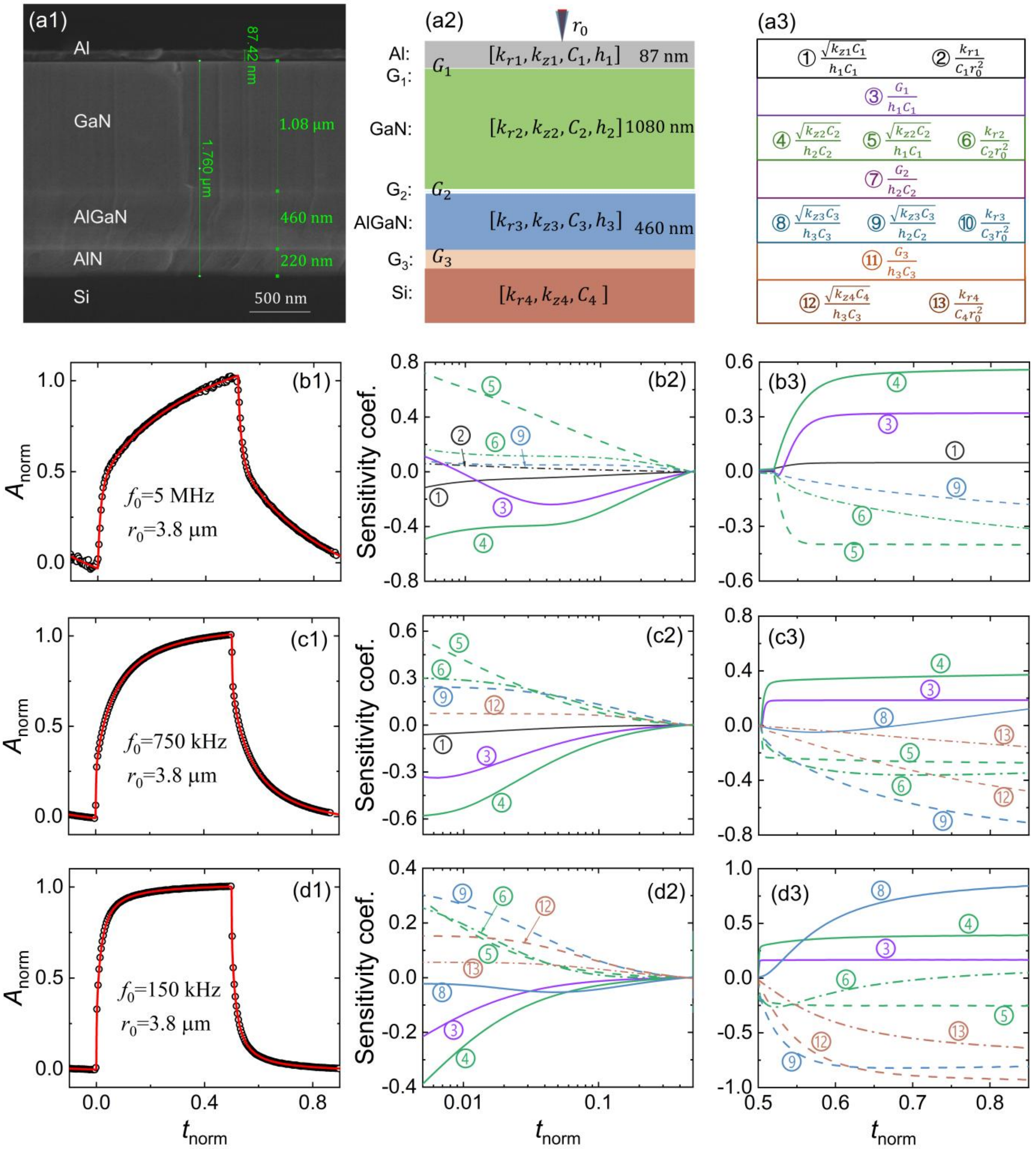

**图 8** 基于 SPS 的 GaN/Si 多层异质结构组合参数分析与热物性反演[55]，Copyright © 2025 American Physical Society

**Figure 8** Combination-parameter analysis and thermophysical-property extraction for a GaN/Si multilayer heterostructure using SPS[55], Copyright © 2025 American Physical Society

无量纲参数分析为这一问题提供了系统化处理方式. 基于热扩散方程的无量纲化，并结合热反射信号的归一化形式和参数敏感性分析，可将原始物理参数重组为若干具有明确物理意义的组合参数[55]。对于薄膜层，纵向热扩散过程可由$\tau_z = h^2C/k_z$表征，表示热量穿透厚度$h$所需的特征时间；面内热扩散过程可由$\tau_r = Cr_0^2/k_r$表征，表示热量扩散至光斑边界的特征时间，反映径向热耗散能力；界面热弛豫过程可由$\tau_G = hC/G$表征，表示单位面积薄膜热容通过界面释放热量的快慢。若与调制频率相乘，则可得到严格

无量纲的组合参数，如 $fh^2C/k_z$、$fCr_0^2/k_r$和$fhC/G$。这些组合参数具有明确的物理意义，分别对应纵向扩散、横向扩散和界面热弛豫三类核心传热机制，并非单纯数学层面的变量合并。

图 8 展示了 Chen 等人[55] 在 SPS 平台上采用组合参数分析研究 GaN/Si 多层异质结构的结果。该样品由约 87 nm Al 传感层、1.08 μm GaN、460 nm AlGaN、220 nm AlN 和 Si 衬底构成(图 8a1)。若直接建立完整多层热模型，体系涉及 19 个物理参数，包括各层热导率、体积热容、厚度、界面热导和光斑参数(图 8a2)。通过无量纲化和组合参数分析，可将这些原始参数归为 13 个组合参数(图 8a3)。其中，AlN 层由于厚度较小且热阻相对较低，可与相邻界面贡献合并为等效热阻，从而进一步降低模型复杂度。组合参数分析不仅降低了反演维度，更重要的是明确了复杂多层体系中哪些参数组合在实验上真正可测。

在该 SPS 案例中，不同调制频率使各组合参数的相对敏感性发生明显变化。在 5 MHz 高频条件下(图 8b1-b3)，热响应主要集中于 Al 传感层和 GaN 近表面区域；750 kHz 时(图 8c1-c3)，热扩散范围进一步进入 GaN 与 AlGaN 缓冲层；150 kHz 低频时(图 8d1-d3)，衬底相关参数的影响权重增加。因此，这里的“层选择性”更准确地理解为不同频率下各层和各组合参数对测量信号的相对敏感性发生变化，而不是逐层独立成像或严格的深度分离。

除频率带来的敏感性变化外，该 SPS 案例还利用周期波形中的加热和冷却阶段增加组合参数的区分信息。近表面热容和界面热导、薄膜纵向扩散、面内扩散以及衬底热物性对波形不同阶段的影响并不完全相同。通过联合拟合多个频率下的完整周期响应，可从候选组合参数中筛选出实验上具有较高敏感性且相关性相对较低的参数，再结合已知层厚、光斑尺寸和传感层面积热容反演具体热物性。

基于上述组合参数反演策略，Chen 等人[55]获得了 GaN/Si 多层结构中多个关键热物性参数：Al/GaN 界面热导$G_1 = 125 \pm 5\ \mathrm{MW/(m^2 \cdot K)}$（由③反演）；GaN 纵向热导率$k_{z2} = 147 \pm 10\ \mathrm{W/(m \cdot K)}$（由④⑤联合反演）；GaN 面内热导率$k_{r2} = 144 \pm 11\ \mathrm{W/(m \cdot K)}$（由⑥反演）；GaN 体积热容$C_2 = 2.7 \pm 0.14\ \mathrm{MJ/(m^3 \cdot K)}$（由④⑤⑥联合反演）；AlGaN 热导率$k_{z3} = 16 \pm 2.5\ \mathrm{W/(m \cdot K)}$（由⑧反演）。同时，低频响应还可约束 Si 衬底的热导率和体积热容等参数。与直接拟合大量单一物理参数相比，这种策略的优势在于先通过无量纲组合参数确定信号真正敏感的对象，再将可测组合参数反演为具有物理意义的单层热导率、体积热容或界面热导。

这一案例更重要的意义在于展示“先判断什么可测，再进行物性反演”的思路。对于多层异质体系，热反射信号往往首先约束由热扩散特征时间、热阻和面积热容形成的组合参数，而不是独立约束所有单层物性。在该具体研究中，SPS 利用不同频率和周期波形构建了多组互补条件；但组合参数先行、通过多工况降低相关性的思想并不依赖于某一特定平台，也可用于 TDTR、FDTR 及其他可改变时间、频率或空间观测条件的方法。

### 4.5 从案例到方法选择：观测信息优先于方法标签

综合上述案例，复杂热反射测量的关键并不是优先选择某一特定技术，而是先明确需要区分哪些热物性参数，再判断现有实验能够提供哪些相互独立的时间、频率、空间或波形信息。本节中 SPS/BO-SPS 案例出现较多，反映的是近年来相关工作的代表性和图例可用性，而不构成方法性能排序。对于同一科学问题，多频 TDTR/FDTR、变光斑或光斑偏移测量、SPS、SDTR 以及跨平台联合方案都可能成立；最终应以参数可辨识性、测量不确定度、实验复杂度和可复现性作为评价依据。

## 5 挑战、趋势与展望

### 5.1 反演不适定性与不确定度量化

第 1.4 节和第 3.2 节已从局部敏感性和参数相关性的角度讨论了热反射参数反演中的可辨识性问题。对于多层异质结构、各向异性材料和复杂器件，仅依靠局部敏感性分析仍难以充分判断反演结果的可靠性。即使实验信号能够与热扩散模型良好拟合，目标函数仍可能存在多个可接受的参数组合；参数间的强相关性、初值选择以及固定输入参数的不确定度也会影响反演结果。因此，复杂体系中的关键问题不再只是获得最优拟合值，而是判断哪些参数能够被实验数据有效约束，并给出相应的不确定范围。

目前，误差传播和 Monte Carlo 采样已用于热反射测量中的参数不确定度评估，贝叶斯反演也为多参数问题提供了更完整的概率描述。贝叶斯方法可将先验信息、实验数据和测量噪声统一到概率框架中，通过后验分布给出参数的可能取值范围及其相关性[91]。相比单一最优值及标准差，这类方法能够更直接地反映参数是否受到数据约束，以及不同参数之间是否存在较强耦合。热反射参数反演也正由单一最优值估计逐步向概率化参数推断拓展。

参数不确定度并不能完全代表测量结果的可信

度。多数反演方法默认所采用的热扩散模型能够充分描述实际传热过程，但当实验尺度接近非扩散输运特征尺度，或体系中存在复杂界面和非平衡效应时，模型本身也可能引入系统偏差。此时，较小的拟合残差并不意味着模型误差可以忽略。对于复杂体系，不确定度分析还应区分参数不确定度与模型失配，并评估当前模型的适用范围，避免利用待测参数补偿模型本身的偏差。

实验设计也应与不确定度分析结合。传统做法通常预先设定调制频率、延迟时间、光斑尺寸或空间偏移量，再对获得的数据进行统一拟合。对于参数耦合较强的体系，增加相似测量条件下的数据未必能够有效提高参数可辨识性。利用敏感性矩阵、参数相关性或 Fisher 信息矩阵，可以在实验前评估不同测量条件对目标参数的约束能力[92]；获得初步数据后，还可根据当前参数的不确定度进一步调整后续实验条件，使实验设计、数据采集、参数反演和不确定度更新形成闭环。相比单纯增加数据量，这种以信息增益为目标的实验策略更适合复杂多参数体系，也为机器学习辅助反演和自适应实验提供了物理基础。

未来热反射数据分析的重点将更多集中于可信参数推断和实验条件的主动优化，而不仅是提高拟合速度或降低残差。将概率反演、模型可信度评估、快速代理模型与自适应实验设计相结合，可根据已有数据动态选择更有信息量的测量条件，在降低参数相关性和不确定度的同时提高实验效率。对于复杂材料和器件，这类方法有望推动热反射测量由离线参数拟合逐步发展为面向可信度和信息增益的闭环测量。

### 5.2 时空分辨率拓展中的尺度效应与模型边界

热反射测量的有效时空分辨能力受到热穿透深度、声子平均自由程、光斑尺寸和金属传感层厚度等多重特征尺度的共同制约。调制频率、光斑尺寸和探测方式的改变不仅影响仪器分辨率，也会改变实验信号对应的主导热输运过程。因此，时空分辨率的进一步提升不能简单理解为向更高频率或更小光斑延伸，而应同时考虑信号来源、输运机制及热扩散模型的适用范围。

在高频测量中，热穿透深度随调制频率升高而减小。当其降低至与 Al 或 Au 传感层厚度相当的尺度时，热量将更多局限于金属传感层及其邻近界面，测量信号对传感层热容和金属/样品界面热导的敏感性增强，而对待测材料本征热物性的敏感性可能下降[94]。减小传感层厚度可以一定程度上缓解这一问题，但也会降低光吸收和热反射信号强度，并增加薄膜连续性、厚度和界面质量带来的不确定性。无传感层热反射技术则通过利用材料自身的光学或电子响应实现直接激励和探测[95-98]，可避免金属传感层及其界面的附加影响，但对材料光学性质、表面状态和信号稳定性提出了更高要求。

当热穿透深度或光斑尺寸接近主要载热声子的平均自由程时，经典扩散近似可能失效，热输运逐步进入准弹道甚至非扩散区间。在此条件下，实验获得的表观热导率可能随调制频率或光斑尺寸变化而发生系统性偏移[99-101]。TDTR 和 FDTR 已被用于分析声子平均自由程分布及重建累积热导贡献[8, 102, 103]，但相关反演通常依赖灰体近似、单弛豫时间等模型假设。对于多层异质结构，界面处还同时存在模式失配、选择性透射和界面散射等过程，使体相声子谱与跨界面输运难以完全解耦[104]。因此，进一步研究跨界面非平衡声子输运，需要将实验反演与非灰体 Boltzmann 输运模型以及第一性原理界面声子输运计算相结合[105]，避免将复杂界面效应简单等效为体材料平均自由程的变化。

低频端的限制则主要来自环境稳定性和信噪比。随着调制频率降低，单周期时间延长，环境温漂、激光功率波动、样品台热稳定性和低频噪声的影响逐渐增强。对于块体材料的低频或稳态绝对热导率测量，3ω 法和稳态量热法等传统方法通常具有更好的稳定性[106, 107]。因此，热反射技术在低频端更适合发挥其非接触、微区分辨和可调热扩散尺度的优势，与电学和量热方法形成互补，而非追求对所有频率区间的覆盖。

空间分辨率的进一步提升则受到远场光学衍射极限的约束。近场光学方法为突破这一限制提供了可能[108]，但当探针—样品间距进入近场范围后，局域电磁场、光子隧穿和局域态密度会显著影响能量沉积和探测信号[109]。此时，热反射信号不再只由热扩散过程决定，而同时依赖探针形貌、探针—样品间距、材料介电函数及电磁边界条件。要从近场信号中可靠提取局部温升和热物性参数，需要建立电磁场与热输运相耦合的多物理场模型，并进一步评估探针扰动和局域光学响应对反演结果的影响。

随着测量尺度继续缩小并逐步面向真实器件，样品几何本身也将成为限制模型适用性的关键因素。曲面结构、厚度非均匀薄膜、微纳器件和柔性电子等体系中的热流路径通常受到局部几何、材料分布和边界条件的共同控制，传统一维或轴对称多层模型难以完整描述其热响应。将有限元正演模型与参数反演相结合，可把实际几何、局部热源和边界条件直接纳入测量模型[110]。这类数值反演框架有望把热反射技术从标准平面样品进一步拓展到复杂工程器件，并为高空

间分辨条件下的局部热物性表征提供更可靠的模型基础。

总体而言，热反射技术的时空分辨率提升将越来越受到“仪器分辨率—输运尺度—模型有效性”三者之间匹配关系的制约。未来的发展重点不应只是追求更高频率或更小光斑，而是建立能够随尺度变化而切换或耦合的多尺度输运模型，并将传感层效应、非扩散输运、近场电磁耦合和复杂几何统一纳入参数反演框架。只有在分辨率提升与模型可信度同步推进的条件下，热反射测量才能进一步拓展到跨界面非平衡输运和真实微纳器件中的局域热传输问题。

### 5.3 器件工况下的多物理场测量与工程应用

随着热反射测量对象由标准薄膜样品扩展至实际器件，实验条件也逐渐由静态表征转向真实工作状态下的原位动态测量。器件运行过程中通常同时存在电学激励、瞬态自热和结构形变等多种物理过程，不同场之间的耦合既会改变实际热输运行为，也可能对热反射信号产生附加影响。因此，面向实际器件的关键问题已不只是提高测量分辨率，而是如何在保持真实工况的同时，实现多物理场响应的有效区分与定量表征。

在宽禁带半导体功率器件中，瞬态自热与载流子输运相互耦合，结温会随开关过程、电流分布和局部功耗动态变化。若测量过程中中断电学偏置，所得温度响应可能难以反映真实工作状态。近年来发展的 on-the-fly 热反射技术通过同步电脉冲与光学探测脉冲，可在器件开关瞬态的特定时间窗口内获取原位结温演化[111]。进一步地，热反射信号还可能包含 Seebeck 效应等电输运过程引入的电致调制分量，通过区分光致热反射与电致热反射信号，有望同时获得温度场和电流密度分布信息[112]。因此，热反射技术正在由单一温度测量向电－热耦合输运表征拓展，而不同物理机制对光学信号贡献的分离仍是需要解决的重要问题。

在柔性电子、聚合物薄膜和先进封装结构中，热输运还常与力学响应耦合。热循环、机械弯折或外加应变会改变界面接触状态和局部结构，进而影响界面热导和面内热扩散；与此同时，热膨胀和热弹性形变也可能对光学信号产生干扰。光热膨胀法可利用表面位移或曲率变化反演热导率、热膨胀系数等参数[113-115]。未来可进一步将热反射测温、光热形变测量与原位力学加载结合，在可控应变或热循环条件下同步分析温度、形变及界面状态变化，为柔性器件和先进封装中的热失效研究提供多物理场实验依据。

从工程应用角度看，热反射技术进一步走向晶圆级扫描、高通量筛查和在线计量，还需要解决采样效率、工艺兼容性、快速反演和测量可靠性等问题。传统 TDTR 依赖机械延迟线扫描，异步光学采样和单发瞬态记录可减少机械扫描；对于 FDTR 和 SPS，多频并行激励和数字锁相检测可提高数据采集效率。另一方面，金属传感层在半导体产线和成品器件检测中可能带来附加界面及工艺兼容性问题，因此基于材料本征光学响应的无传感层测量具有重要潜力，但仍需解决表面状态、光学非均匀性及热反射系数标定等问题。与此同时，物理信息神经网络、代理模型和预训练数据库等方法有望加快复杂结构的参数反演，但仍需结合物理约束、不确定度评估和标准样品验证，以保证结果的可靠性。

总体而言，面向实际器件的热反射测量正在经历由标准样品向真实工况、由单一热场向多物理场、由串行实验室测量向高通量工程表征的转变。未来发展的重点不仅在于提高时空分辨率，更在于实现真实工况下多物理场信号的可分离定量测量，并同步提升采样效率、工艺兼容性和参数反演可信度，从而推动热反射技术进一步服务于功率器件、柔性电子、先进封装及半导体制造过程中的在线热计量。

### 5.4 自然语言驱动的数据分析与智能实验

除测量分辨率和反演算法外，热反射技术的进一步推广还受到分析流程复杂和隐性专业知识依赖的限制。TDTR、FDTR 和 SPS 等方法通常涉及数据整理、样品层结构定义、参数设置、拟合区间选择、敏感性分析及不确定度传播等多个步骤。即使已有成熟代码或专用软件，复杂多层结构的分析仍较依赖使用者经验，不利于流程标准化、高通量处理和跨学科应用。

基于 LLM 的 AI 智能体为这一问题提供了不同于传统机器学习反演的技术路径。已有研究表明，工具增强型 LLM 已开始应用于科学用户设施、自动化实验和复杂仪器操作[116-119]。对于热反射测量，更合理的定位并不是由 LLM 直接生成热物性数值，而是由其承担任务理解、流程编排和工具调用，再由经过验证的热扩散模型、拟合程序和不确定度模块完成具体计算。

据此，可构建“自然语言交互层—AI 智能体编排层—确定性物理计算层”的分层架构。实验人员通过自然语言描述样品结构、已知参数、数据位置和分析目标，智能体将其转换为结构化配置，并依次调用数据预处理、模型计算、敏感性分析、参数拟合和不确定度评估模块，同时保留完整的配置与执行记录。由此，热反射数据分析可由依赖专业软件操作逐步转

向以科学意图为驱动的可复现工作流。

进一步地，当智能体与数据采集和仪器控制接口连接后，还可与 5.1 节所述自适应实验设计形成闭环。系统根据当前数据更新参数敏感性和不确定度，并据此选择新的调制频率、延迟时间、光斑尺寸或空间偏移条件，从而提高后续测量的信息增益和实验效率。需要强调的是，科学计量场景对可靠性的要求较高，此类系统应建立在经过验证的领域代码、严格的参数与单位检查、数据和操作溯源以及明确的人工确认机制之上。未来，热反射智能实验的发展重点应是形成可验证、可追溯且关键操作保留人工确认的分析与实验闭环，而不是以自动化程度本身作为性能指标。

## 6 总结

本文围绕基于热反射的微纳尺度热物性测量技术，系统梳理了 TTR、TDTR、FDTR、SSTR、SDTR 和 SPS 等主流测试手段的物理基础、实验构型、信号采集方案和适用场景。这些方法共享光热激励、热扩散和热反射探测的物理基础，但主要观测变量不同：TTR 记录单脉冲瞬态衰减，TDTR 以泵浦–探测延迟时间解析超快时域响应，FDTR 以调制频率扫描获取频率域幅相信息，SSTR 利用准稳态温升，SDTR 及光斑偏移方法引入横向空间信息，而 SPS 记录周期方波激励下的完整温度波形。各类方法并不是简单的替代关系，而是在不同热扩散尺度、观测维度和参数可辨识性条件下形成的互补测量手段。

在实际应用中，方法选择应综合考虑样品几何尺度、目标热物性、可获得的先验参数和仪器带宽。对于薄膜面外热导率和界面热导，TDTR 与 FDTR 具有成熟应用基础；对于面内各向异性，SDTR 及多种光斑偏移方法可提供直接空间约束；对于低热导或强耦合体系，可根据所需时间尺度和信息维度采用低频 FDTR、多工况 TDTR、SPS、扩展瞬态测量或跨平台联合方案。无论采用哪种方法，测量可信度都依赖合理的热扩散模型、准确的结构参数、敏感性分析和不确定度评估，因此不宜将某一方法简单视为复杂参数反演的通用最优解。

未来热反射技术的发展可概括为四个相互关联的方向。第一，继续拓展时间和空间分辨率，以研究超快热弛豫、准弹道声子输运和局部微纳结构中的热扩散行为。第二，加强与电学偏置、力学加载、近场光学和器件原位测试的结合，使热反射测量更贴近真实工作状态。第三，数据分析将从单一最优参数拟合进一步转向包含敏感性分析、全局可辨识性评估、不确定度量化和物理约束的可信反演框架。第四，人机交互方式有望由源代码和专用软件进一步发展为自然语言驱动的智能体接口，由 AI 负责理解实验意图、编排分析流程和调用工具，而关键数值计算和高风险仪器动作仍由经过验证的确定性程序与明确的安全约束执行。在此基础上，敏感性和不确定度评估还可与数据采集形成闭环，推动热反射实验向可对话、可追溯和自适应的智能实验发展。

总体而言，热反射表征技术已经成为微纳尺度热物性研究的重要工具。其核心优势不仅在于非接触、高时间分辨率和微区测量能力，也在于能够利用调制频率、延迟时间、光斑尺寸和空间位置等实验变量主动改变热扩散路径和参数敏感性。随着实验系统、标定方法、可信反演模型以及自然语言交互和自动化控制逐步融合，热反射测量有望从依赖少数专业人员操作的实验技术进一步发展为更加开放、可复现、可追溯和自适应的热计量平台，并在功能薄膜、异质界面、功率器件、先进封装和半导体工艺监测中发挥更重要的作用。

## 参考文献